\let\latexaddtocontents\addtocontents 
\documentclass[a4paper,onecolumn,11pt,noarxiv]{quantumarticle}
\let\addtocontents\latexaddtocontents
\pdfoutput=1
\usepackage[a4paper,top=2cm,bottom=2cm,left=3cm,right=3cm,marginparwidth=1.75cm]{geometry}
\usepackage{microtype}
\usepackage[english]{babel}
\usepackage[utf8]{inputenc}
\usepackage[T1]{fontenc}
\usepackage{eso-pic}

\usepackage{amsmath, amssymb, amsfonts, mathtools, bbm}
\usepackage{physics}
\usepackage{graphicx}
\usepackage{subcaption}
\usepackage{wrapfig}
\usepackage{float} 
\usepackage{braket}
\usepackage[shortlabels]{enumitem}
\usepackage[dvipsnames]{xcolor}
\usepackage[title]{appendix}
\usepackage{dsfont} 
\usepackage{tikz}
\usepackage{rotating} 
\usepackage{tikz-cd}
\usepackage{multicol}
\usepackage{algorithm}
\usepackage{algpseudocode}
\usepackage[version=4]{mhchem}
\algrenewcommand\algorithmicrequire{\textbf{Input:}}
\algrenewcommand\algorithmicensure{\textbf{Output:}}
\usepackage[colorlinks=true, allcolors=blue]{hyperref}
\hypersetup{
	colorlinks=true,  
	linkcolor=blue,   
	citecolor=blue,   
	urlcolor=blue     
}

\newtheorem{example*}{Example}

\newtheorem{remark*}{Remark}

\usepackage[backend=bibtex,sorting=none,giveninits=true]{biblatex}
\usetikzlibrary{quantikz2}
\usepackage{circuitikz}
\usepackage{multirow}
\DeclareUnicodeCharacter{2061}{}

\usepackage{relsize}

\begin{document}


\title{CovAngelo: A hybrid quantum-classical computing platform for accurate and scalable drug discovery}
\author{Linn Evenseth}
\author{Kamil Galewski}
\author{Witold Jarnicki}
\author{Piero Lafiosca}
\author{Vyom N. Patel}
\author{Grzegorz Rajchel-Mieldzioć}
\author{Martin \v{S}imka}
\author{Michał Szczepanik}
\author{Emil Żak}

\affiliation{BEIT Sp. z o. o., ul.\ Wadowicka 8A, 30-415 Krak{\'o}w, Poland}
\date{\today}

\maketitle

\begin{abstract}
We present a computational platform for modeling chemical reactions in complex molecular environments, focused on ligand–protein binding in drug discovery. The platform implements our new quantum-in-quantum-in-classical (QM/QM/MM) multiscale embedding model that integrates molecular dynamics with a quantum-information-enhanced density matrix embedding theory and quantum chemistry solvers, including explicit solvent. Quantum-information metrics are utilized to generate entanglement-consistent orbitals, enabling a high-accuracy description of strongly correlated regions. The framework supports multiple computational backends, including multi-CPU, NVIDIA multi-GPU architectures, and quantum hardware (IQM, IonQ, IBM) integrated under CUDA-Q, and is designed for compatibility with future fault-tolerant quantum systems.

The new platform’s capabilities are demonstrated by modeling covalent docking of zanubrutinib to Bruton's tyrosine kinase via a Michael addition mechanism, computing the full reaction energy profiles and energy barriers at a reduced computational cost relative to existing methods. As a 2nd-generation anticancer agent, zanubrutinib serves as a proof of concept for covalent inhibitor discovery. Accurate first-principles reaction barrier estimations provided by our method can contribute to reducing false positive and negative rates in drug discovery pipelines. Scalability is validated through benchmarks on GPU clusters, cloud-based CPU infrastructures. We demonstrate integration with quantum devices (up to 20 qubits), alongside resource estimates for fault-tolerant quantum computing, indicating potential speedups of up to 20x. Beyond single reactions, the platform supports the construction of reaction networks in chemical metric space, facilitating ligand screening and systematic exploration of reactive pathways.
\end{abstract}

\section{Introduction}
Accurate computation of binding energies, reaction barriers, and electronic and structural properties of protein-ligand (PL) complexes remains a central challenge in computational chemistry and computer-aided drug design (CADD)~\cite{GhaziVakili2025,Petrova2013,Chen2023}. Molecular recognition is steered by subtle electronic effects, including electron correlation, polarization, charge transfer, dispersion interactions, hydrogen bonding, salt bridges, and, in the case of covalent inhibitors, chemical bond formation. Although these contributions represent only a small fraction of the total electronic energy, they govern the energetics of chemical reactivity and biological function, rendering the task computationally challenging~\cite{wei2024structure}. The energetic differences that discriminate selectivity of a ligand in a protein binding pocket often occur at the sub-kcal/mol level and therefore require accurate quantum-mechanical treatment. Even modest errors can have dramatic consequences: for example, an overestimation of a reaction barrier by 3 kcal/mol changes the predicted dissociation constant of a covalent inhibitor 150-times, potentially leading to false-negative predictions and propagating costly errors throughout the drug-discovery pipeline.

Achieving such a sub-kcal/mol accuracy typically requires high-level quantum-chemical methods whose computational cost grows steeply with system size, making realistic biomolecular simulations prohibitively expensive. This challenge is particularly acute for PL complexes, which may contain thousands of atoms embedded in heterogeneous environments of protein residues, solvent molecules, and cofactors. As a result, conventional quantum chemistry faces a curse of dimensionality when applied to drug-discovery systems~\cite{Santagati2024}. For binding free-energy calculations and robust scoring of PL systems, thorough sampling of the conformational space of both the complex and its environment, typically via classical molecular dynamics, is critical for predictive accuracy~\cite{Ross2023FEPAccuracy}. Consequently, high-level quantum chemical descriptions must be paired with ensemble averaging to prevent model imbalances.

\paragraph{Our proposal.}
We address the problem of accurate and fast modeling of PL interactions for drug discovery. For this purpose, we developed a multiscale computational platform combining first-principles quantum chemistry with classical molecular mechanics to capture both the electronic structure of the reactive region and the complex environment of proteins and solvent. In our approach, a quantum core containing the chemically active region is embedded within a larger quantum subsystem, which is itself immersed in a semi-classical dynamical protein–solvent environment. This hierarchical embedding strategy mitigates the computational curse of dimensionality while preserving an accurate quantum-mechanical description of the binding (reaction) center. High-accuracy electronic structure calculations are performed on the core region using a new type of electronic orbitals, which are optimized using quantum-information metrics, while the surrounding system is simulated using efficient classical high-performance computing architectures, including GPU clusters and wafer-scale engine accelerators~\cite{santos2024breaking, perez2024breakingmold, oppelstrup2025beyond}. 

To further extend the perspective to accessible system sizes, we integrate our computing engine with current and future fault-tolerant quantum computing architectures, unified under an efficient CUDA-Q framework~\cite{CUDAqGithub}. Quantum algorithms offer the potential for polynomial-to-exponential reductions in computational resources compared with classical techniques for electronic-structure problems, providing a path toward tractable simulations of strongly correlated active sites. Thus, although our platform today operates on existing hybrid quantum-classical computing architectures, it is designed to be compatible with future fault-tolerant quantum computers. On that end, we integrate our in-house quantum algorithms for first-principles electronic structure simulations, which offer up to 20$\times$ speedup compared to other methods~\cite{Deka2025}.

First-principles quantum chemistry offers key advantages in this context. Unlike empirical or semi-empirical approaches, ab initio methods allow systematic control over model accuracy and provide transferable descriptions of molecular interactions across diverse chemical systems. In contrast, empirical force fields and fitted interaction models commonly used for hydrogen bonding, entropy, covalent, and non-covalent interactions are often limited by the scope and quality of their training data. Such limitations can hinder the discovery of novel inhibitors whose chemistry lies outside existing parametrization regimes.
Despite that, due to the limited scalability of quantum-chemical methods in modeling PL complexes, other tools often rely on these low-cost density functional theory (DFT) or semi-empirical approaches, such as PM6, which sacrifice predictive accuracy and transferability. Higher-level wavefunction methods, including coupled-cluster (CC) theory, multiconfigurational self-consistent field methods (MCSCF), and density matrix renormalization group (DMRG) techniques, provide systematically improvable descriptions of electron correlation, but become rapidly intractable as system size or accuracy requirements increase. As a result, reliable calculations of binding free energies, activation barriers, and electronic properties for drug discovery remain uncommon in industrial practice. However, accurate prediction of these quantities across chemically diverse ligand and pocket datasets is essential for high-quality virtual screening campaigns, lead optimization, covalent inhibitor design, and generation of high-fidelity datasets for machine-learning (ML)  models~\cite{Pollice2021,Zhavoronkov2019,Chen2023}. 

Our multiscale framework is designed to overcome the scalability and transferability limitations of existing approaches. The primary objective of our design is the generation of high-quality quantum-mechanical data suitable for training ML models and generative artificial intelligence (AI) systems for \textit{de novo} drug discovery.
Our targets are thus next-generation ML potentials and foundation models for chemistry that require training data that are physically consistent, transferable, and derived from robust quantum-mechanical descriptions - an idea that has been recently cultivated across the community~\cite{NEURIPS2023_09f8b246}.

The core of our method is based on first-principles quantum chemistry, delivering high-quality reaction barrier predictions at the cost of computational time. Even with multiscale embedding and optimized core orbitals, a single ligand–protein evaluation takes 30 minutes to a few hours on a laptop, depending on the level of theory. To address this, we enable adjustable compute levels for rapid screening and are developing a physics-informed scoring function for fast virtual screening. The current implementation is therefore best suited for high-quality screening and lead optimization. Our platform is designed to answer a central question: which receptors will a candidate drug bind to, and at what rate? It achieves this by accurately capturing pocket composition and geometry, including explicit solvent and protein environments, while sampling their conformations.

\paragraph{Impact.} Modern drug discovery follows a funnel-shaped pipeline in which millions of candidate compounds are computationally screened and experimentally evaluated to identify a small number of viable drug candidates. Despite advances in molecular docking, high-throughput screening, and ML-based scoring functions, early-stage filtering remains plagued by high false-positive rates~\cite{Shoichet2004}. A major source of this limitation is the insufficient physical fidelity of commonly used computational models. Docking scores typically neglect explicit electron correlation, environmental polarization, and reaction barriers associated with covalent bond formation~\cite{boike2022advances,Chen2023}. In particular, activation barriers and binding free energies strongly influence the potency and selectivity of covalent inhibitors, but are rarely computed with sufficient accuracy. As a result, compounds that appear promising \textit{in silico} frequently fail in downstream biochemical assays.

Improved predictive models for binding energies and reaction barriers can have a measurable impact across the discovery pipeline. Even modest reductions in false-positive rates can significantly decrease experimental screening costs and accelerate the identification of viable leads. Given that the total cost of bringing a single drug to market is commonly estimated to exceed $\$$2 billion, improvements in early-stage computational filtering can translate into substantial economic and social benefits~\cite{Petrova2013}. Similarly, hit-to-lead computational stages are affected by inadequate model quality and computational cost, especially for covalent inhibitors.

Although the present work focuses on PL complexes, the proposed methodology applies broadly to chemical systems in which strongly correlated active regions interact with complex environments. These include catalytic reactions in enzymes, heterogeneous catalysis at surfaces and interfaces, electrochemical processes in batteries, and photochemical reactions in materials. Many chemically and industrially relevant systems, such as the FeMo cofactor (Fe$_7$MoS$_9$C) of nitrogenase for nitrogen fixation, reactive intermediates in cytochrome P450 enzymes~\cite{Goings2022}, transition-metal complexes used in catalysis, and artificial photosynthesis~\cite{Nishioka2023,Mori2025} for water splitting or CO$_2$ reduction, feature strongly correlated electronic structures coupled to complex environments. These systems exemplify chemically complex active sites in which strong electron correlation and environmental effects jointly determine catalytic reactivity~\cite{Anda2016}. Accurate modeling of these systems largely remains beyond the reach of conventional computational methods but is essential for rational catalyst and material design and today represents a major underdeveloped niche~\cite{Li2019,Legeza_2026}. At the same time, these systems represent major areas of interest in modern industry and align with current environmental and sustainability policies~\cite{Fu2024}. Accurate computation of thermodynamic reaction barriers and profiles for these processes could significantly impact materials design by replacing early-stage laboratory screening with predictive computer simulations, thereby accelerating discovery and reducing development costs~\cite{Wang2025}.
Additionally, a reliable reaction energy profile provides first-principles insight into the mechanisms of chemical reactions.

\subsection{Main computational limitations and how do we overcome them}
The design of molecules with desired biological features that are capable of modulating the activity of a target protein is an inverse problem~\cite{GhaziVakili2025}, in which the underlying chemical search space is enormous. For this reason, CADD has historically relied on semi-empirical and fitted models to handle this complexity - an approach that nonetheless carries fundamental limitations.
These limitations are due to the difficulty of simultaneously achieving chemical accuracy, computational scalability, and consistency across chemically diverse systems. Notably, one of the most serious issues concerns ligand datasets associated with a single protein binding pocket that are not modeled with uniform error, which leads to systematic biases and a degradation of predictive performance metrics.

\paragraph{Robust QM/MM embedding strategies.}
DFT is a standard tool for biomolecular modeling, often achieving errors of ~2-3 kcal/mol in benchmarks~\cite{Liang2025}. However, modern functionals remain empirically parameterized and can be computationally demanding for large systems. A transferable description of PL interactions requires systematic treatment of static and dynamic correlation, as well as explicit inclusion of the environment. In practice, this necessitates QM/MM partitioning with a consistent treatment of the interface and environment-induced corrections to the quantum Hamiltonian.

Accurate treatment of electron correlation often requires extending the quantum region beyond the immediate reaction site, especially in the presence of $\pi$-delocalization or dispersion-driven binding. The relevant region typically includes the ligand and nearby residues ($\sim$50–300 atoms), which limits practical calculations to semi-empirical or low-cost DFT methods, reducing predictive accuracy and transferability. These methods are also unsuitable for generating high-quality training data due to their empirical nature. In contrast, wavefunction-based methods offer systematic improvability but scale poorly, making, e.g., CCSD prohibitive for systems with hundreds of orbitals.

To address this, quantum embedding techniques such as Density Matrix Embedding Theory (DMET)~\cite{Wouters2016} have gained traction. These approaches treat a chemically relevant fragment at a high level while representing the environment via an effective bath, significantly reducing computational cost. Embedding frameworks have therefore become attractive for chemical modeling in drug discovery and are increasingly integrated into software platforms~\cite{Bensberg2025,Macetti2021,Rossmannek2023,Battaglia2024,Bickley2025,Ma2024,Shajan2025}. In this work, we adopted a DMET quantum embedding method modified to meet the requirements of modeling chemical reaction and covalent PL interactions.

\paragraph{Limitations of modern computing architectures.}
Even with embedding techniques, the size and complexity of realistic PL systems often exceed the capabilities of current classical computing architectures. This limitation is particularly severe when strong electron correlation or chemical bond formation must be treated explicitly~\cite{Ma2024}. For example, achieving sub-chemical accuracy (1 kcal/mol or better) for reaction barrier energies can already be prohibitively expensive for isolated ligands containing on the order of 200 atoms. Incorporating the surrounding protein environment further increases the computational cost, as it typically includes dozens of amino acids and solvent molecules.

In this work, we address these limitations in two complementary directions. First, we present the aforementioned correlation-aware orbital representations with QM/QM/MM embedding techniques. These approaches reduce the number of degrees of freedom required to achieve a given level of accuracy. Second, we integrate with modern heterogeneous computing architectures, optionally combining classical high-performance (multi-GPU) simulations with quantum computing resources.

\paragraph{Near-Term and Fault-Tolerant Quantum computing integration.}
The classical components of our platform rely on established electronic-structure methods such as DMRG, CCSD, and CASCI, complemented by our efficient implementation of the size-consistent Brillouin-Wigner Second Order Perturbation Theory method (sc-BW2)~\cite{carter2023repartitioned}, which provides favorable performance characteristics for large-scale calculations. The quantum component of the computational pipeline (illustrated in Fig.~\ref{fig:pipeline}) is designed to interface with future fault-tolerant quantum computing architectures. For these platforms, we employ our proprietary quantum algorithms for electronic-structure simulations. 
Many fault-tolerant quantum simulation proposals focus on isolated active sites, neglecting coupling to the molecular environment, and risking discrepancies with experiment. Our design explicitly incorporates environment coupling. This approach allows for scaling consistently with computational resources and benefit from quantum algorithmic speedups~\cite{Santagati2024,Shajan2025,Shajan2025,Goings2022,Deka2025}.

Finally, in line with the constantly increasing computing power of quantum devices~\cite{Shajan2025}, we integrate our workflow with modern QPUs within CUDA-Q framework~\cite{CUDAqGithub}. We enable several backends, including superconducting qubit-based architectures (IBM, IQM) and trapped ion (IonQ), joined with our proprietary circuit optimization techniques.

\paragraph{Related approaches.}
Several computational frameworks for PL modeling employ hybrid QM/MM methodologies, in which a chemically active region is treated quantum mechanically while the surrounding protein and solvent environment are described using classical force fields. Open-source implementations include CP2K~\cite{laino2005efficient,laino2006efficient}, ChemShell~\cite{metz2014c}, LiChem~\cite{kratz2016lichem}, and modular platforms such as ASH~\cite{Bjornsson2026} and FreeQuantum~\cite{Ma2024}. 

Commercial software packages offer related capabilities, including QSite (Schrödinger), QUELO/QuValent platform developed by QSimulate, while emerging companies such as Qubit Pharmaceuticals and FullQubit explore hybrid quantum–classical approaches to molecular simulation~\cite{Li2024}. 
Commercial quantum-chemistry packages, such as  Q-Chem~\cite{Shao2014} or Orca~\cite{ORCA}, also offer basic QM/MM calculation capabilities.
Recent research has also investigated hierarchical QM/QM/MM embedding schemes in which a high-level correlated method is applied to a localized quantum core embedded within a lower-level quantum environment~\cite{Bensberg2025,Macetti2021,Rossmannek2023,Battaglia2024,Bickley2025,Ma2024,Shajan2025}. However, available implementations generally lack scalable strategies for correlated QM-in-QM embedding and do not integrate naturally with emerging fault-tolerant quantum computing architectures.

In particular, existing proposals do not yet enable sufficiently systematic and computationally tractable simulations of chemical bond formation in covalent PL docking calculations~\cite{Chen2023}. To illustrate the need for accurate and correlated QM-in-QM methodologies, we present a case study on a particularly challenging chemical system: an electrophilic addition reaction leading to the C–S bond formation in a $\pi$-delocalized system. As a representative example, we investigate a biologically important Bruton’s tyrosine kinase protein target together with a recently approved FDA inhibitor designed to target cancer-associated mutations of this protein.

\paragraph{Case study: Michael addition for covalent docking simulations.}

\begin{figure}
    \centering
    \includegraphics[width=\linewidth]{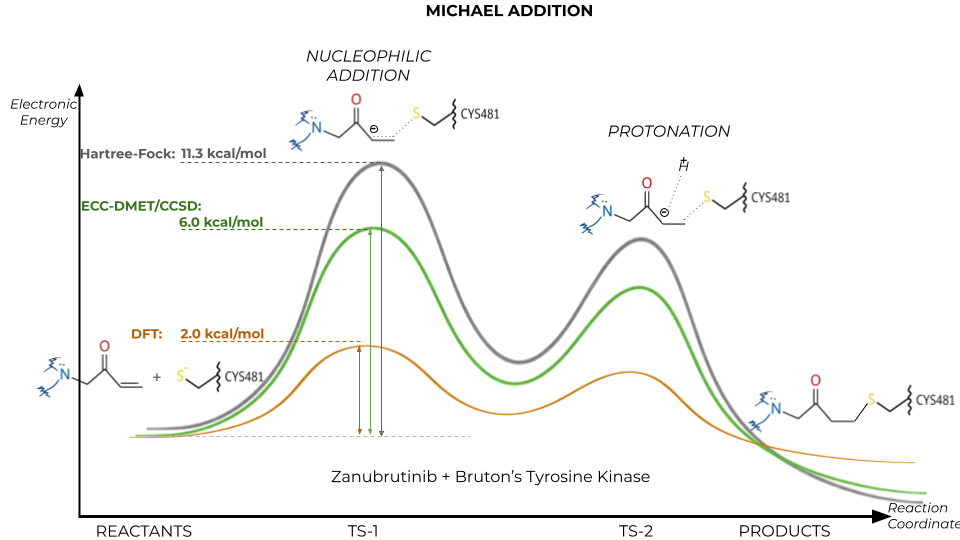}
        \includegraphics[width=0.6\linewidth]{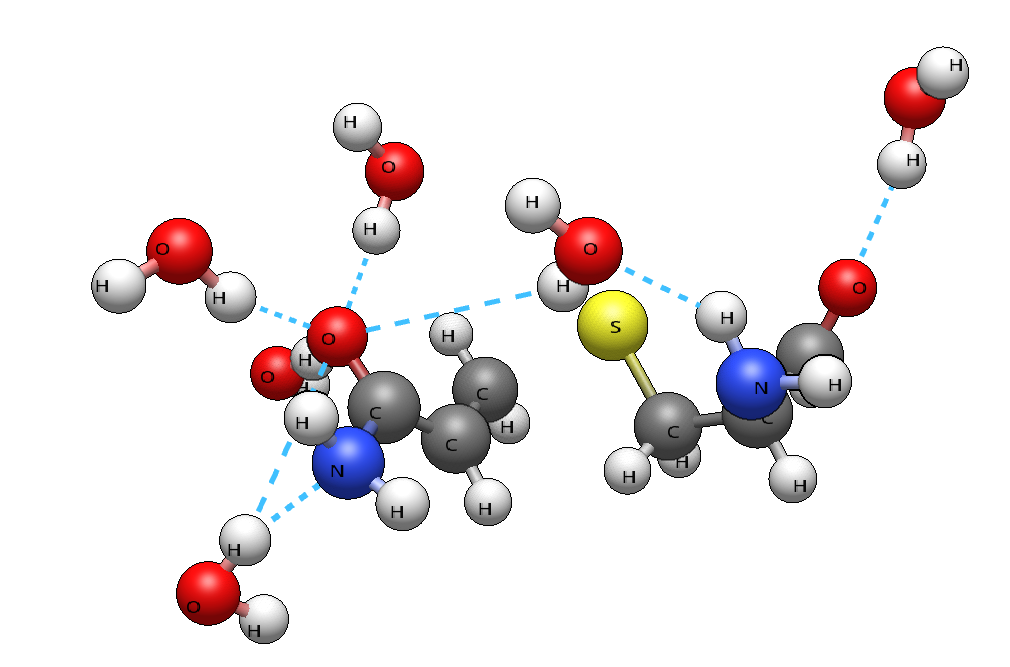}
    \caption{Top: Reaction energy profile for Michael addition~\cite{Roseli2022,Liu2023}. The reaction involves the nucleophilic attack of the sulfur atom of Cys481 on the electrophilic $\beta$-carbon atom of zanubrutinib's acrylamide warhead. The electron-withdrawing effect of the adjacent carbonyl group polarizes the C=C double bond, making the $\beta$-carbon electron-deficient and susceptible to attack by nucleophiles like the cysteine thiol or, more potently, the thiolate anion. Barrier heights calculated within our QM/QM/MM model for several methods are shown: Hartree-Fock (grey), DFT/$\omega$B97X-D3BJ (orange), and our implementation of the new ECC-DMET method (green). aug-cc-pVDZ basis set was used for all methods.
    Bottom: Example geometry of the quantum region, showing explicit water molecules and hydrogen bonds. Here are shown acrylamide warhead of zanubrutinib and cysteine residue of BTK receptor. }
    \label{fig:michael}
\end{figure}

Bruton’s tyrosine kinase (BTK) is a central component in the B-cell receptor signaling pathway and an established therapeutic target for multiple types of cancer ~\cite{Nawaratne2024}. Covalent inhibitors targeting BTK, such as zanubrutinib, exploit the nucleophilic cysteine residue Cys481 located in the ATP-binding pocket of the kinase ~\cite{Guo2019, Nawaratne2024}. These inhibitors contain electrophilic warheads that form an irreversible covalent bond with the thiol group of Cys481 via a Michael addition reaction~\cite{Roseli2022,Liu2023}, as shown in Figure~\ref{fig:michael}. Covalent inhibition offers several pharmacological advantages compared to reversible inhibition, including irreversible inactivation of the target, long residence time, and improved target occupancy even after plasma clearance of the drug~\cite{Chen2023,Singh2025}.

Although molecular docking is widely used in CADD to predict ligand binding modes and interactions, modeling covalent inhibitors presents additional challenges~\cite{Sliwoski2014}. Covalent bonding must account for not only the initial noncovalent binding of the ligand but also for the formation of a new chemical bond between the ligand warhead and the target residue ~\cite{Singh2011, London2014, Wen2019}. Most existing covalent-bonding approaches rely on empirical scoring functions and geometric approximations that may not fully capture the electronic effects involved in covalent bond formation, electron correlation in particular  ~\cite{Scarpino2018, Wen2019,Liu2023}. In this case study, reported in section~\ref{sec:case-study}, we investigated the covalent interaction between zanubrutinib and Cys481 in BTK using the ECC-DMET technique. We calculated the Michael addition reaction energy profile for zanubrutinib+BTK binding and demonstrated that with our method, increased accuracy can be achieved with fewer computational resources compared to other techniques, reducing runtime to minutes instead of hours. The proof-of-concept we provide in this case study provides mechanistic insight into the structural determinants of BTK inhibition and highlights the value of utilizing higher-level electronic structure methods in the study of targeted covalent inhibitors~\cite{Wei2022}. 

We demonstrate scalability of this proof-of-concept through benchmarks on modern NVIDIA's GPU bundles (Hopper H100 and Blackwell B200) and multi-CPU AWS instances, as well as on today's quantum computing devices with up to 20 qubits (IQM Garnet), and give resource estimates for future fault-tolerant machines. The dedicated platform we developed provides a user interface for modeling a single selected chemical reaction, here Michael addition, as well as building networks of chemical reactions between clusters in chemical metric space (sec.~\ref{sec:code}).

\section{Computational model and methodology}
We propose a computational QM/QM/MM embedding model to simulate intra- and inter-molecular interactions, specifically ligand behavior in a protein–solvent environment. The method enables the calculation of energies and electronic properties of molecules, including ligands, proteins, and their complexes, to predict binding affinities, particularly in cases involving chemical bond formation where reaction barrier heights are critical. Our implementation builds upon the central ideas of DMET, but in an adapted and generalized form. In contrast to conventional quantum-in-quantum (QM/QM) DMET formulations~\cite{Wouters2016}, our method uses quantum-information metrics to select and optimize fragment and bath orbitals with a correlated DMET reference state~\cite{Sekaran2023}. As a result, the orbital space required for the correlated calculation can be substantially reduced relative to standard, unoptimized DMET approaches, leading to computational savings. 

\begin{figure}[!h]
    \centering
    \includegraphics[width=0.7\linewidth]{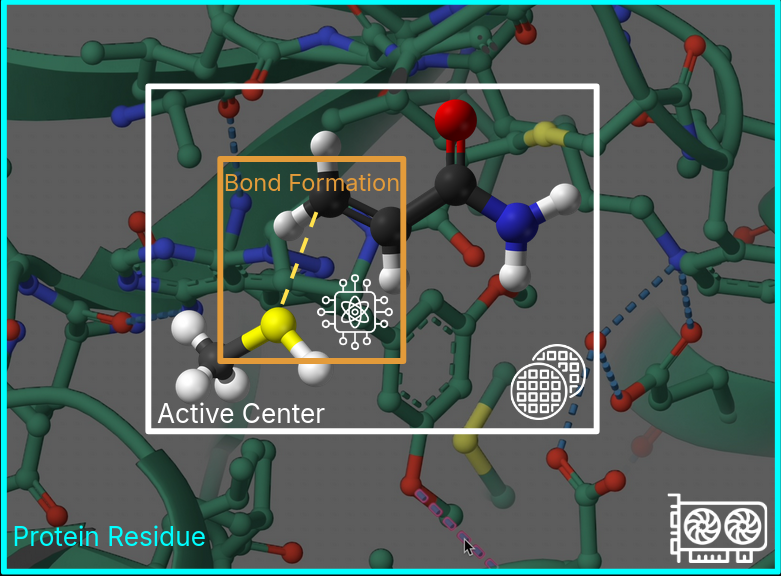}
    \caption{Division of molecular system into classical Protein Residue (semi-classical mechanics with classical computers), Active Center (quantum-mechanics with classical computers), and Bond Formation region (quantum-mechanics with quantum computers).}
    \label{fig:P-L-reaction}
\end{figure}

For the quantum core (impurity) solver, we provide three backend options. These include classical CPU/GPU-based quantum chemistry methods (CCSD, CASCI, sc-BW2~\cite{carter2023repartitioned}, and DFT), a GPU-accelerated quantum circuit simulator implemented with CUDA-Q that exploits modern NVIDIA architectures such as A100, H100, and B200 GPUs, allowing circuit simulations with over 100 qubits in cuTensorNet mode. The third option we provide is execution on quantum processing units offered by quantum hardware manufacturers (IQM, IONQ, IBM) using in-house optimized quantum circuits~\cite{Tulowiecki2024_DiagonalStick, Szczepanik2025}. 

The solver backend is designed to extend along the technological roadmap toward fault-tolerant quantum computing (FTQC). Because FTQC is expected to alleviate the exponential scaling bottleneck associated with electronic structure calculations~\cite{babbush2018, Goings2022}, our computational pipeline is structured to accommodate future large-scale quantum devices. To this end, we have developed and integrated an FTQC algorithm for computing electronic ground-state energies and carried out quantum resource estimation, including extrapolation of the resources required to reach a specified target accuracy, as described in sec.~\ref{sec:FTQC}. Our algorithm achieved up to a 20-fold reduction in T-gate counts compared with existing approaches, potentially shortening the transition from near-term quantum simulations to practical FTQC-based electronic structure~\cite{Deka2025}.

\begin{figure}
    \centering
    \includegraphics[width=\linewidth]{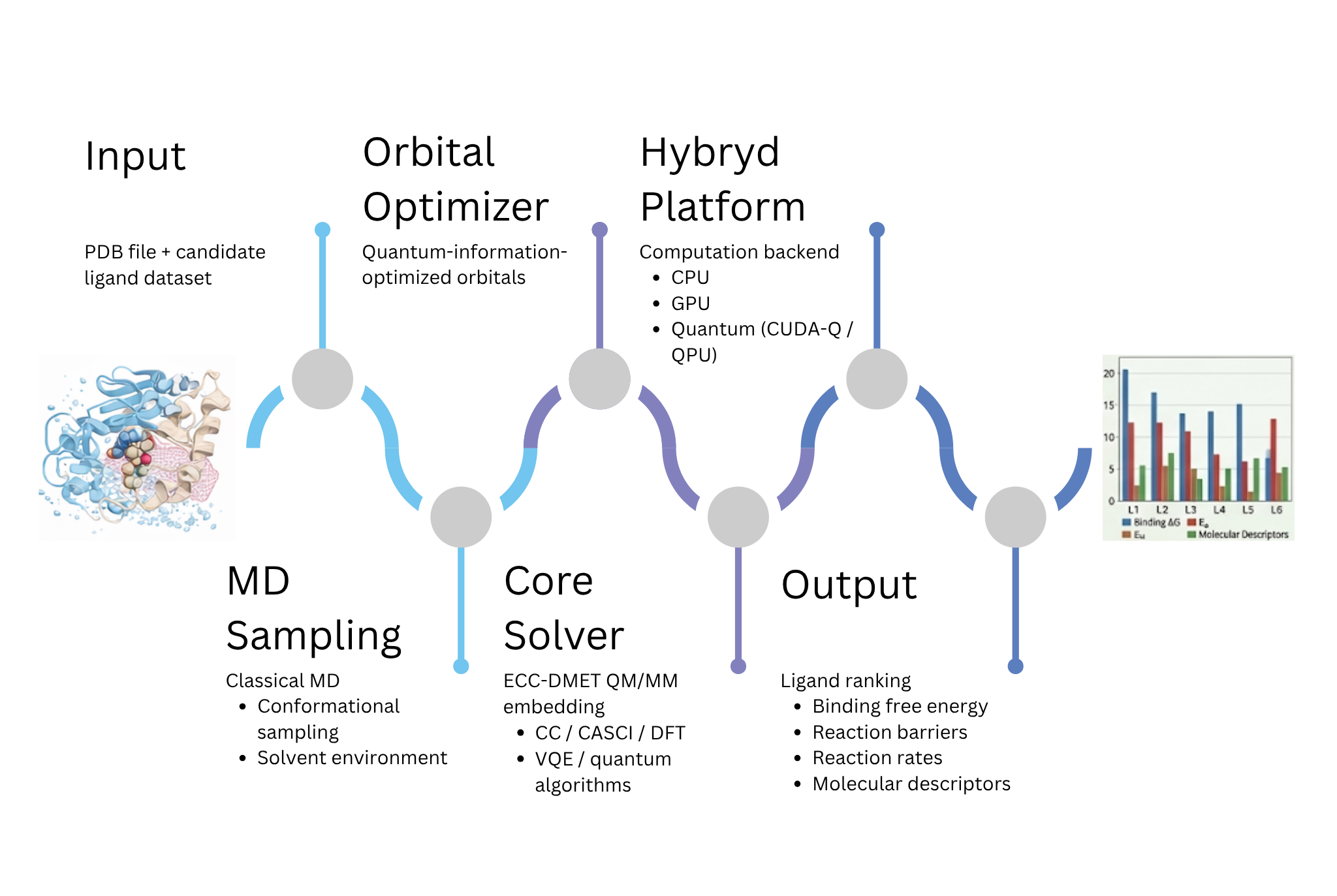}
    \caption{Hybrid QM/QM/MM Computational Pipeline for Modeling Protein–Ligand Reactions.}
    \label{fig:pipeline}
\end{figure}

Our computational pipeline, illustrated in Figure~\ref{fig:pipeline}, consists of a molecular dynamics simulation module coupled with a quantum solver. In an example mode of operation for ranking protein-target inhibitor ligands, the pipeline accepts a protein data bank (PDB) file along with a dataset of ligand structures (SMILES or MOL2) and performs high-quality virtual screening, returning a ranked list of ligands based on their covalent binding affinity.

\subsection{Molecular Dynamics}
\begin{figure}
    \centering
    \includegraphics[width=\linewidth]{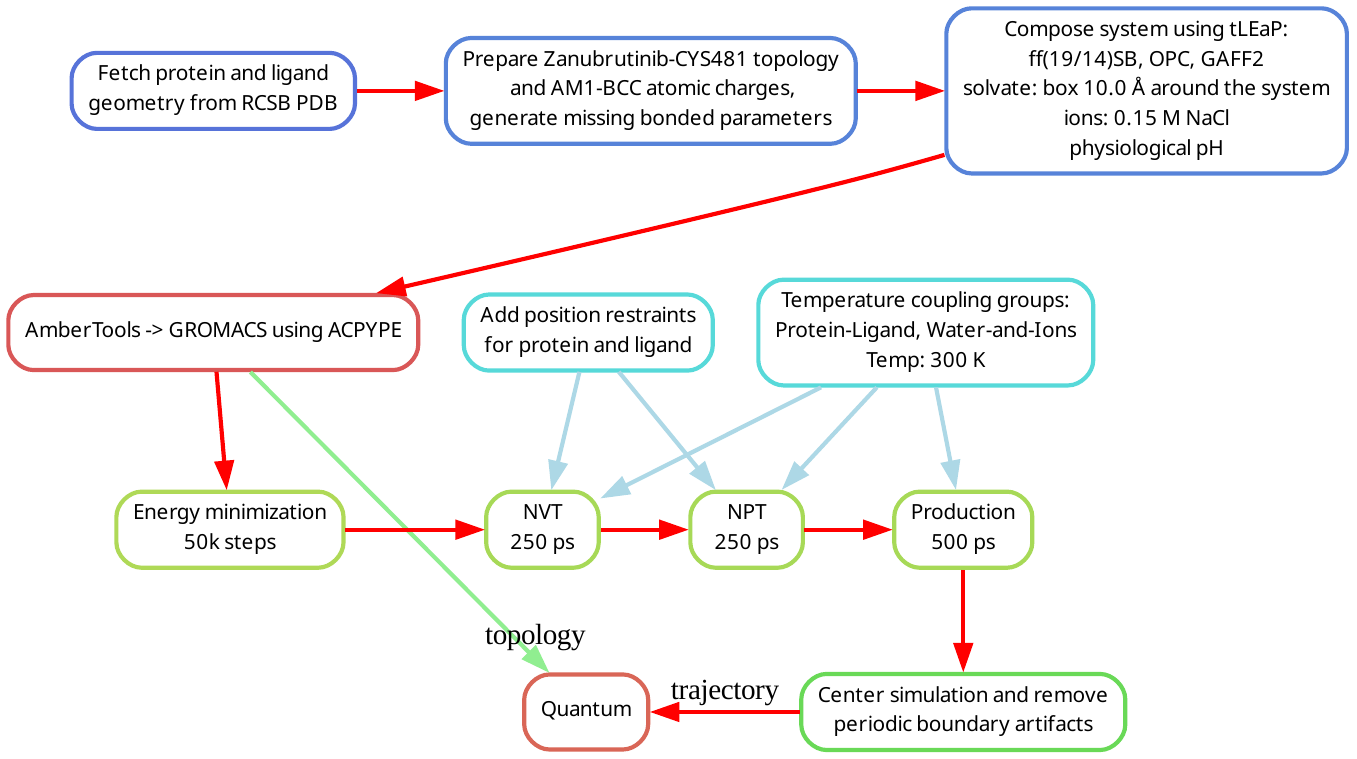}
    \caption{Schematic representation of the molecular dynamics simulation workflow. The red trajectory indicates the system's temporal evolution. Example parameter values were set to case study of zanubrutinib-CYS481 binding, discussed further in sec.~\ref{sec:case-study}}
    \label{fig:md}
\end{figure}

The first stage of the computational pipeline involves molecular dynamics (MD) simulations, utilizing the \texttt{GROMACS}~\cite{Abraham2015,GROMACS2026Manual} package. Prior to the simulations, the topology and coordinate files are converted to the appropriate format using \texttt{ACPYPE}~\cite{SousaDaSilva2012}. The simulation protocol consists of the following steps:
\begin{itemize}
    \item Energy minimization to remove steric clashes and relax the system.
    
    \item NVT equilibration, during which the system temperature was gradually increased to target temperature.
    
    \item NPT equilibration, allowing the system's pressure to equilibrate.
    
    \item A production MD run.
\end{itemize}
All simulations are performed under periodic boundary conditions with a default 2~fs integration time step. A detailed schematic of the workflow is illustrated in Fig.~\ref{fig:md}, with specific parameters' values set to our case study simulation of zanubrutinib-CYS481 binding, discussed further in sec.~\ref{sec:case-study}.
Production MD run delivers geometries passed on to the QM/MM module to calculate ensemble-averaged quantities.

\subsection{QM/MM coupling model}
For a chosen subset of geometries provided by MD simulations, we perform a QM/MM calculation using one of the available quantum solvers. The total energy of the QM/MM system is written as~\cite{Senn2009}:

\begin{equation}
    E_{QM/MM} = E_{QM}+E^\mathrm{int}_{QM-MM}+E_{MM}
\end{equation}
where, $E_{QM}$ and $E_{MM}$ denote the energy of the quantum region and the energy of the classical region, respectively, while $E^\mathrm{int}_{QM-MM}$ corresponds to the interaction of the two regions. The interaction energy is further partitioned into two terms:
\begin{equation}
    E^\mathrm{int}_{QM-MM} = E^\mathrm{ele}_{QM-MM} + E^\mathrm{vdw}_{QM-MM}.
\end{equation}
In the last equation, $E^\mathrm{ele}_{QM-MM}$ is the electrostatic part of the interaction energy, and it is treated at the quantum level using electrostatic embedding~\cite{Senn2009}. The second term describes non-electrostatic van der Waals contributions, which are modeled using a 6-12 Lennard-Jones potential~\cite{LJpotential}. 

Let $R_A$, $Q_A$ denote the position and charge of the $A$-th nucleus, with the first $N_{QM}$ nuclei belonging to the QM region. Additionally, let $\eta$ be the total number of electrons in the QM region and $N$ the total number of spatial orbitals.  The full Hamiltonian operator, by setting $E_{MM}=0$, can then be expressed as
\begin{equation}
    H = \sum_{p,q=1}^{N}\sum_{\sigma \in \{\alpha,\beta\}}\tilde{h}_{pq}a_{p\sigma}^\dag a_{q\sigma} + \frac{1}{2}\sum_{p,q,r,s=1}^{N} \sum_{\sigma,\tau \in \{\alpha,\beta\}}g_{pqrs}a_{p\sigma}^\dag a_{r\tau}^\dag a_{s\sigma} a_{q\tau},+ \tilde{V}_{nuc}
    \label{eq:hemb0}
\end{equation}
where 
\begin{equation}
    \tilde{V}_{nuc} = \sum_{A>B}^{N_{QM}}\frac{Q_AQ_B}{R_{AB}}+\sum_{A=1}^{N_{QM}}\sum_{B=N_{QM}+1}^{N_{QM}+N_{MM}}\frac{Q_AQ_B}{R_{AB}}+\sum_{A=1}^{N_{QM}}\sum_{B=N_{QM}+1}^{N_{QM}+N_{MM}}4\epsilon_{AB}\left\{\left(\frac{\sigma_{AB}}{R_{AB}}\right)^{12}-\left(\frac{\sigma_{AB}}{R_{AB}}\right)^{6}\right\}
    \label{eq:QMMM-nuclear}
\end{equation}
and
\begin{equation}
    \tilde{h}_{pq} = h_{pq}  -\sum_{A}^{N_{MM}}\bra{p}\frac{Q_A}{|\boldsymbol{R}_A-\boldsymbol{r}|}\ket{q}.
\end{equation}
Here, $h_{pq}$ and $g_{pqrs}$ denote the standard 1- and 2-electron integrals, $R_{AB}=|\boldsymbol{R}_A-\boldsymbol{R}_B|$, $\boldsymbol{r}$ is a position of the electron and $\epsilon_{AB}$, $\sigma_{AB}$ are Lennard-Jones potential parameters. For consistency, these parameters are taken to be the same as those used in MD simulation.

We implemented several methods to define the QM region. In the simplest approach, the quantum region consists of residues that have at least one atom inside a sphere defined by a user-specified center and radius, e.g., protein pocket center. Alternatively, instead of providing a single center, the user can specify a set of atoms. In that case, the quantum region consists of all residues within the specified distance from these atoms. For a more customized choice, we also enable using \texttt{MDAnalysis}~\cite{MichaudAgrawal2011,Gowers2016} queries directly. Our method also enables using Conductor-like Polarizable Continuum Model (C-PCM) \cite{barone1998quantum}.

When dealing with PL systems, it naturally happens that the QM/MM region boundary crosses some chemical bonds. To properly describe bonds between the QM and MM region, we resort to the link atoms approach~\cite{Field1990,Singh1986,Senn2009,Lin2006}. For each broken single bond, we introduce a hydrogen atom (link atom), which is not present in the MD calculation. Let $\boldsymbol{R}_Q$ and $\boldsymbol{R}_C$ denote positions of atoms $Q$ from the quantum region and $C$ from the classical region in the broken bond and $\hat{\boldsymbol{R}}_{Q-C} = \frac{\boldsymbol{R}_Q-\boldsymbol{R}_C}{|\boldsymbol{R}_Q-\boldsymbol{R}_C|}$. The position of the link atom is given by~\cite{Lin2006}:
\begin{equation}
    \boldsymbol{R}_{linker} = \boldsymbol{R}_Q + l_{QH}\hat{\boldsymbol{R}}_{Q-C}
\end{equation}
where $l_{QH}$ is a typical bond length~\cite{crc2005} between the atom from the quantum region and hydrogen, as shown in tab.
\begin{table}[h]
    \centering
    \begin{tabular}{c|c|c|c|c|c|c|c|c|c}
        Element &F&O&N&C&Cl&Br&I&P&S  \\
        \hline
        $l_{QH}$ [\AA]&$0.92$&$0.96$&$1.02$&$1.09$&$1.28$&$1.41$&$1.61$&$1.42$&$1.34$
    \end{tabular}
    \caption{Typical covalent bond lengths between and hydrogen and various elements found in small gas-phase molecules~\cite{crc2005}.}
\end{table}
To avoid the problem of over-polarization by a classical charge too close to the QM region, the charges from atoms directly bound to the active region are set to zero, also known as the Z1 method~\cite{Z1scheme,Lin2006}.

For a given MD frame, the energy barrier is calculated as the difference between transition state geometry and  the pre-complex geometry:
\begin{equation}\label{eq:energy_diff}
    \Delta E_i = E^{TS}_i-E^{pre-complex}_i.
\end{equation}
In order to obtain the corresponding geometries, we perform an energy scan, varying the distance between two user-specified atom, one from the ligand and one the from the protein. For each distance, the energy is minimized with respect to all other degrees of freedom inside the QM region, while the positions of MM atoms are frozen. 
Note that in this approach, the MM energy $E_{MM}$ remains constant and does not contribute to energy barrier $\Delta E_i$. All geometry optimizations are performed using \texttt{geomeTRIC}~\cite{Wang2016} software accessed via \texttt{pyscf.geomopt.geometric\_solver} module. The default method for geometry optimization is DFT with $\omega$B97X-D3BJ functional using aug-cc-pVDZ basis with density fitting. 
We also include explicit solvent molecules in the quantum-chemical calculation to account for hydrogen bond formation, polarization, and stabilization of the TS as show in Fig.~\ref{fig:michael}. In future releases, we plan to extend the QM/MM framework by incorporating Intrinsic Reaction Coordinate (IRC) calculations~\cite{Ishida1977} to obtain a more accurate energy profile along the entire reaction path.


The final energy barrier estimation is obtained by averaging the values obtained from different MD geometries
\begin{equation}
    \Delta E = \frac{1}{M}\sum_{i=1}^{M}\left(E_{i}^{TS}-E_{i}^{pre-complex}\right),
\end{equation}
where $M$ is the number of MD geometries samples.

\subsection{Quantum-chemical model for the active center}

\subsubsection{Quantum-in-Quantum embedding}
The quantum-chemical model for the active center (cf. Fig.~\ref{fig:P-L-reaction}) is defined by its division into a quantum core region and a quantum environment (QM/QM embedding). The active center itself is embedded in a classical (protein+solvent) environment and described at the QM/MM level discussed in the previous section. The purpose of the QM/QM division is to treat the core region using a high-accuracy quantum-chemistry method implemented on CPU/GPU or QPU backends, while modeling the surrounding region using a correlated wavefunction method of moderate computational cost. The effect of the protein environment is included in both layers of the quantum embedding model through an electrostatic and dispersion QM/MM coupling as given in eq.~\eqref{eq:QMMM-nuclear}.

To achieve this, we design a new embedding framework based on DMET. DMET is a wavefunction-based quantum embedding approach designed to treat strongly correlated subsystems of large quantum systems with high accuracy while maintaining the general computational cost manageable. In DMET, an expensive correlated electronic-structure method is applied only to a small active region of the system, while the entire system is described at a lower level of theory, typically mean-field. This multilevel strategy enables accurate calculations for systems whose full many-body treatment would otherwise be computationally prohibitive.

\subsubsection{Summary of our contribution}\label{sec:summary_contribution}
In contrast to standard DMET, our approach utilizes a correlated reference wavefunction rather than a mean-field one, enabling a more accurate description of global electron correlation. While conventional DMET constructs bath orbitals from a mean-field one-particle density matrix, we instead introduce quantum-information-optimized orbitals (QIOs), designed to balance spatial locality with maximal separability of the fragment-plus-bath subsystem from the environment under a correlated reference (Alg.~\ref{alg:ECCDMET}). A singular value decomposition of the fragment-environment block of the density matrix defines the optimal bath subspace within an iterative orbital-optimization loop, where both fragment and bath orbitals are updated self-consistently. Incorporating quantum-information metrics, including single-orbital entropies, mutual information, and cumulants, incurs only modest computational overhead. In return, we obtain an orbital basis that reflects both chemical structure and entanglement patterns. Such orbitals define an effective system partitioning that captures the dominant correlation effects within a compact cluster Hamiltonian, thereby making it suitable for accurate many-body solvers.

\begin{algorithm}[H]
\caption{ECC-DMET algorithm.}
\label{alg:ECCDMET}
\begin{algorithmic}[1]
\Require Starting point $\zeta_0$; fragment set $F$ (and $\mathcal{B}_F$); thresholds:$\tau_B$,$\epsilon_J$,$\epsilon$; weights $\vec{\omega^{(0)}},\vec{\omega^{(1)}}$
\Ensure Bath space $\mathcal{B}$, cluster Hamiltonian $\tilde{H}$ in optimized orbitals basis $\mathcal{C}$, the total DMET energy of the system $E_{tot}$.

\State Initialize orbital rotation: $\mathbf{U}(\zeta_0)\leftarrow \mathbf{I}$; choose fragment orbital set $\mathcal{B}_F$.
\State Compute the reference correlated state $\ket{\Psi_0}$ and evaluate quantum-information quantities: $\{S_i\}$ and $\{I_{ij}\}$, as well as RDMs $\gamma$ and $\Gamma$.
\State Set $t\rightarrow t+1$ and set the orbital rotation using an incremental unitary:
\[
    \mathbf{U}(\zeta_t)\leftarrow \mathbf{U}(\zeta_{t-1})\,\Delta\mathbf{U},
\]
using Covariance Matrix Adaptation Evolution Strategy or gradients $\nabla_{\zeta}\mathbf{U}$, if available.
\State Preselect bath candidates by ranking environment orbitals using the following scoring function:
\[
\mathcal{J}_j^{(0)}
=
\omega_0^{(0)}\mathcal{W}_j
+\omega_1^{(0)}S_j
+\omega_2^{(0)} \sum_{i,k\in F}\big|\Lambda_{ijik}\big|
+\omega_3^{(0)} \sum_{i\in F} G_{ij},
\qquad j\in \mathcal{E},
\]
where
\[
    \mathcal{W}_j = \sum_{i\in F} I_{ij},
    \qquad
    S_j = -\mathrm{Tr}\!\left[\rho^{(1)}_j\ln\rho^{(1)}_j\right], \qquad j=1,2,...,D_{E}
\]
choosing $D_{\tilde{B}}$ largest components to produce candidate bath space $\mathcal{B}_{\tilde{B}}$.
\State Perform SVD of the (candidate-bath)-fragment coupling block of 1-RDM:
\[
    \mathbf{\gamma}_{\tilde{B}F} = \mathbf{Q}\,\mathbf{\Sigma}\,\mathbf{V}^{\dagger}.
\]
and select bath orbitals from the singular spectrum: keep columns $\{\mathbf{Q}_{j}\}$ such that $\sigma_j^2>\tau_B$, to define the bath space $\mathcal{B}$:
\begin{equation}
        \mathcal{B}=\mathrm{span}\!\left(\left\{\mathbf{Q}_{:j}\right\}_{j=1}^{D_B}\right),
        \qquad D_B \geq D_F.
    \end{equation}
\State   Repeat steps 3--5 until convergence of the objective function
    \begin{equation}
        \mathcal{J}^{(1)}(\zeta_t')
        =
        -\omega_0^{(1)}\sum_{j\in F} S_j
        -\omega_1^{(1)}\sum_{j\in F}\sum_{k\in B} I_{jk}
        +\omega_2^{(1)}\sum_{s\in C}\sum_{k\in \mathcal{E}} I_{sk}
        +\omega_3^{(1)}\left\|\gamma_{C\mathcal{E}}\right\|^2
        +\omega_4^{(1)}\left\|\Lambda_{C\mathcal{E}}\right\|^2
        ,
    \end{equation}
    i.e., $|\mathcal{J}(\zeta_t)- \mathcal{J}(\zeta_{t-1})|< \epsilon_{\mathcal{J}}$
    or reaching the maximum number of iterations.
    \State Run impurity DMET with cluster Hamiltonian defined by eq.~\eqref{eq:hemb}:
    \begin{equation}
    \hat{H}_{emb}=\hat{P} \hat{H}\hat{P}
\end{equation}
where $\hat{P} = \sum_{j=1}^{D_F}\ket{\tilde{f}_j}\bra{\tilde{f}_j}\otimes \sum_{l=1}^{D_B}\ket{\tilde{e}_j}\bra{\tilde{e}_j}$ is the projector onto the $D_F+D_B$ \textit{fragment+bath} orbitals. Return the total energy $E$.
\end{algorithmic}
\end{algorithm}

\subsubsection{ECC-DMET}\label{sec:ECC-DMET}
Our ECC-DMET protocol retains the basic logic of DMET~\cite{Wouters2016} but replaces embedding with a mean-field reference state by a correlated construction of the impurity space. Rather than relying exclusively on a single-determinant reference and a hand-picked set of fragment orbitals, we define the embedding from localized orbitals and use a correlated quantum-information metric to identify the environment degrees of freedom that remain entangled with the active center. In this way, the fragment, bath, and inactive spaces are determined systematically from the electronic structure of the system itself. This is particularly important for systems in which covalency, near-degeneracy, spin polarization, charge transfer, long-range correlation, or electron delocalization prevent a reliable by-hand selection of fragment orbitals and the description of the whole system by a mean-field reference wavefunction.

We begin the construction of our quantum-chemical model by embedding the active center in the surrounding protein and solvent environment, whose principal effect on the active center is incorporated through an external embedding field. The active center is described quantum-mechanically, and it defines the DMET system, in which we distinguish a fragment space containing the chemically relevant reaction center. The Hilbert space for the active center is partitioned as
    \begin{equation}
        \mathcal{H} = \mathcal{H}_\mathcal{F}\otimes \mathcal{H}_\mathcal{B} \otimes \mathcal{H}_\mathcal{E},
        \label{eq:hilbert-paritioning}
    \end{equation}
where $ \mathcal{H}_\mathcal{F}$ denotes the fragment space, $\mathcal{H}_\mathcal{B} $ the bath space, and $\mathcal{H}_\mathcal{E}$ the inactive core and environment space. The fragment contains the orbitals expected to carry the dominant static correlation and the key charge and spin rearrangements. The bath contains the subset of environment orbitals that is entangled with the fragment and must therefore be treated explicitly together with it. The remaining orbitals are integrated out and enter the embedded problem only through an effective one-body contribution. 

The embedded electronic ECC-DMET Hamiltonian is constructed in the fragment-plus-bath space, \(\mathcal{C}=\mathcal{F}\cup\mathcal{B}\), and written as
\begin{equation}
    H_{emb} = \sum_{p,q\in \mathcal{C}}\left(\tilde{h}_{pq}+\sum_{r,s\in \mathcal{E}} (g_{pqrs}-g_{psrq}) \gamma_{rs}^{\mathcal{E}\mathcal{E}}\right)\tilde{d}_p^\dag \tilde{d}_q + \frac{1}{2}\sum_{p,q,r,s\in \mathcal{C}}g_{pqrs}\tilde{d}_p^\dag \tilde{d}_{r}^\dag \tilde{d}_s \tilde{d}_q+ \tilde{V}_{nuc},
    \label{eq:hemb}
\end{equation}
where $\gamma_{rs}^{\mathcal{E}\mathcal{E}}$  denotes 1-electron RDM of environment ($\mathcal{E}$) orbitals, $g_{pqrs}$ are 2-electron integrals in chemists' notation, and $
\tilde V_{nuc}$ contains Coulomb interaction between nuclei and the Lennard-Jones dispersion interaction. Orbitals $\tilde{d}_p^\dag= \mathbf{U}(\zeta)d_p^\dag$ are constructed by rotating fragment localized orbitals $d_p^\dag$ by a unitary derived from quantum-information metrics quantifying fragment, bath, and environment orbital entanglement. The Schr\"odinger equation constructed with the active-space Hamiltonian written in eq.~\eqref{eq:hemb} can be written as
\begin{equation}
\hat H_{\mathrm{emb}}\lvert\Psi\rangle = E\lvert\Psi\rangle,
\qquad
\hat H_{\mathrm{emb}} = \hat P\hat H\hat P ,
\label{eq:projection}
\end{equation}
where $
\hat P
= \sum_{j=1}^{D_F} \lvert f_j\rangle\langle f_j\rvert
\otimes
\sum_{\ell=1}^{D_B} \lvert \tilde e_\ell\rangle\langle \tilde e_\ell\rvert
$ is a projector onto $(D_F + D_B)$ orbitals; fragment ${|f_j\rangle}$ and bath ${|\tilde e_\ell\rangle}$. 
Eq.~\eqref{eq:projection} can be solved for the electronic energy levels with several solvers and backends, as discussed in sec.~\ref{sec:core-solver}.

\begin{figure}[!ht]
    \centering
    
    \begin{minipage}{0.3\linewidth}
        \centering
        \includegraphics[width=\linewidth,angle=90]{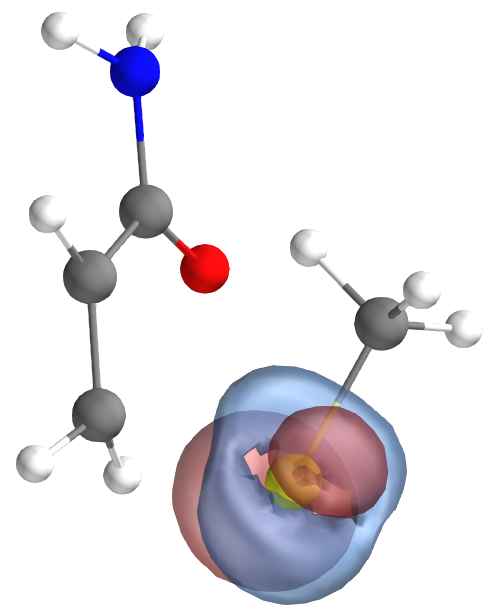}
        \vspace{0.3cm}
        
        \includegraphics[width=\linewidth,angle=90]{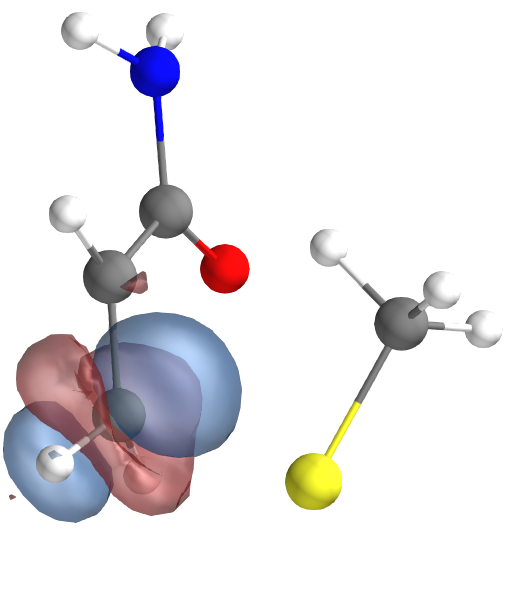}
    \end{minipage}
    \hspace{3cm}
    \begin{minipage}{0.3\linewidth}
        \centering
        \includegraphics[width=\linewidth,angle=90]{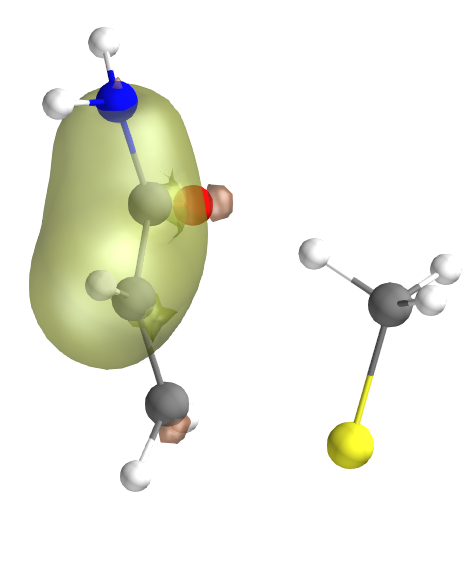}
        \vspace{0.3cm}
        
        \includegraphics[width=\linewidth,angle=90]{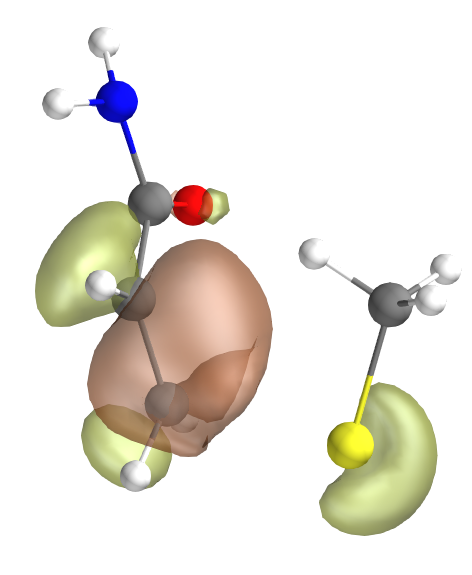}
        \vspace{0.3cm}
        
        \includegraphics[width=\linewidth,angle=90]{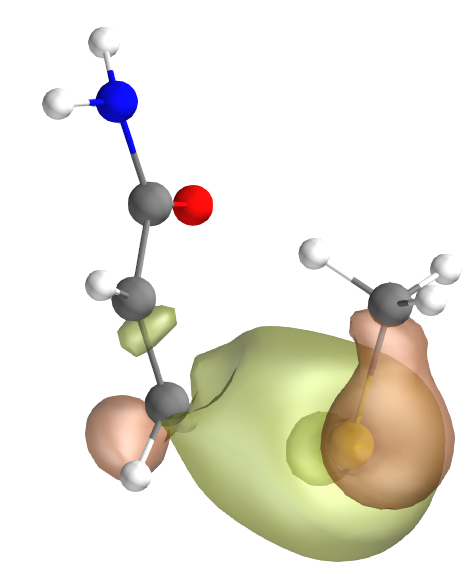}
    \end{minipage}

    \caption{Example fragment (right, red-blue) and bath (left, yellow-brown) orbitals for acrylamide - methanethiol complex at transition state.}
    \label{fig:fragment-bath-orbitals}
\end{figure}

The quality of \(\hat{H}_{\mathrm{emb}}\) depends critically on the choice of orbitals used to construct it. 
We therefore do not work directly with delocalized canonical molecular orbitals. 
Instead, the occupied orbitals and virtual orbitals are first transformed into a localized representation $\lbrace d^{\dagger}\rbrace$ related to canonical molecular orbitals $\lbrace a^{\dagger}\rbrace$ through a linear transformation 
\begin{equation}
    a_m^{\dagger}=\sum_{k=1}^{D}U^{(L)}_{km}d_k^{\dagger}
    \label{eq:localization}
\end{equation}
Localization yields a chemically transparent orbital picture in which individual orbitals can be associated with the active center, nearby ligands, lone pairs, or acceptor and donor functions in the immediate environment. An example set of localized orbitals used in our work to model Michael addition in covalent binding to BTK receptors is shown in Figure~\ref{fig:fragment-bath-orbitals}.

Our approach generalizes the idea of localized orbitals to account for their entanglement. We therefore optimize the $\mathbf{U}^{(L)}$ unitary given in eq.~\eqref{eq:localization} to form an orbital basis that reflects the correlation structure within the system while retaining, to a reasonable extent, the localization property, with details described further on.  

\paragraph{Fragment orbitals selection.}
Choosing fragment orbitals often has to rely on chemical intuition. In our method, we avoid the need for this step, which often depends on manual inspection of the system. 
Instead, the orbitals $\lbrace d^{\dagger}_j\rbrace$  are ranked using the following scoring function:
    \begin{equation}
        \mathcal{J}_j^{(0)}(\zeta)
        =
        \omega_0^{(0)}\,\mathcal{W}_j
        +\omega_1^{(0)}\,S_j
        +\omega_2^{(0)} \sum_{i,k\in F}\big|\Lambda_{ijik}\big|
        +\omega_3^{(0)} \sum_{i\in F} G_{ij},
        \qquad j\in \mathcal{E},
        \label{eq:scoring-functional-frag}
\end{equation}
where
\[
\mathcal{W}_j = \sum_{i\in F\cup \mathcal{E}} I_{ij}, 
\quad
S_j = -\mathrm{Tr}[\rho^{(1)}_j\ln\rho^{(1)}_j],
\]
\[
\Lambda_{ij,kl} = \Gamma_{ij,kl} - (\gamma_{ik}\gamma_{jl}-\gamma_{il}\gamma_{jk}),
\quad
G_{ij} = \int d^3r\, |\phi_i^*(r)\phi_j(r)|.
\]
Here, $I_{ij}$ is the mutual information based on von Neumann entropy. 

The scoring function given in eq.~\eqref{eq:scoring-functional-frag} is parametrized by $N(N-1)/2$ parameters $\zeta$ defining orbital rotation, which can be optimized as discussed later. Non-negative weights $\omega^{(0)}_0,\cdots,\omega^{(0)}_3$ are determined in benchmark calculations, optimized for a given class of problems. Choosing $D_{F}$ largest components of $ \mathcal{J}_j^{(0)}(\zeta)$ produces a candidate fragment space composition $\mathcal{B}_{\mathcal{F}}$.
Naturally, our design also enables the traditional formulation, without orbital optimization, with localized orbitals generated with Boys~\cite{boys1960construction,foster1960canonical}, Pipek-Mezey \cite{pipek1989fast}, and other popular methods~\cite{li2014localization}.

\paragraph{Bath orbitals selection.}
In standard DMET, bath is constructed by diagonalizing the environment block of the 1-RDM $\mathbf{\gamma}^{\mathcal{E}\mathcal{E}}$.
However, our general model adopts a somewhat more refined approach that captures electron correlation at a multipartite level, in contrast to one-particle RDM-based criteria. The improved quality of the orbital selection metric is consistent with the correlated nature of the reference wavefunction, which we employ to maintain model quality consistency.
Accordingly, we construct a scoring functional similar to that given in eq.~\eqref{eq:scoring-functional-frag}, but with $\mathcal{W}_j=\sum_{i\in F} I_{ij}$ for selecting those bath orbitals that are most strongly entangled with the whole fragment.
The candidate environment subspace for bath construction is formed by selecting the ${D}_\mathcal{B}$ largest components of the environment scoring vector $\vec{\mathcal{J}}^{(0)}$ to form the candidate bath space $\mathcal{B}_{B'}$. In this way, we score candidate bath orbitals based on their multipartite entanglement and electron correlation with the fragment and the environment.

Bath orbitals are then constructed from the environment-fragment block of the 1-body RDM
    \begin{equation}
        \mathbf{X}(\zeta) \coloneqq \gamma_{\tilde{B}F}(\zeta),
    \end{equation}
represented with the following block structure
    \begin{equation}
        \gamma
        =
        \begin{pmatrix}
            \gamma_{FF} & \gamma_{F\tilde{B}'} & \gamma_{FE} \\
            \gamma_{\tilde{B}'F} & \gamma_{\tilde{B}'\tilde{B}'} & \gamma_{\tilde{B}'E} \\
            \gamma_{EF} & \gamma_{E\tilde{B}'} & \gamma_{EE}
        \end{pmatrix},
    \end{equation}
where fragment is denoted by $F$, candidate bath subspace by  $\tilde{B}'$, and remaining environment by $E$.
Optionally, one can apply a Householder transformation~\cite{Shajan2025} to $\gamma$ to maximally decouple the fragment orbitals from the environment. The subsequent step involves computing an SVD
    \begin{equation}
        \mathbf{X} = \mathbf{Q}\mathbf{\Sigma}\mathbf{V}^{\dagger},
    \end{equation}
    and defining the bath space as $\mathcal{B} = \mathrm{span}\!\left(\left\{\mathbf{Q}_{:j}\right\}_{j=1}^{D_B}\right)$. In doing so, we choose the bath size $D_B$ using a singular-value criterion, i.e, select the smallest $D_B$ such that the discarded singular weight is below a threshold:
    \begin{equation}
        \sum_{j>D_B}\sigma_j^2 < \tau_B,
    \end{equation}
    where $\{\sigma_j\}$ are singular values of $\mathbf{\Sigma}$. With a correlated DMET reference state, the MacDonald's theorem~\cite{MacDonald_1933} is no longer valid, and the number of bath orbitals can exceed the number of fragment orbitals.
The correlated reference DMET state that replaces the standard mean-field reference is a multi-configuration state written in second-quantized form as
\begin{equation}
\lvert\Psi_0\rangle
=\sum_{n_1,\ldots,n_N}\psi_{n_1\cdots n_N}\lvert n_1\cdots n_N\rangle,
\end{equation}
which in our case is obtained from DMRG, CCSD, or MP2 calculations. Thus, all supplementary quantum-information quantities used for constructing the ECC-DMET Hamiltonian given in eq.~\eqref{eq:hemb} are derived from a correlated wavefunction.

\begin{figure}[H]
    \centering
    \includegraphics[width=0.7\linewidth]{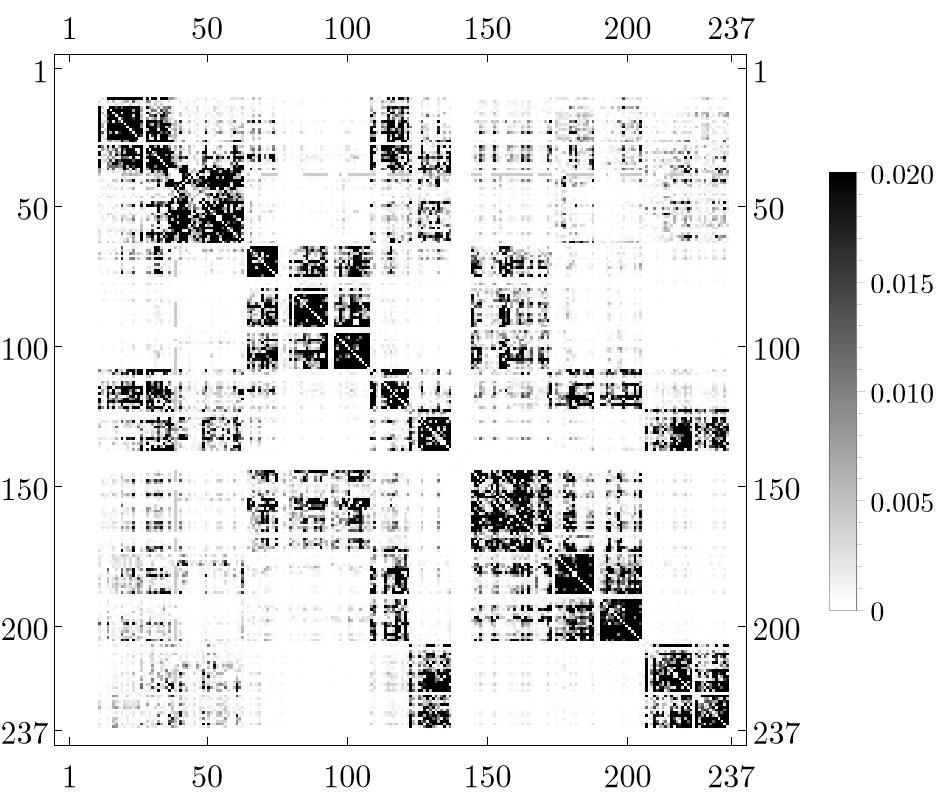}
    \caption{Mutual information matrix constructed with Boys localized orbitals of acrylamide - methanethiol complex at transition state, for CCSD/aug-cc-pVDZ state. The first 11 core orbital rows and columns are set to 0. For clarity, the adopted greyscale casts as black all elements greater than 0.02 black.
    }
    \label{fig:MI_Boys}
\end{figure}

To give an example, we display in Fig.~\ref{fig:MI_Boys} the mutual information matrix for an acrylamide-cysteine fragment of the zanubrutinib-BTK receptor binding site, at transition state geometry.

The most straightforward practical way to choose fragment (bath) orbitals is by ranking localized orbitals by their total summed mutual information with all other orbitals (row-summed MI), i.e., we set $\omega_0^{(0)}=1$ and all other weights $0$ in the scoring function in eq.~\eqref{eq:scoring-functional-frag}. Following the initial preselection of fragment and bath orbitals, we optimize their shape.

\paragraph{Quantum-information optimized (QIO) orbitals.}
\label{sec:QIO}
Simultaneously with the selection of fragment and bath orbitals, these orbitals can be optimized through a unitary rotation such that the relevant quantum-information metrics are minimized, thereby removing as much of the energy penalty associated with the system partitioning as possible. 
During the optimization, we keep the orbitals localized, as explained in app.~\ref{appendix:localization}.
The orbital rotation unitary can be written as:
\begin{equation}
   \tilde{d}_m^{\dagger}(\zeta)=\sum_{k=1}^{D}\mathbf{U}_{km}(\zeta)a_k^{\dagger}
    \label{eq:unitary}
\end{equation}
The unitary $\mathbf{U}(\zeta_t)$ is optimized iteratively, through a sequence of updates enumerated by $t=0,1,2,...$ and calculations of QI quantities, as summarized in algorithm~\ref{alg:ECCDMET}. In a given optimization step, having the reference state $\ket{\Psi_0}$ and the current candidate orbital rotation unitary $\mathbf{U}(\zeta_{t})$, we update QI objects $\gamma,\Gamma,\mathbf{S},\mathbf{I},\Lambda$. Optionally, for every optimization step $T$, a high-level solver (full DMET calculation) can be called to recompute $\gamma,\Gamma,\mathbf{S},\mathbf{I},\Lambda$ and the total energy, to control the quality of the formed basis. In other cases, that is $t\mod T \neq 0$, we recompute rotated QI quantities through linear transformations:
            \[
                \gamma(\zeta_t') = \Delta\mathbf{U}^{\dagger}\,\gamma(\zeta_t)\,\Delta\mathbf{U},
                \qquad
                \Gamma(\zeta_t') = (\Delta\mathbf{U}\otimes\Delta\mathbf{U})^{\dagger}\,\Gamma(\zeta_t)\,
                (\Delta\mathbf{U}\otimes\Delta\mathbf{U}),
            \]
along with derived non-linear quantities $\Lambda(\zeta_t')$, $\vec{S}(\zeta_t')$, and $\mathbf{I}(\zeta_t')$. At each iteration $t$ a new orbital rotation unitary increment is formed:
\begin{equation}
     \mathbf{U}(\zeta_t)\leftarrow \mathbf{U}(\zeta_{t-1})\,\Delta\mathbf{U}.
\end{equation}
  
Next, we evaluate an \textit{information leakage} functional to penalize the residual correlation between the
    cluster $C$ and the environment $\mathcal{E}$:
    \begin{equation}
        \mathcal{J}^{(1)}(\zeta_t')
        =
        -\omega_0^{(1)}\sum_{j\in F} S_j
        -\omega_1^{(1)}\sum_{j\in F}\sum_{k\in B} I_{jk}
        +\omega_2^{(1)}\sum_{s\in C}\sum_{k\in \mathcal{E}} I_{sk}
        +\omega_3^{(1)}\left\|\gamma_{C\mathcal{E}}\right\|^2
        +\omega_4^{(1)}\left\|\Lambda_{C\mathcal{E}}\right\|^2       
        \label{eq:leakage-functional}
    \end{equation}
    where $\gamma_{C\mathcal{E}}$ and $\Lambda_{C\mathcal{E}}$ denote chosen 1-RDM and cumulant blocks, coupling $C$ and  $\mathcal{E}$. If the unitary rotation $\mathbf{U}(\zeta)$ lowers the value of the information leakage functional given in eq.~\eqref{eq:leakage-functional}, $\mathcal{J}^{(1)}\zeta_t') < \mathcal{J}^{(1)}(\zeta_t)$ we accept $\Delta\mathbf{U}$ and set $\zeta_t \leftarrow \zeta_t'$.
    
\begin{figure}[ht]
    \centering
    \includegraphics[width=\linewidth]{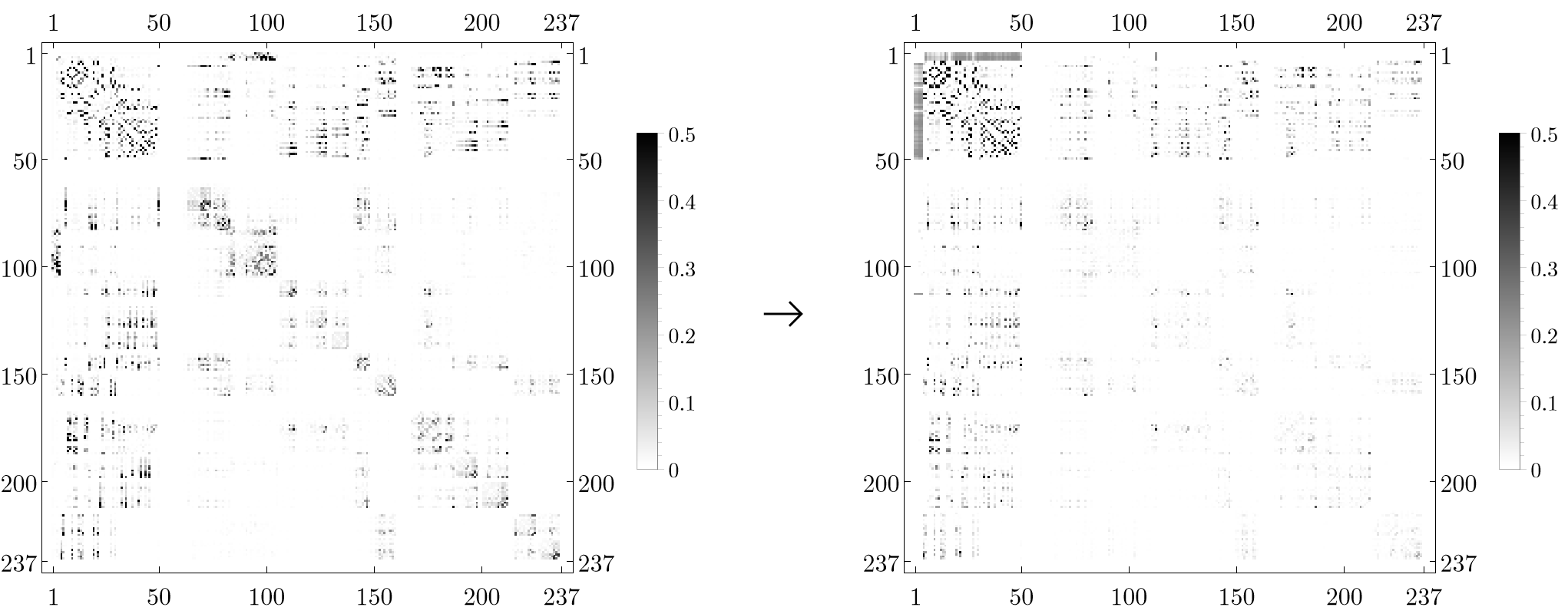}
    \caption{Mutual information matrices before and after optimization. 50 fragment orbitals correspond to the upper left block of the matrix. Elements greater than 0.5 are marked in black. The decrease in overall correlations in the environment block of the matrix, along with the increase in overall correlations in the fragment block, is associated with a lower electronic energy.
    }
    \label{fig:MI_matrix_change}
\end{figure}

We observed that minimization of the leakage functional correlates positively with energy minimization, i.e. $E_t \le E_{t-1}$, for systems studied, see sec.~\ref{sec:results}.
We repeat the unitary optimization cycle until convergence of the objective function  $|\mathcal{J}^{(1)}(\zeta_t)- \mathcal{J}^{(1)}(\zeta_{t-1})|< \epsilon_{\mathcal{J}}$ or when the maximum number of iterations is reached.
Otherwise, we reject and damp the direction in which the unitary optimization parameters propagate. The unitary defined in eq.~\eqref{eq:unitary} is represented in expontential form $\mathbf{U}=\exp(-K)$, where $K=\sum_{i,j}\zeta_{ij}(a^{\dagger}_ia_j-a^{\dagger}_ja_i)$, and $\zeta_{ij}$ is a set of $N(N-1)/2$ real parameters. 
Optimal parameters are found using the Covariance Matrix Adaptation Evolution Strategy (CMA-ES)~\cite{Hansen_2001}, in which samples are drawn from a guess distribution. This distribution is updated based on the values of the goal function. 
Notably, this algorithm does not require gradients. An example mutual information matrix transformation with the optimized unitary $\mathbf{U}$ is shown in Fig.~\ref{fig:MI_matrix_change}, for a subsystem in the acrylamide-cysteine transition state complex.

\subsubsection{System partition}
The choice of fragment-bath orbitals defines system partitioning (cf. Eq.~\eqref{eq:hilbert-paritioning}) and requires prior assumptions for fragment selection. We consider the active center (ligand, explicit solvent, and nearby residues) within a quantum-in-quantum embedding framework, partitioned into fragments and environments. Although the number of possible partitions grows exponentially (down to single-atom resolution), practical choices rely on chemically motivated partitions that reflect the covalent structure, interaction strengths, and electron delocalization. In principle, our method enables the determination of an optimal partition, albeit at significant computational cost. 

\begin{figure}
    \centering
    \includegraphics[width=0.5\linewidth]{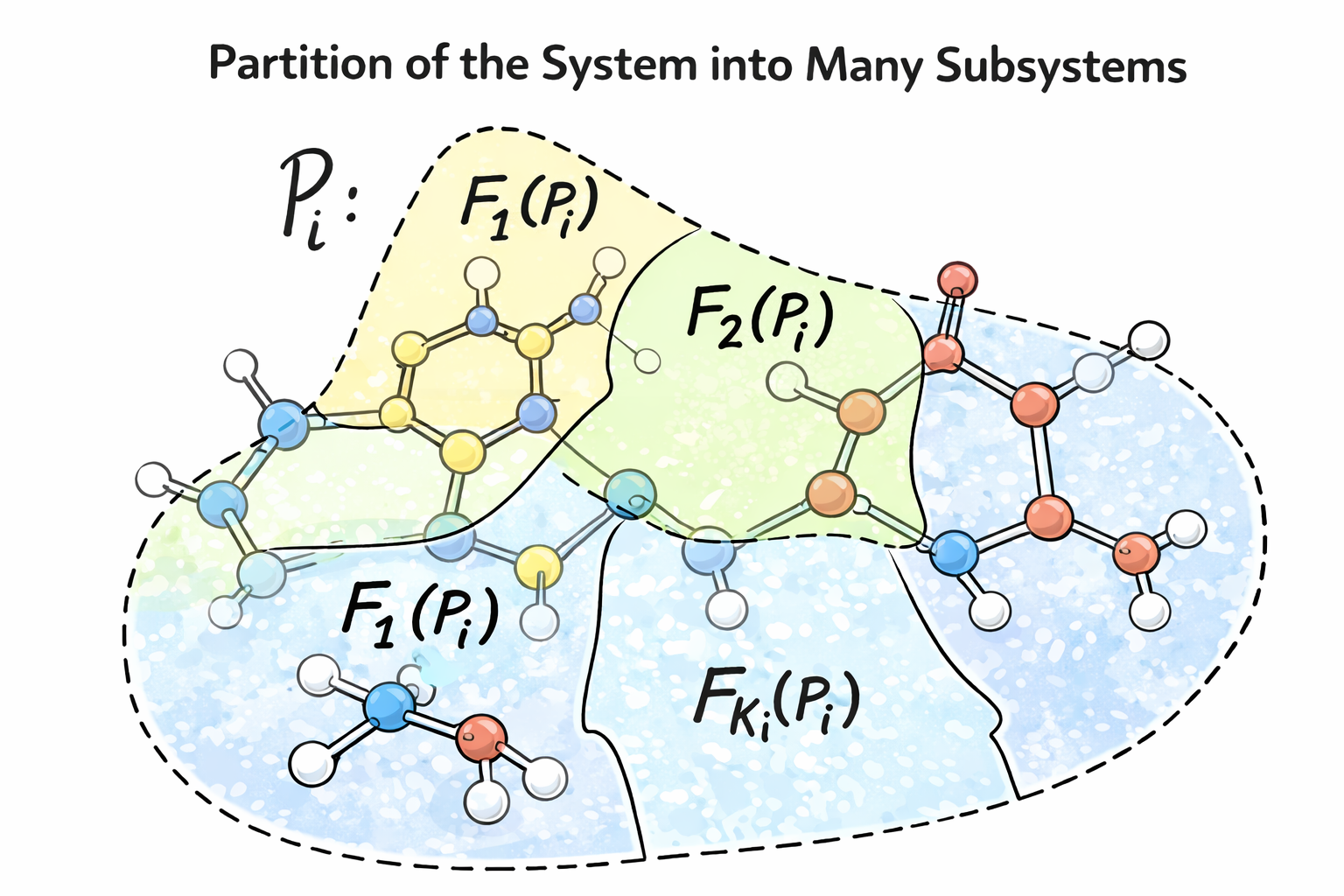}
    \caption{Schematic for system partitioning into fragments for a selected partitioning scheme $P_i$.}
    \label{fig:partitions}
\end{figure}

For each partition $\mathcal P_i$ of the system $\mathcal{S}$, we define fragments $\mathcal F^{(P_i)}_{k_i}$ as subsets associated with that partition.
The complete set of partitions is $ \mathcal{P}=\{\mathcal P_i\}_{i=1}^{N_P}$, where each $P_i$ consists of $K_i$ fragment–environment pairs $(\mathcal F^{(i)}_l,\mathcal{E}^{(i)}_l)$ covering $\mathcal{S}$ (Fig.~\ref{fig:partitions}). For a selected partition, we denote $(\mathcal F_l^{(P_i)},\mathcal{E}_l^{(P_i)})\equiv(\mathcal F,\mathcal{E})$ and define the fragment-orbital set $
B_F=\{\lvert\phi^{(F)}_j\rangle\mid j\in \mathcal F\}$,
with the size $\lvert B_F\rvert=D_F,$
where $\lvert\phi^{(F)}\rangle=\mathbf U^{(F)}\lvert\phi\rangle.$

Each embedded fragment problem is solved independently, and the resulting one- and two-particle reduced density matrices are combined to evaluate global properties such as the total energy and particle number. To mitigate inconsistencies in the total electron count arising from fragment partitioning, a global chemical potential is introduced, $\hat{H}_{emb} \rightarrow \hat{H}_{emb} - \mu_{\text{glob}}\sum_{l\in \mathcal F}\tilde{d}^{\dagger}_l\tilde{d}_l,$. To improve consistency with the correlated reference state, we additionally match selected one- and two-particle correlation potentials: $\hat{c} = \sum_{i
\leq j}^Nc_{ij}^{(1)}\tilde{d}^{\dagger}_{i}\tilde{d}_{j}+ \sum_{p,q,r,s}c_{pqrs}^{(2)}\tilde{d}^{\dagger}_{p}\tilde{d}_{q}\tilde{d}^{\dagger}_{r}\tilde{d}_{s}$.
This potential is optimized iteratively to enforce the correct total particle number and proper inter-fragment correlations. 
In the present single-fragment embedding scheme, property matching is unnecessary, although in the future our framework will be extended to multi-fragment optimization.

\subsection{Quantum core solvers}\label{sec:core-solver}
The embedded ECC-DMET Hamiltonian is constructed in the space of fragment and bath orbitals and takes the form of an active-space quantum chemistry Hamiltonian. The orbitals belonging to the fragment and bath define the correlated subsystem, while the remaining occupied and virtual environment orbitals are treated as inactive. The resulting embedded problem can then be solved using accurate many-body methods such as full configuration interaction (FCI), complete active space self-consistent field (CASSCF), coupled-cluster (CC) theory, or density matrix renormalization group (DMRG).

However, the size of this computational problem often exceeds the capabilities of modern classical computing architectures, particularly when high accuracy is required or when the system exhibits significant complexity and strong electron correlation. This situation frequently arises in the case of protein-ligand docking, especially when the binding process involves the formation of a chemical bond. Achieving sub-chemical accuracy (approximately 1 kcal/mol or better) for reaction barrier energies is often prohibitively expensive even for isolated ligands, as they may contain up to 200 atoms. The challenge becomes even greater when the immediate protein environment is included, as it typically consists of dozens of amino-acid residues together with surrounding solvent molecules.

For this reason, we present three possible choices for quantum core solver architectures: classical CPU/GPU-based implementations, quantum circuit simulators with tensor-network backends using CUDA-Q coupled and optimized for performance with the most powerful GPU units (A100, H100, B200), and quantum-hardware, where we use our own improved version of VQE with UCCSD ansatz within the ADAPT-VQE framework, originally introduced in Ref.~\cite{Peruzzo2014}.

\subsubsection{Classical architectures}
In principle, any many-body quantum chemistry method can be used to get the energy associated with the embedded ECC-DMET Hamiltonian. In the current implementation, we support four different solvers for this task, which differ by computational complexity and accuracy: second-order M\o ller-Plesset perturbation theory (MP2, $\mathcal{O}(N^5)$, low-accuracy), the size-consistent Brillouin-Wigner perturbation theory which is a promising size-consistent and size-extensive method that solves the MP2 issues with small-gap systems (sc-BW2, $\mathcal{O}(N^5)$, improved accuracy), Coupled-Cluster Singles and Doubles (CCSD, $\mathcal{O}(N^6)$, high accuracy), Configuration Interaction Singles and Doubles (CISD, $\mathcal{O}(N^6)$, high accuracy) and Full Configuration Interaction (FCI, exponential scaling, exact with the chosen orbital basis). Note that the scaling is expressed in terms of $N=D_F+D_B$, which is the dimension of the embedded Hamiltonian (see eq.~\eqref{eq:hemb}). Hence, if the size of the embedded Hamiltonian is kept small, it is possible to afford high-level quantum chemistry methods also for large systems with low computational effort, provided we choose the active fragment correctly. 

In future releases, we plan to extend the classical solver to more accurate wavefunctions, like $N$-Electron Valence state second-order Perturbation Theory (NEVPT2)~\cite{Angeli2002PCNEVPT2} and Complete Active Space methods second-order Perturbation Theory (CASPT2)~\cite{Finley1998CASPT2}, which allows for the description of both static and dynamic electron correlation.

\subsubsection{Near-term quantum devices}
For the quantum computing calculation of electronic structure on near-term devices, we utilize Variational Quantum Eigensolver, a hybrid algorithm used for estimating the ground state energy of a given Hamiltonian operator $\hat{H}$. The quantum circuit prepares a parametrized state
\begin{equation}
    \ket{\psi(\boldsymbol{\theta})} = U(\boldsymbol{\theta})\ket{\psi_{ref}}
\end{equation}
where $\ket{\psi_{ref}}$ is a reference initial vector, which in the case of an electronic Hamiltonian is often taken to be Hartree-Fock state. Using the variational principle, one can approximate the ground state energy from the above by
\begin{equation}
    E_{VQE} = \min_{\boldsymbol{\theta}} \bra{\psi_{ref}}U(\boldsymbol{\theta})^\dag \hat{H}U(\boldsymbol{\theta})\ket{\psi_{ref}}.
\end{equation}
We use the Unitary Coupled Clusters Singles and Doubles ansatz (UCCSD), which is based on classical coupled clusters theory. Let $I_{occ}$ and $I_{vir}$ correspond to sets of indices for occupied and virtual orbitals, respectively. The parametrized unitary can be expressed as
\begin{equation}
    U(\boldsymbol{\theta}) = e^{\hat{T}(\boldsymbol{\theta})-\hat{T}(\boldsymbol{\theta})^\dag},
\end{equation}
where the excitation operators $\hat{T}(\boldsymbol{\theta})$ read
\begin{equation}
    \hat{T}(\boldsymbol{\theta}) = \sum_{i\in I_{occ}}\sum_{a \in I_{vir}}\theta_{i;a} a^\dag_i a_a + \sum_{i,j\in I_{occ}}\sum_{a,b\in I_{vir}}\theta_{ij;ab} a^\dag_i a^\dag_j a_a a_b.
\end{equation}
The creation and annihilation operators are mapped onto qubits using Jordan-Wigner mapping. We utilize binary encoding, with the spin degree of freedom encoded on the least significant bit.

To estimate the ground state energy with the VQE methods, the Hamiltonian is decomposed as a sum of Pauli strings, which are straightforward to measure on a quantum computer:
\begin{align}
    \hat{H} = \sum_{i}\alpha_i P_i \qquad \text{and} \qquad P_i = \bigotimes_{k=1}^{N_{qubit}} \sigma_{i_k}. 
\end{align}
Here, $\sigma_i$ denotes Pauli matrices with $\sigma_0 \equiv \mathbb{I}$ and $N_{qubit}$ is the number of qubits in the system. In order to reduce the number of measurements, we use a greedy algorithm to group Pauli strings into qubit-wise-commuting groups. While in theory any group of commuting operators can be measured simultaneously, this requires implementing an additional basis change at the end of the circuit, which in general requires 2-qubit gates. However, using only qubit-wise-commuting operators requires only single-qubit gates. This simpler approach was already able to reduce the number of necessary measurements by a factor of around 3, for the considered Hamiltonians.

\begin{figure}[h]
    \centering
    \includegraphics[width=0.99\linewidth]{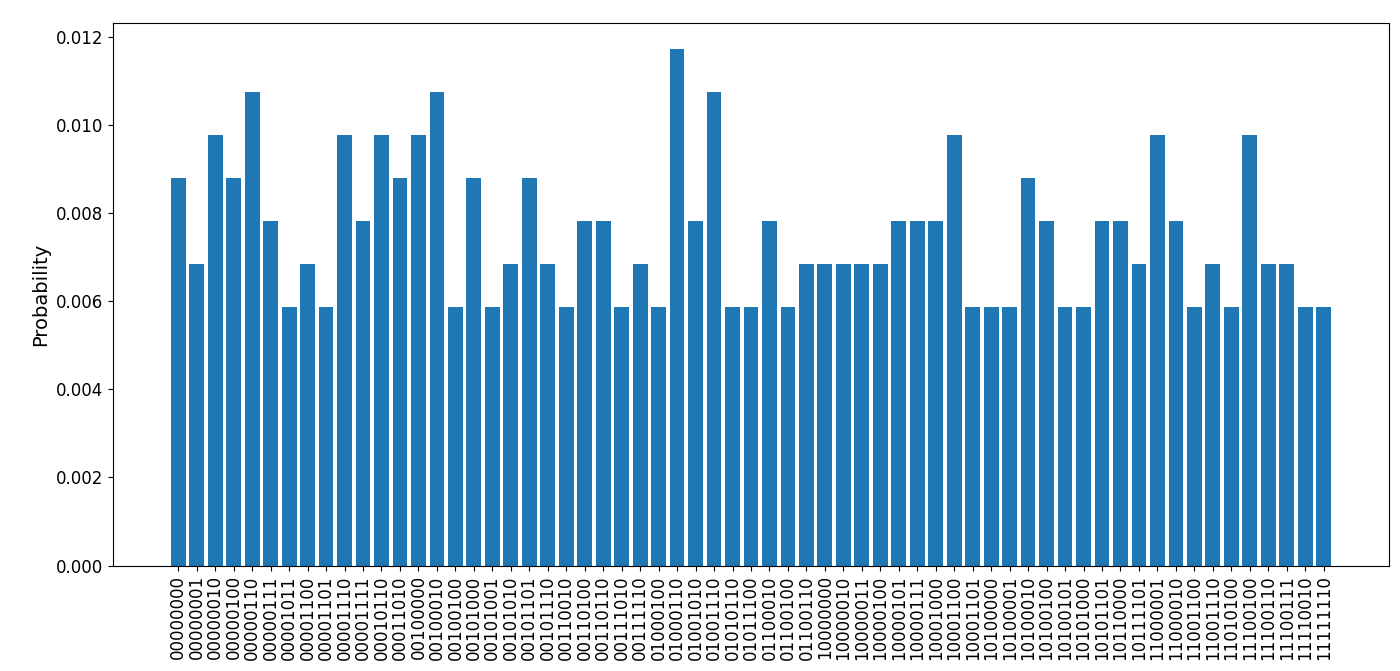}
    \caption{Example distribution obtained by sampling the state prepared by UCCSD. For clarity, only states with probability higher then $ 0.05$ are shown.}
    \label{fig:iqm_distribution}
\end{figure}

We demonstrated our method on IQM's Garnet 20 qubit superconducting QPU by estimating the energy of a simplified ligand-protein system (Acrylamide+Methanethiolate) at transition state geometry. To include the effect of solvation while keeping the size of the system suitable for current QPU's, we use the polarizable continuum model (PCM) with $\epsilon=4$. The Hamiltonian used for the VQE was constructed with the DMET method with 2 fragment orbitals, leading to $8$-qubit quantum circuit. For the classical optimization, we have used the Constrained Optimization BY Linear Approximation (COBYLA) method as implemented in the CUDA-Q library. Two choices of orbitals were compared: chemically motivated and quantum-information-optimized as described in sec.~\ref{sec:summary_contribution}. The results compared to FCI energy are shown in Table \ref{tab:qpu_resuts}. For each circuit, the quantum state was sampled with $1024$ shots. Fig.~\ref{fig:iqm_distribution} shows an example distribution obtained by sampling the state prepared by the UCCSD circuit.
\begin{table}[h]
    \centering
    \begin{tabular}{c|c|c}
            & Chemical & Optimized \\
            \hline
         $E_{FCI}$ [a.u.]&  -683.048&	-683.050 \\
         $E_{VQE}$ [a.u.]& -680.105  &	-680.992 \\
         $\Delta E$ [a.u.] & 2.943 &	2.058 \\
    \end{tabular}
    \caption{Ground state energy calculated with VQE on Garnet QPU compared to FCI results for chemically motivated and optimized orbitals. Here $\Delta E = E_{FCI}- E_{VQE}$.}
    \label{tab:qpu_resuts}
\end{table}
As can be seen from Table~\ref{tab:qpu_resuts}, our optimization procedure improves the energy estimation. However, there is still a significant difference between VQE results and FCI energies, which can be attributed to the large error rates of current quantum computers.

In addition to the standard UCCSD ansatz, we have tested on simulators various other methods of state preparation, such as hardware-efficient ansatz~\cite{Leone_2024}, symmetry-preserving ansatz~\cite{Gard2020}, as well as ADAPT-VQE framework~\cite{Grimsley2019}. However, we found the UCCSD ansatz to be most reliable.

There are several possibilities for optimizing the circuits even further. For instance, the simple implementation on UCCSD ansatz as implemented in CUDA-Q could be optimized for a given quantum architecture, reducing the number of required native gates. Additionally, qubit number reduction techniques such as entanglement forging~\cite{Eddins_2022} or qubit tapering~\cite{Sergay2017} can be utilized. In addition to circuit optimization, the embedded Hamiltonian definition could possibly be improved by incorporating downfolding techniques~\cite{Huang2023} into the DMET Hamiltonian construction, leading to improved accuracy.

\subsubsection{Quantum Computer Simulators}
Apart from accessing real QPU's, our method incorporates simulation of quantum computation using \texttt{CUDA-Q} and \texttt{Qiskit} libraries. For testing and benchmarking quantum circuits on classical machines, we use NVIDIA's \texttt{CUDA-Q} simulators. The \texttt{CUDA-Q} library provides a few different simulators, for instance, based on state vector simulation or tensor network methods, available for both CPU and GPU architectures. We have also performed benchmark comparisons with Qiskit's AerSimulator on GPU backend.

For internal testing and benchmarks, we have used a local machine with a GeForce 4060 GPU as well as modern NVIDIA architectures such as A100, H100, and B200 GPUs, available via the NVIDIA Brev platform. All calculations were performed with a noiseless simulator and exact sampling (shots $=-1$ in \texttt{CUDA-Q}).

\begin{figure}[h]
    \centering
    \includegraphics[width=0.5\linewidth]{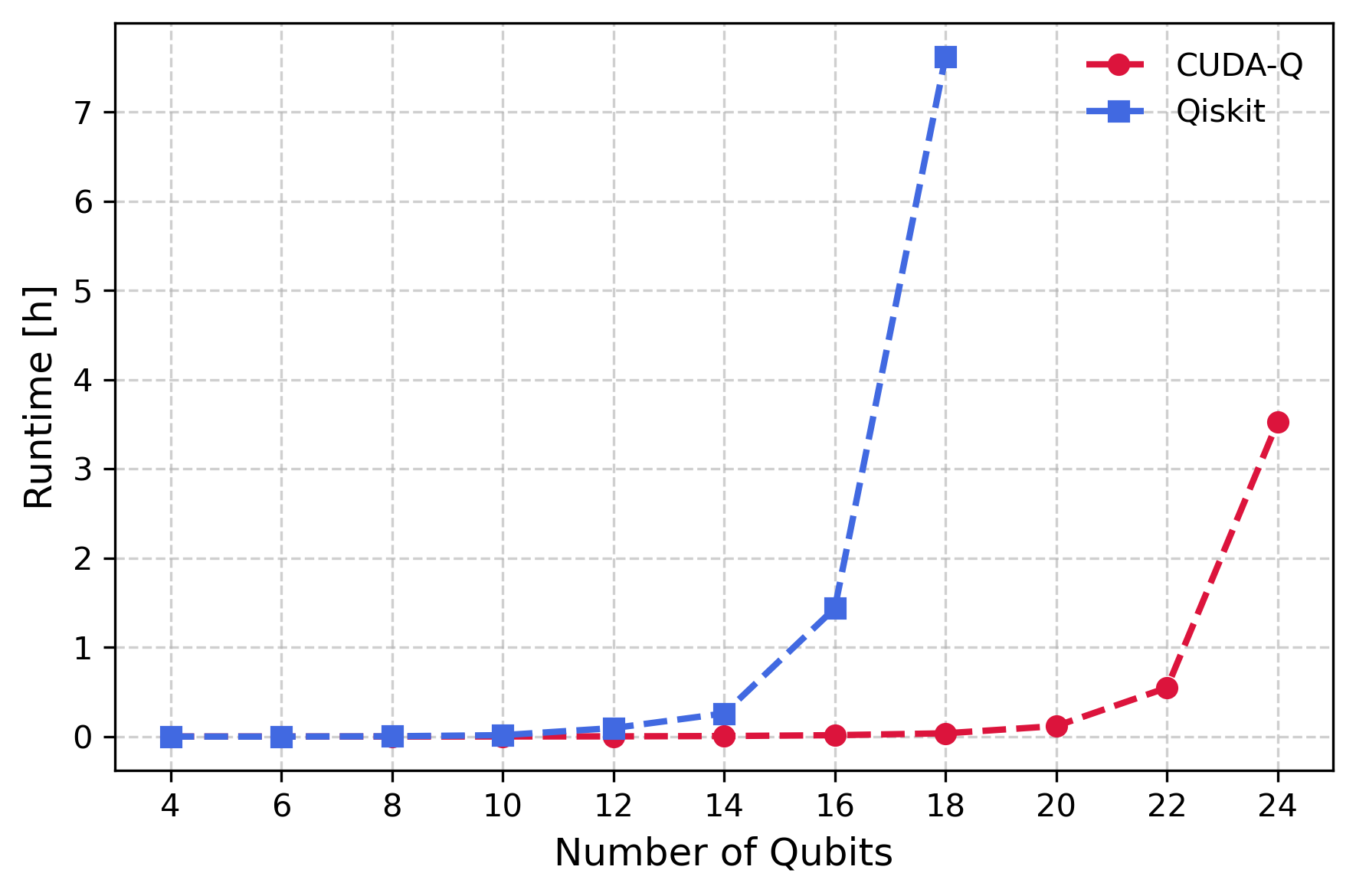}
    \caption{Comparison of VQE-UCCSD runtimes on transition state geometries with 6-31++G** basis, for different sizes of active space. Natural orbitals obtained from the MP2 solution were used for active space selection.}
    \label{fig:qiskit_cudaq_comparison}
\end{figure}

We compared VQE-UCCSD energies and runtimes for CUDA-Q and Qiskit simulators by estimating the energy barrier for a simplified protein-ligand system (Acrylamide+Methanethiolate) with the PCM solvation model. The results are shown in Fig.~\ref{fig:qiskit_cudaq_comparison}.

\begin{figure}[h]
    \centering
    \includegraphics[width=0.5\linewidth]{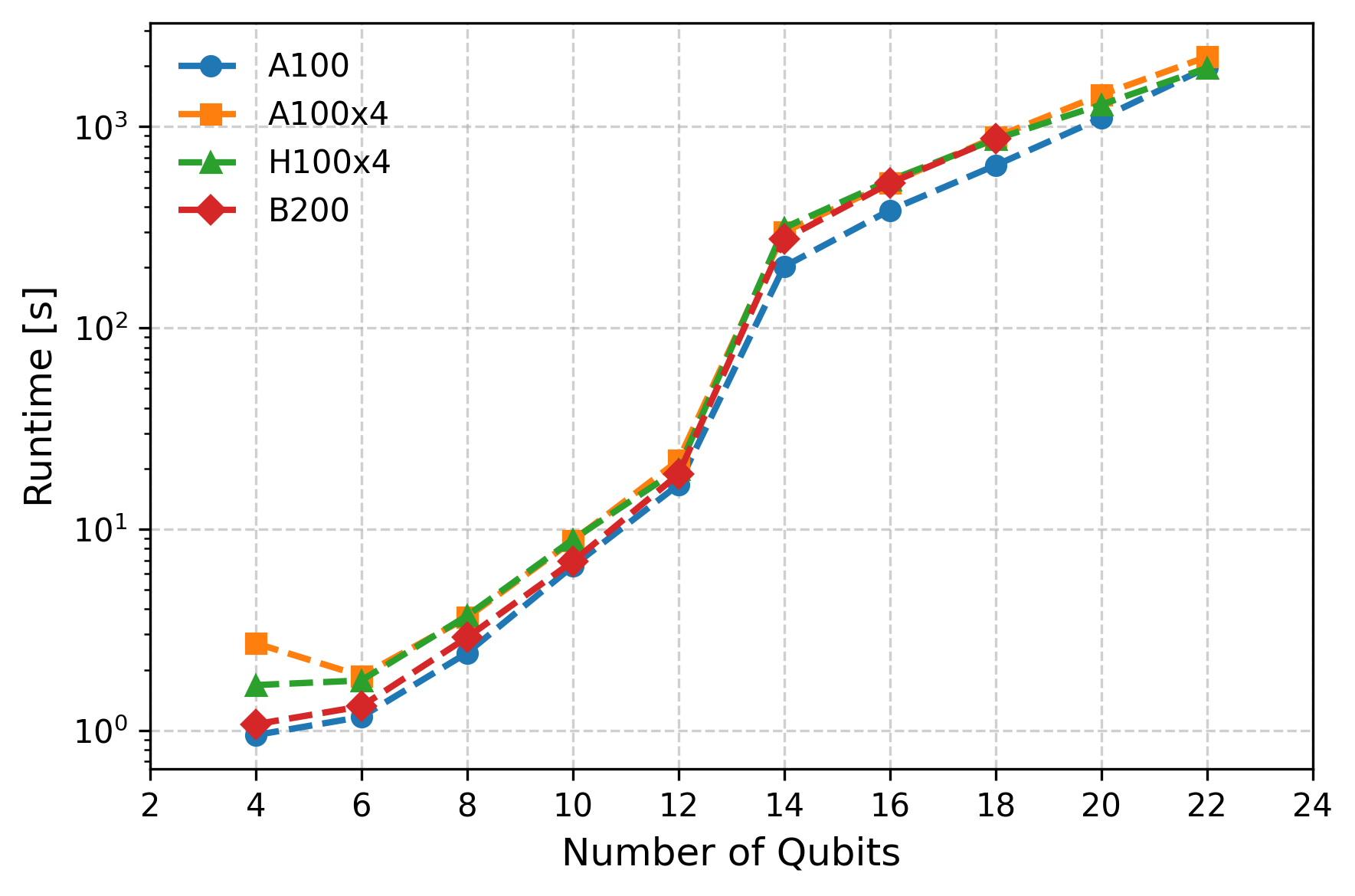}
    \caption{Comparison of runtimes with different GPUs}
    \label{fig:benchmark_gpus}
\end{figure}

Additionally, a comparison of the runtimes of VQE-UCCSD on different GPUs was performed. The results are shown in Fig.~\ref{fig:benchmark_gpus}.
As can be seen, for the size of the system in question, there was no significant difference between runtimes on different devices. 

\begin{figure}[h]
    \centering
    \includegraphics[width=0.5\linewidth]{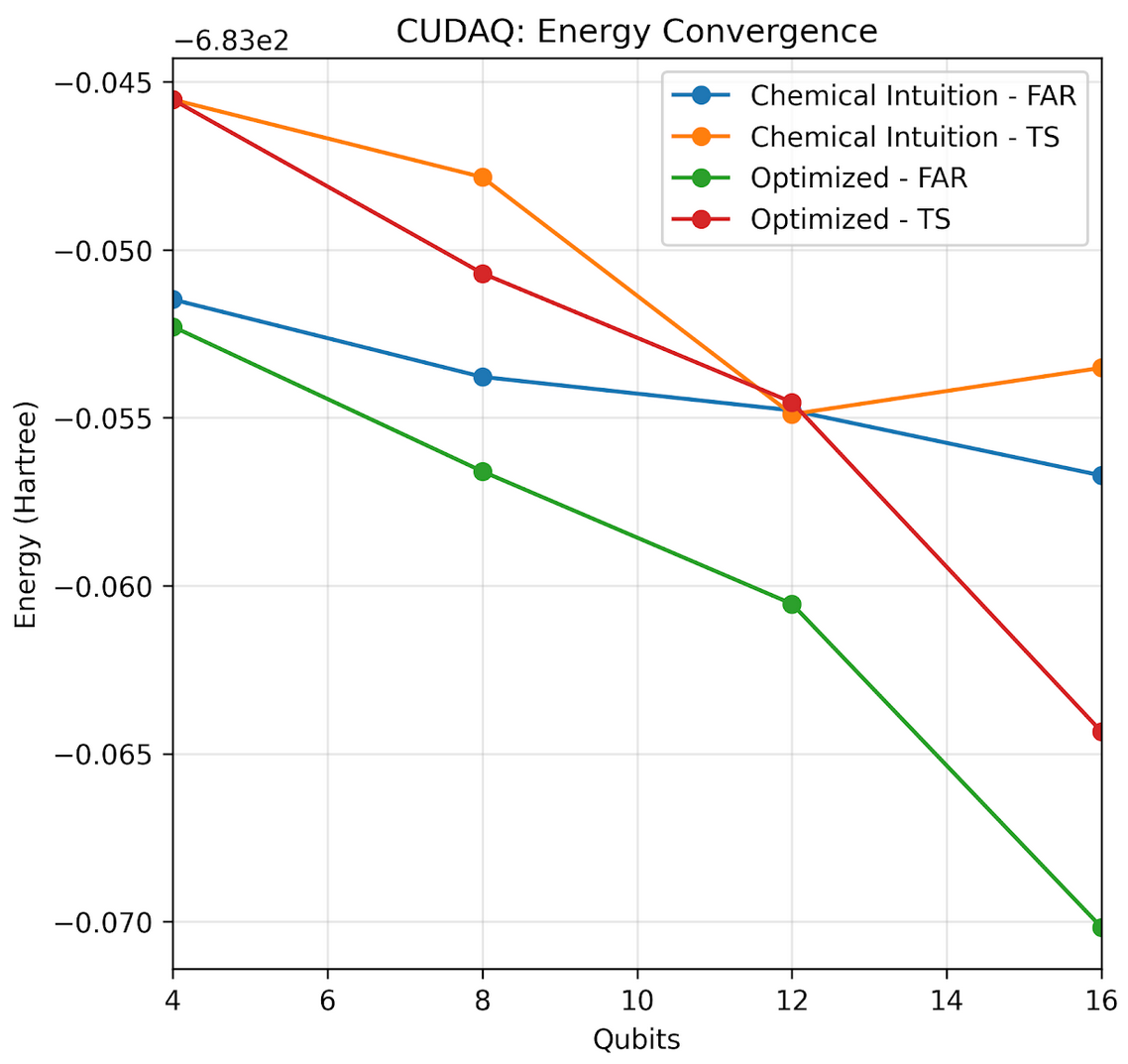}
    \caption{Ground state energy estimation for chemically motivated and ECC-optimized orbitals estimated with VQE-UCCSD. The aug-cc-pVDZ basis set was used. The circuits were simulated with CUDA-Q state-vector simulator on NVIDIA's A100 GPU. TS denotes transition state geometry, and FAR corresponds to the geometry with spatially separated reactants.}
    \label{fig:cudaq_orbitals}
\end{figure}

We have also performed the comparison of chemically motivated and quantum-information-optimized orbitals, shown in Fig.~\ref{fig:cudaq_orbitals}. Optimized orbitals provide a more accurate and stable energy estimation for the considered system.

\subsubsection{Fault-Tolerant Quantum Computers}
\label{sec:FTQC}
For the embedded cluster Hamiltonian produced by the correlated DMET/QM/MM workflow, the fault-tolerant quantum algorithm follows the following pipeline:
\[
    H_{emb} \;\longrightarrow\; \text{double factorization} \;\longrightarrow\; \text{block encoding} \;\longrightarrow\; \text{qubitized QPE}.
\]
The first step exposes the algebraic structure of the electronic Hamiltonian, the second converts that structure into a unitary oracle, and the last step extracts eigenvalues using Quantum Phase Estimation (QPE). This decomposition is useful conceptually because it separates the chemistry-specific preprocessing from the generic quantum subroutines, and it also makes clear where the dominant resource bottleneck enters.

\paragraph{Electronic Hamiltonian and double factorization.}
The starting point is the second-quantized electronic Hamiltonian of the embedded cluster from eq~\eqref{eq:hemb},
\begin{equation}
    H = \sum_{p,q=1}^{2N}\sum_{\sigma \in \{\alpha,\beta\}}\tilde{h}_{pq}a_{p\sigma}^\dag a_{q\sigma} + \frac{1}{2}\sum_{p,q,r,s=1}^{2N} \sum_{\sigma,\tau \in \{\alpha,\beta\}}g_{pqrs}a_{p\sigma}^\dag a_{r\tau}^\dag a_{s\sigma} a_{q\tau},
\end{equation}
where \(a^{\dagger}_{i\sigma}\) and \(a_{j\sigma}\) are fermionic creation and annihilation operators, \(h_{ij}\) are one-electron integrals, and \(g_{ijkl}\) are the spin-summed two-electron coefficients. The one-body term contains kinetic energy and electron-nuclear attraction, while the two-body term captures electron-electron repulsion. For \(N\) spatial orbitals, the tensor \(g_{ijkl}\) contains \(O(N^4)\) entries, so a direct fault-tolerant implementation is not attractive.

To expose additional structure, we first apply a general double-factorized (DF) representation of the two-electron tensor,
\begin{equation}
    g_{ijkl} \approx \sum_{r=0}^{R-1} A_{r,ij} A_{r,kl},
    \qquad
    A_{r,ij} = A_{r,ji},
\end{equation}
where \(R\) is a truncation rank and each \(A_r\) is a symmetric \(N \times N\) matrix.
For each $r$, different factors $A_{r,ij}$ are matrices in indices $i$ and $j$. 
We diagonalize those matrices as
\begin{equation}\label{eq:A_diagonalization}
    A_{r,ij} = \sum_t \lambda_{r,t}\, u^{(r)}_{t,i}\, u^{(r)}_{t,j}.
\end{equation}
It is convenient to define the one-body operator
\begin{equation}
    \mathrm{One}(A_{r}) \coloneqq  \sum_{ij} A_{r,ij} \sum_{\sigma \in \{\alpha,\beta\}} a^{\dagger}_{i\sigma} a_{j\sigma}.
\end{equation}
Up to the chosen DF truncation error and an additive scalar shift that can be tracked classically, the Hamiltonian can then be reorganized into the form
\begin{equation}
    H \approx H_{\mathrm{DF}}
    \coloneqq  \mathrm{One}(h') + \sum_{r=0}^{R-1} \mathrm{One}(A_r)^2 + \mathrm{const},
\end{equation}
where \(h'\) denotes the effective one-body matrix obtained after collecting the standard contraction terms.
This is the representation used in the fault-tolerant pipeline below: instead of working with a dense rank-4 tensor, we work with a list of structured one-body matrices and their squares.

\paragraph{Block encoding of the factorized Hamiltonian.}
To use \(H_{\mathrm{DF}}\) inside a fault-tolerant algorithm, we embed it into a larger unitary via a block encoding. A unitary \(U_A\) acting on \(a\) ancilla qubits and the system register is an \emph{\((\alpha,a,\epsilon)\)-block encoding} of an operator \(A\) if
\begin{equation}
    \left\|
    A - \alpha ( \bra{0}^{\otimes a}\otimes I) U_A (\ket{0}^{\otimes a}\otimes I) \right\| \le \epsilon.
\end{equation}
In the exact case \((\epsilon=0)\), the operator \(A/\alpha\) appears as the upper-left block of \(U_A\).

For the DF Hamiltonian, the block encoding is built in the usual linear-combination-of-unitaries (LCU) style from \textsc{Prepare}/\textsc{Select} primitives over the factor label \(r\), the eigenmode index \(t\), orbital indices, and the associated coefficients. 
This is precisely the regime in which modern QROM constructions become useful: one needs coherent table lookups for structured data, but never needs to materialize the full \(O(N^4)\) tensor. In particular, the optimized Walsh-Hadamard QROM constructions from our earlier work~\cite{qrom_beit} fit naturally into this data-loading layer. Once this construction is in place for the factorized Hamiltonian, we obtain a block encoding \(U_H\) of $H_{DF}$ with some normalization factor \(\lambda\).

\paragraph{Qubitized quantum phase estimation.}
Given an exact \((\lambda,a,0)\) block encoding \(U_H\) of the Hamiltonian, qubitization promotes it to a walk operator \(W\) whose phases encode the spectrum of \(H\). Concretely, for each eigenpair \(H\ket{\psi_k} = E_k \ket{\psi_k},\)
there is a corresponding qubitized eigenstate \(\ket{\Phi_k}\) such that the eigenphase \(\theta_k\) of \(W\) satisfies
\begin{equation}
    \cos \theta_k = \frac{E_k}{\lambda}.
\end{equation}
Quantum phase estimation is then applied to controlled powers \(W^{2^j}\), from which \(\theta_k\) is estimated and the physical energy is recovered as
\begin{equation}
    E_k = \lambda \cos \theta_k.
\end{equation}
The key scaling point is that the number of controlled walk-operator applications required to achieve energy precision \(\varepsilon\) scales linearly with \(\lambda/\varepsilon\) (up to the usual logarithmic factors in success probability and register size). Thus, once a block encoding exists, the dominant cost driver is the normalization factor \(\lambda\).

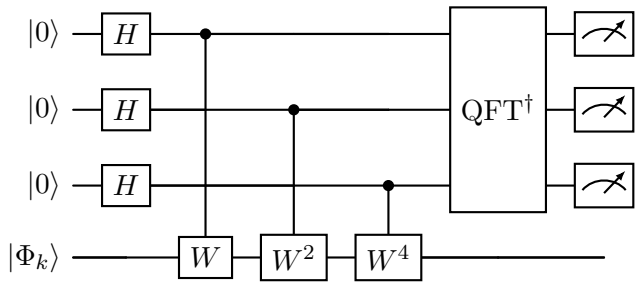
\begin{figure}[t]
    \centering
    \begin{quantikz}[row sep=0.30cm, column sep=0.38cm]
        \lstick{$\ket{0}$}        & \gate{H} & \ctrl{3} & \qw      & \qw      & \gate[wires=3]{\mathrm{QFT}^{\dagger}} & \meter{} \\
        \lstick{$\ket{0}$}        & \gate{H} & \qw      & \ctrl{2} & \qw      & \ghost{\mathrm{QFT}^{\dagger}}         & \meter{} \\
        \lstick{$\ket{0}$}        & \gate{H} & \qw      & \qw      & \ctrl{1} & \ghost{\mathrm{QFT}^{\dagger}}         & \meter{} \\
        \lstick{$\ket{\Phi_k}$}   & \qw      & \gate{W} & \gate{W^{2}} & \gate{W^{4}} & \qw                               & \qw
    \end{quantikz}
    \caption{Illustrative qubitized QPE circuit. The bottom wire denotes the combined work register on which the qubitized walk operator \(W\) acts; this includes the system register together with the ancillas required by the block encoding. A three-bit phase register is shown for concreteness.}
    \label{fig:qpe_from_block_encoding}
\end{figure}

\paragraph{Symmetry-optimized Double Factorization.}
The previous paragraph makes the optimization target transparent. Standard DF is typically chosen to reproduce the two-electron tensor accurately, but the fault-tolerant cost is governed more directly by the block-encoding normalization,
\begin{equation}
    \lambda_{\mathrm{DF}} = \frac{1}{2}  \sum_{r=0}^{R-1} \Lambda_r^2 + \Lambda_{-1},
    \,\,\,\, \text{with} \,\,\,\,  \Lambda_r = \|A_r\|_{\mathrm{nuc}} = \sum_t |\lambda_{r,t}|\,\,\,\, \text{and} \,\,\,\,
    \Lambda_{-1} = \|h'\|_{\mathrm{nuc}},
\end{equation}
where $\lambda_{r,t}$ comes from the diagonalization of the matrix $A_{r}$, as in eq.~\eqref{eq:A_diagonalization}. 
Here, $\|\cdot\|_{\mathrm{nuc}}$ is the nuclear (Schatten) norm of a matrix, i.e., the sum of its singular values. 
For a symmetric matrix, it coincides with the sum of the absolute values of its eigenvalues.

In other words, a low tensor-factorization error does not by itself guaranty a low-cost qubitized simulation. The real bottleneck is the effective Hamiltonian norm entering the block encoding. This is exactly where the symmetry-optimized DF construction becomes useful. The idea is to modify the Hamiltonian \emph{before} factorization by adding an operator that vanishes identically on the target electron-number sector,
\begin{equation}
    \widetilde H = H + \left(\sum_{ij\sigma} \xi_{ij} a^{\dagger}_{i\sigma} a_{j\sigma} +
    \kappa \right) (N_e - n_e),
\end{equation}
where \(N_e = \sum_i E_{ii}\) is the electron-number operator, \(n_e\) is the target particle number, \(\xi\) is a symmetric matrix, and \(\kappa\) is a scalar. Because \( (N_e-n_e)\ket{\Psi}=0 \) on the physically relevant sector, \(\widetilde H\) and \(H\) are equivalent for phase estimation on that sector. At the same time, the shift changes the one- and two-body tensors to
\begin{equation}
    \widetilde h_{ij} = h_{ij} - n_e \xi_{ij} + \kappa \delta_{ij},
    \qquad
    \widetilde g_{ijkl} = g_{ijkl} + \frac{1}{2} \left(\xi_{ij}\delta_{kl} + \delta_{ij}\xi_{kl}\right),
\end{equation}
while preserving the usual \(8\)-fold permutation symmetry of the two-electron tensor~\cite{Deka2025}.
Thus, factorization
\begin{equation}
    \widetilde g_{ijkl}  \approx \sum_{r=0}^{R-1}  \tilde{A}_{r,ij}  \tilde{A}_{r,kl},
\end{equation}
forms the associated effective one-body tensor
\begin{equation}
    \widetilde h'_{ij}  = \widetilde h_{ij} + 2 \sum_k \widetilde g_{ijkk},
\end{equation}
and optimizes the shift parameters and factor matrices to reduce the resulting block-encoding scale,
\begin{equation}
    \lambda_{\mathrm{SODF}} = \frac{1}{2} \sum_{r=0}^{R-1} \| \tilde{A}_r\|_{\mathrm{nuc}}^2 +
    \|\widetilde h'\|_{\mathrm{nuc}}.
\end{equation}
A convenient optimization objective is a penalized reconstruction problem of the form
\begin{equation}
    \mathrm{Err}(\kappa,\xi,A) = \sum_{ijkl} \left| \widetilde g_{ijkl} - \sum_{r=0}^{R-1}  \tilde{A}_{r,ij}  \tilde{A}_{r,kl}
    \right|^2 \,\, \text{and} \,\,\,
    \mathrm{Cost}(\kappa,\xi, \tilde{A}) = c\cdot\,\mathrm{Err}(\kappa,\xi, \tilde{A}) + \lambda_{\mathrm{SODF}},
\end{equation}
initialized at the standard DF solution \((\kappa,\xi)=(0,0)\) and $c$ is a fitted constant.

The practical advantage is that this is a purely classical preprocessing improvement: the downstream quantum algorithm remains unchanged. What changes is the normalization factor \(\lambda\), and therefore the number of controlled walking steps required in the estimation of the phase.

For our workflow, this yields a clean division of labor. The correlated ECC-DMET/MM procedure produces a compact, chemically meaningful cluster Hamiltonian; standard DF turns that Hamiltonian into a structured form suitable for fault-tolerant simulation; block encoding and qubitization provide the generic quantum subroutines; and symmetry-optimized DF targets the main resource bottleneck by reducing the effective Hamiltonian norm. This combination preserves the physics of the active region while directly lowering the leading cost of fault-tolerant quantum simulation.

\paragraph{Classical-Quantum Integration and Computational Stack.}
The overall workflow can be broken down into three key abstraction layers. At the top, the application layer defines the target quantity, for example, a binding affinity, reaction barrier, or relative stability along a catalytic pathway, and selects representative geometries from molecular dynamics. The next layer builds the physical model through QM/MM partitioning, treatment of the protein and solvent environment, and definition of the chemically active region. This is followed by the embedding layer, where the correlated DMET/QIO machinery compresses the active region into an embedded cluster Hamiltonian that retains the dominant correlation effects in a much smaller orbital space. 

This cluster Hamiltonian acts as a key interface
between all solver backends: the same object can be passed to a classical quantum-chemistry solver on CPU/GPU, to a GPU circuit simulator, to current quantum hardware, or to the FTQC layer, where it is transformed by double factorization, norm-reduction preprocessing, block encoding, and qubitized phase estimation. At the hardware level, this maps naturally onto conventional CPU/GPU nodes, accelerator hardware, and, ultimately, fault-tolerant quantum processors. 

This layered picture also clarifies how distinct methodological innovations contribute to the same objective: molecular modeling determines which geometries and environments matter, embedding reduces the size of the quantum problem while preserving the essential chemistry, and FTQC reduces the cost of solving the final many-body Hamiltonian once classical methods become the bottleneck. Because the information flow can also go in the reverse direction, FTQC resource estimates can be fed back to refine fragment definitions, bath thresholds, and orbital optimizations.
This yields a unified classical-quantum workflow in which the chemical model is prepared classically, and the hardest electronic-structure step is handed off to fault-tolerant quantum hardware only when it is genuinely advantageous.

\subsection{Code design and interface}
\label{sec:code}
The platform is designed as a command-line tool accessed through the \texttt{angelo} script. Users interact with the system by providing a \emph{simulation suite}, defined as a directory containing configuration files and optional input data required for the workflow.

A typical simulation suite contains the following files:
\begin{itemize}
    \item \texttt{quantum.yaml}: the primary configuration file describing the molecular system and computational methods used by the quantum solver. A suite may contain multiple such files, selectable via command-line arguments.
    \item \texttt{classical.yaml} (optional): configuration describing the molecular dynamics workflow used to generate simulation snapshots.
    \item Additional geometry or topology files (optional): structures or force-field data required by the solvers.
\end{itemize}

\subsubsection{Workflow orchestration}

\begin{figure}
    \centering
    \includegraphics[width=\linewidth]{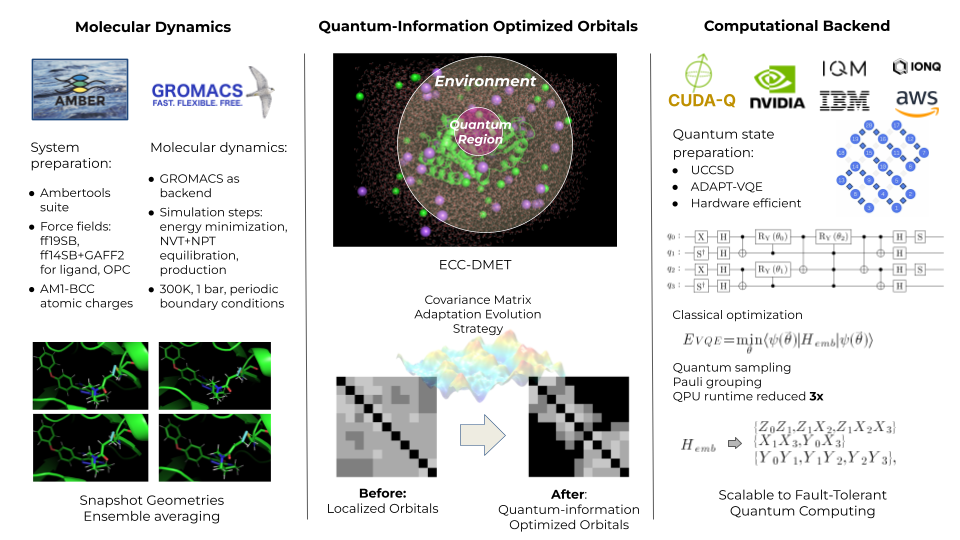}
    \caption{A scheme of our computational workflow, with Molecular Dynamics module shown in the left panel, quantum-information-optimized core solver for QM/QM/MM model based on DMET in the middle panel, and backend options in the right panel.}
    \label{fig:qm-mm-scheme}
\end{figure}

The computational workflow is orchestrated using the \texttt{Snakemake} workflow management system~\cite{Koster2012,Molder2021}. Snakemake represents pipelines as directed acyclic graphs (DAGs) of rules, enabling automatic dependency tracking, reproducible execution, and efficient parallelization.
Individual workflow stages, including molecular dynamics propagation, quantum calculations, preprocessing, and postprocessing, are implemented as Snakemake rules. These modular components can be combined through configuration presets, allowing users to construct complex simulation pipelines by selecting the required steps in the configuration file. The workflow engine automatically determines dependencies and recomputes only missing or outdated results.
A sketch of the workflow is displayed in Figure~\ref{fig:qm-mm-scheme}.

\subsubsection{Configuration structure}

The \texttt{quantum.yaml} configuration file contains two main sections: \texttt{system} and \texttt{methods}.

The \texttt{system} section defines the molecular structure used by the quantum solver. The system may either be specified as a standalone molecular description (including conformers and parameters) or as a snapshot extracted from a molecular dynamics trajectory. In the latter case, the simulation suite must also include a configuration for the molecular dynamics workflow.

The \texttt{methods} section lists the computational techniques used to evaluate electronic energies. Multiple methods can be specified, each with configurable parameters and backend implementations. Supported examples include RHF, BW2, MP2, CCSD, and DMET.

\subsubsection{Software environment and dependencies}

The workflow is implemented in a Python-based scientific computing environment that integrates classical molecular modeling tools with quantum chemistry and quantum computing frameworks.

\paragraph{Classical molecular modeling.}
System preparation and molecular dynamics simulations rely on established packages such as \texttt{AmberTools} and \texttt{GROMACS}. Additional tools, including \texttt{OpenMM}, \texttt{MDAnalysis}, \texttt{RDKit}, \texttt{OpenBabel}, \texttt{pdbfixer}, and \texttt{PyMOL} support molecular manipulation, structural preprocessing, and trajectory analysis. Semiempirical quantum chemistry calculations are provided via \texttt{MOPAC}.

\paragraph{Quantum chemistry and quantum computing frameworks.}
Classical computations using standard quantum chemistry methods are performed using \texttt{PySCF} \cite{pyscf1,pyscf2} together with the \texttt{pyscf-dispersion} module. Quantum-enabled simulations are supported through libraries such as \texttt{qiskit}/\texttt{qiskit\_aer} and \texttt{CUDA-Q}/\texttt{cudaq\_solvers}, which support both CPU- and GPU-based simulation. The mapping of the electronic structure problem to qubit Hamiltonians is performed using \texttt{OpenFermion}.

\subsubsection{Deployment and execution environments}

The workflow is designed to run on environments ranging from local workstations to high-performance computing clusters. Through Snakemake's native support for distributed execution, simulations can scale across multi-node systems or cluster schedulers.

Deployment of the software stack is automated using \texttt{Ansible}, allowing reproducible installation on remote infrastructure such as AWS instances, GPU-enabled cloud environments (e.g., Brev), or any machine accessible via SSH.
    
\begin{figure}
    \centering
    \includegraphics[width=\linewidth]{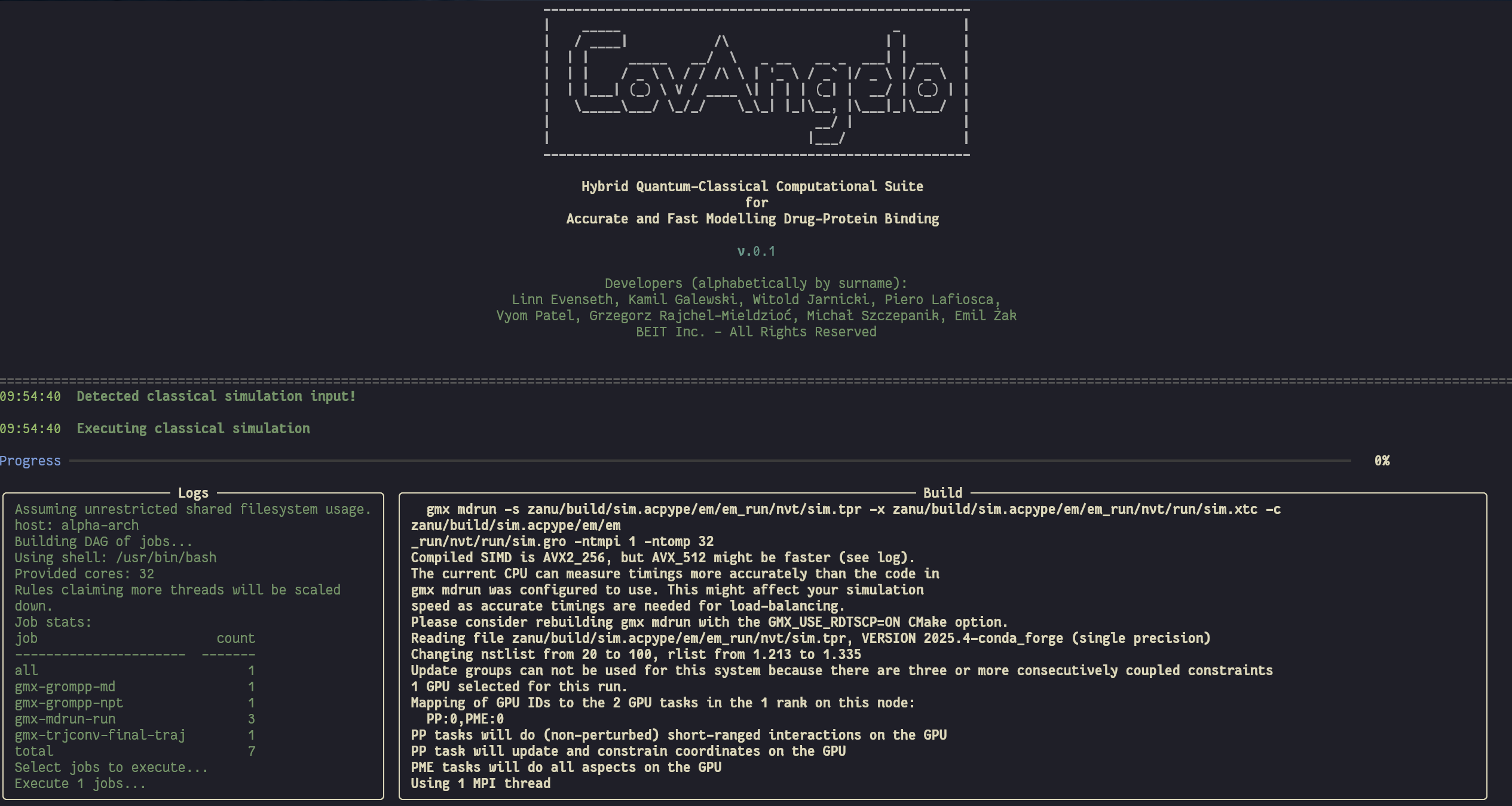}
    \caption{Screenshot of the terminal user interface of our tool}
    \label{fig:screenshot}
\end{figure}

\begin{figure}[!htbp]
    \centering
    \includegraphics[width=\linewidth]{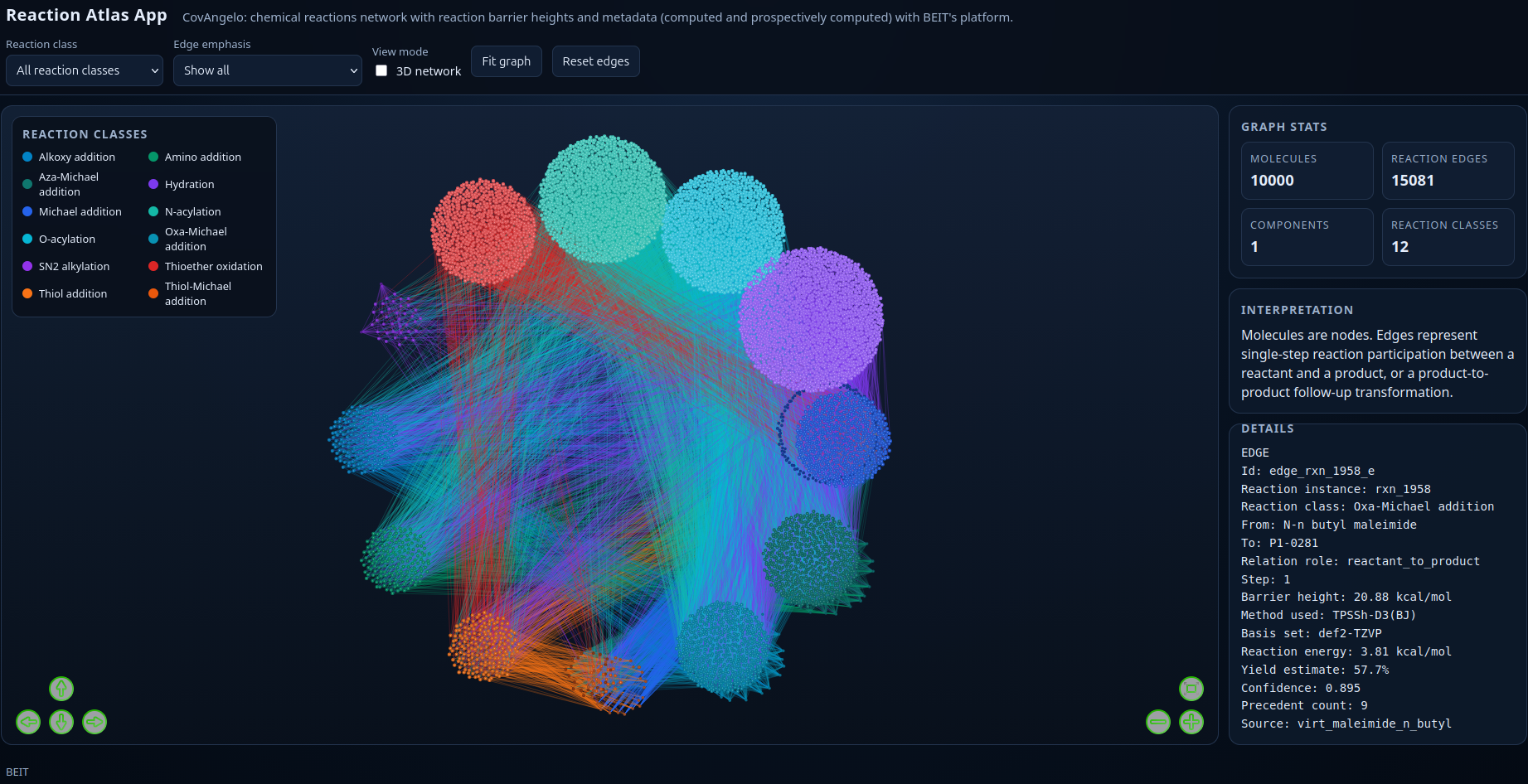}
    \caption{Snapshot of MolZart: a tool for chemical reactions modeling using CovAngelo backend.
    }
    \label{fig:Snapshot-molzart}
\end{figure}

\paragraph{Chemical Reactions Network Simulator.}
Our computational workflow can be used for quantum-accurate virtual screening of covalent inhibitors and modeling chemical reactions. For this purpose, we have developed MolZart, a user-friendly interface built on top of the CovAngelo backend, designed for large-scale predictions of chemical reaction barrier heights and exploration of reaction networks, as shown in Fig.~\ref{fig:Snapshot-molzart}. It enables high-quality virtual screening of covalent inhibitors and other bond-forming events by combining rigorous mechanistic modeling with a machine learning layer for enhanced speed and accuracy. MolZart allows to efficiently evaluate reactivity and selectivity across parametrized chemical spaces, supporting \textit{de novo} drug discovery workflows with first-principles-reliable and scalable results.

\begin{figure}[!htbp]
    \centering
    \includegraphics[width=0.7\linewidth]{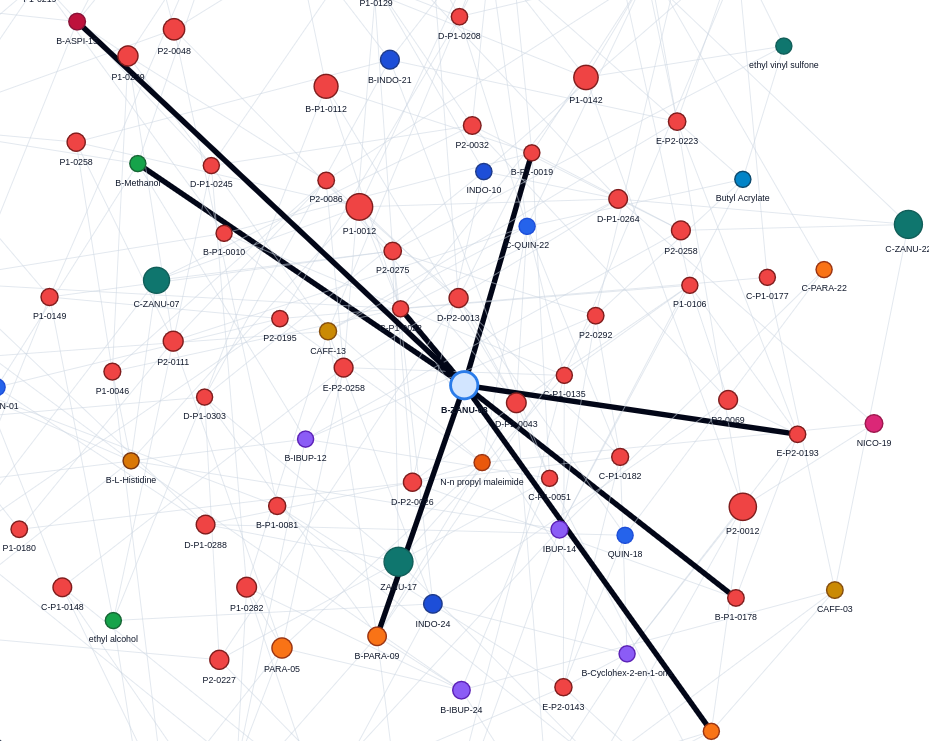}
    \caption{Snapshot of reaction graph in MolZart, where zanubrutinib is connected via edges with various nucleophilic compounds, for which barrier height is calculated.
    }
    \label{fig:Snapshot-molzart-network}
\end{figure}
\begin{figure}[!htbp]
    \centering
    \includegraphics[width=\linewidth]{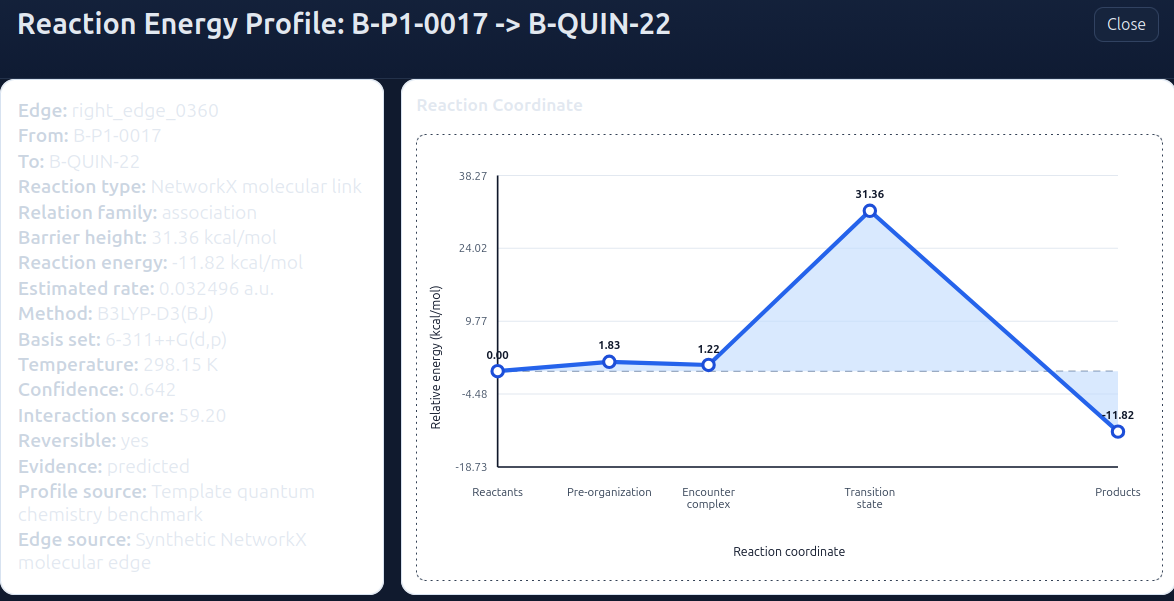}
    \caption{Example reaction energy profile generated with MolZart, sampling pre-organization, pre-complex formation, transition state, and product points relative to free reactants.
    }
    \label{fig:reaction-profile}
\end{figure}

\section{Case Study: Michael addition in covalent docking of zanubrutinib to Bruton's Tyrosine Kinase}
\label{sec:case-study}

\subsection{Introduction}

\paragraph{Covalent Docking in Computer-Aided Drug Design.}
Computer-aided drug design (CADD) is an essential step in modern drug discovery, as it can facilitate more efficient use of resources by reducing the time and cost of experimental testing in a drug design and discovery pipeline ~\cite{Sliwoski2014}. In CADD, molecular docking is a widely used method for predicting ligand binding modes and estimating interaction energies with a protein target ~\cite{Sliwoski2014}. However, conventional docking algorithms do not explicitly include the possibility of forming a covalent bond between a ligand and a protein, and therefore require methodological extensions when targeting inhibitors that bind covalently ~\cite{Scarpino2018,Wen2019}. To address this challenge, several covalent docking tools have been developed, including AutoDock4, CovDock, FITTED, GOLD, ICM-Pro, Schrodinger, and MOE, each employing distinct strategies for handling covalent bond formation ~\cite{Scarpino2018,Wen2019}.

Covalent docking is more complex than conventional docking, where a ligand is placed in a binding pocket without forming any bonded interactions. Not only must covalent docking methodologies first perform a noncovalent placement of the ligand in the binding pocket, but they must also form a new chemical bond between the ligand warhead and the target residue. Covalent bond formation alters ligand topology and depends on protonation state, reaction geometry, and intrinsic electrophile–nucleophile reactivity. These are features that standard noncovalent scoring functions are not designed to perform and evaluate~\cite{London2014, Scarpino2018}. Consequently, most covalent docking workflows rely on predefined reactive residues and reaction types or employ geometric approximations, thereby limiting automation and transferability ~\cite{London2014,Wen2019}. A major limitation of current approaches is their dependency on empirical scoring functions that approximate covalent bond energetics at the molecular mechanics level, potentially neglecting key electronic effects governing bond formation ~\cite{Scarpino2018}. Hybrid methods that incorporate quantum-mechanical calculations have demonstrated improved accuracy~\cite{Li2024}, and benchmarking efforts have highlighted the need for reaction-specific evaluation and standardized datasets for validating covalent docking tools ~\cite{Wei2022,Wen2019}. 

\paragraph{Bruton's Tyrosine Kinase.}
Bruton’s tyrosine kinase (BTK) is a non-receptor cytoplasmic tyrosine kinase belonging to the Tec family kinases (TFK) and is predominantly expressed in B lymphocytes~\cite{Sousa2022,Nawaratne2024}. BTK plays a central role in the B-cell receptor (BCR) signaling cascade, where its activation triggers phosphorylation of phospholipase C gamma 2 (PLC$\gamma$2), calcium mobilization, and downstream activation of the nuclear factor kappa-light-chain-enhancer of activated B cells (NF-kB) and mitogen-activated protein kinase (MAPK) pathways, promoting B-cell proliferation and survival~\cite{Guo2019,Nawaratne2024}. As an ATP-dependent kinase, BTK binds ATP within its catalytic binding pocket and mediates phosphorylation of downstream substrates. This ATP-binding site contains a cysteine residue (Cys481) that can be covalently targeted by irreversible inhibitors~\cite{Nawaratne2024}. The important physiological function of BTK is highlighted by the discovery of loss-of-function mutations that alter B-cell maturation, resulting in frequent and severe infections, a condition called X-linked agammaglobulinemia~\cite{Nawaratne2024}. Conversely, increased BTK activity can contribute to the pathogenesis of B-cell malignancies such as chronic lymphocytic leukaemia and mantle cell lymphoma, establishing BTK as a highly attractive therapeutic target~\cite{Nawaratne2024,Guo2019}.
 
\paragraph{Development of Covalent BTK Inhibitors.}
The first approved BTK inhibitor for clinical use, ibrutinib, is an irreversible covalent inhibitor that targets Cys481 in the ATP-binding pocket of BTK. This covalent binding of ibrutinib to Cys481 ensures sustained target inhibition, even after short systemic exposure, due to this irreversible modification of the catalytic domain ~\cite{Nawaratne2024}. However, ibrutinib exhibits off-target effects by inhibiting other kinases, such as EGFR, among other members of the TEC family~\cite{Nawaratne2024, Guo2019}. 

\begin{figure}
    \centering
    \includegraphics[width=1\linewidth]{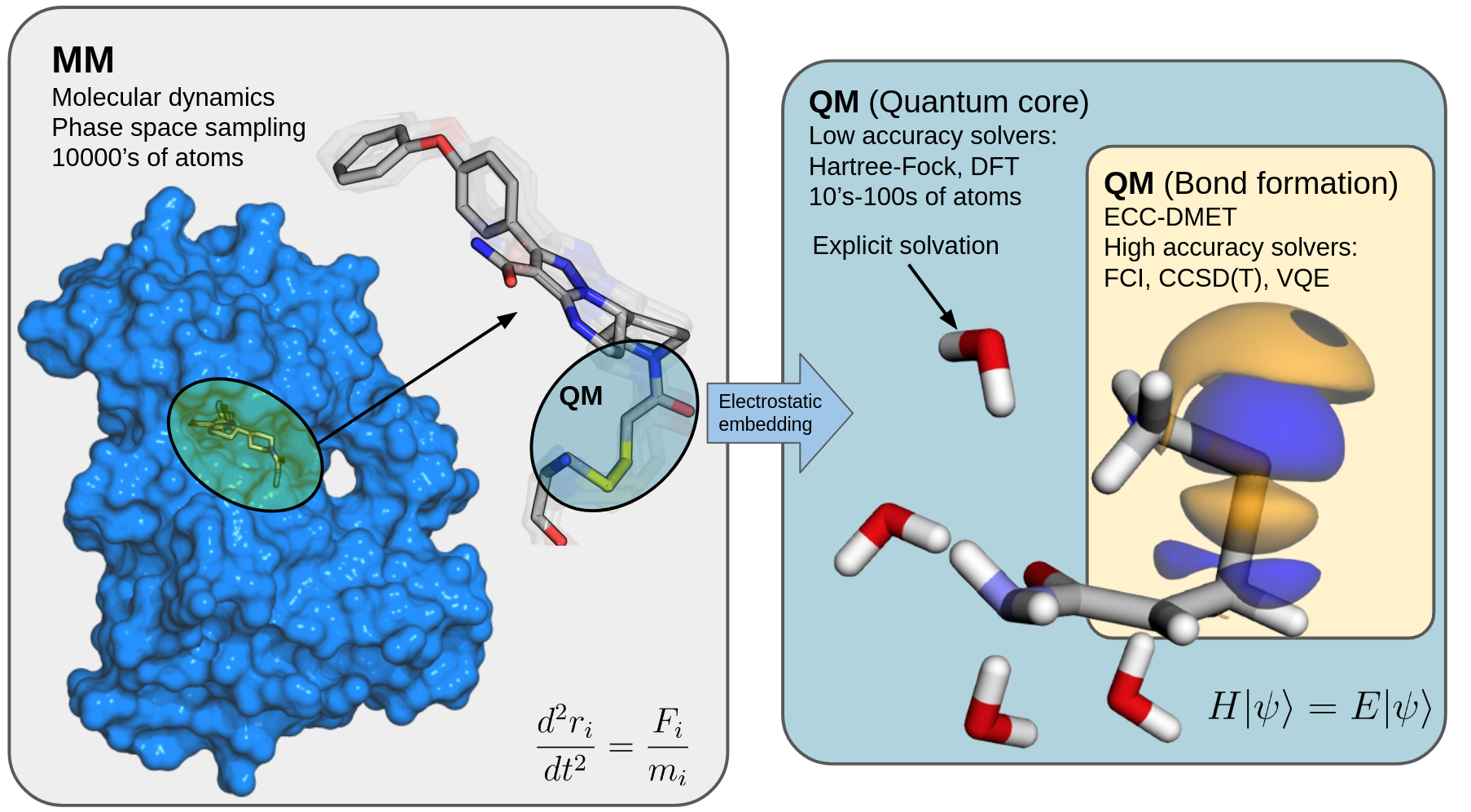}
    \caption{Our QM/QM/MM framework applied to study covalent docking of zanubrutinib to
Bruton’s Tyrosine Kinase.}
\end{figure}

To improve selectivity, second-generation inhibitors such as acalabrutinib and zanubrutinib were developed~\cite{Nawaratne2024}. These BTK inhibitors were developed through structure-guided medicinal chemistry optimization to retain covalent binding to Cys481 while improving kinase selectivity relative to first-generation inhibitors. Zanubrutinib was found to demonstrate enhanced selectivity and sustained BTK target occupancy over acalabrutinib, as well as first-generation inhibitors ~\cite{Nawaratne2024, Guo2019}.

\paragraph{Mechanism and Advantages of Covalent Inhibition.}
Covalent inhibitors are characterized by the formation of a covalent bond between an electrophilic group, also termed a “warhead”, on the ligand and a nucleophilic residue on the target protein~\cite{Scarpino2018,Chen2023}. In kinase-targeting covalent inhibitors, cysteine residues are frequently exploited due to their nucleophilic thiol side chain, and for zanubrutinib, an acrylamide-based warhead undergoes a Michael addition reaction to form the covalent bond with a cysteine residue in the binding pocket ~\cite{Scarpino2018, Chen2023, Roseli2022}.
 
Covalent inhibition generally binds to the target via a two-step mechanism in which 1) initial noncovalent binding positions the electrophile in proximity to the target residue, and 2) a covalent bond is formed to the nucleophilic amino acid ~\cite{Scarpino2018}. The efficiency of inhibition, therefore, depends on both the initial binding affinity and the rate of covalent bond formation, resulting in prolonged target engagement. Compared to non-covalent inhibitors, covalent inhibitors can achieve sustained pharmacodynamic effects even after plasma clearance, reduced dosing frequency, and improved target occupancy. In kinases with highly conserved ATP-binding pockets, the presence of Cys481 in BTK, which is not universally conserved across the human kinome, provides an additional selectivity that reversible inhibitors cannot exploit ~\cite{Guo2019, Nawaratne2024}. However, acquired mutations at Cys481 can also occur and represent a challenge, as this would prevent irreversible inhibition ~\cite{Nawaratne2024}.
 
For BTK, structural confirmation of covalent bond formation with Cys481 underscores the necessity of explicitly modeling both the pre-reactive complex and the covalently bound state~\cite{Guo2019}. In such situations, MD simulations with reactive force fields, such as ReaxFF~\cite{vanDuin2001ReaxFF} carry substantial limitations for the use in covalent docking, especially when chemical accuracy is required.

\subsection{Reaction profile and transition state}\label{sec:irc}

\paragraph{System Preparation and Molecular Dynamics Characteristics.}
The initial geometry for the zanubrutinib-BTK complex was obtained from the RCSB Protein Data Bank (PDB)~\cite{Berman2000} using the crystal structure with PDB identifier \texttt{6J6M}.
System preparation was performed using AmberTools~\cite{Case2023} with the following parameters described in app.~\ref{app:md_details}.
All simulations were performed under periodic boundary conditions using a 2~fs integration time step. The final set of MD geometries, corresponding to a selected subset of timesteps, is passed to the quantum solver module for the calculation of ensemble-averaged quantities.

\paragraph{Reaction Profile.}
To investigate the zanubrutinib-BTK complex reaction barrier, we resorted to a reduced-size model composed of acrylamide and methanethiolate anion.
We determined the reaction mechanism on this system with Intrinsic Reaction Coordinate (IRC) calculations~\cite{ishida1977intrinsic} as implemented in \texttt{ORCA} software~\cite{ORCA}. We ran the calculations with the Density Functional Theory (DFT) \cite{hohenberg1964inhomogeneous,kohn1965self} using the range-separated hybrid $\omega$B97X functional \cite{chai2008systematic} combined with Grimme's DFT-D3~\cite{grimme2010consistent} dispersion correction using Becke-Johnson (BJ) damping \cite{grimme2011effect}. The resulting $\omega$B97X-D method can reproduce the geometries and energetics of accurate ab initio wave function methods accurately \cite{smith2013range,awoonor2020quantum}.
IRC calculations were performed using the augmented correlation-consistent double-$\zeta$ Dunning (aug-cc-pVDZ) basis set~\cite{dunning1989a,kendall1992a,woon1993a}, in combination with the aug-cc-pVDZ/C auxiliary basis set for density fitting~\cite{ORCA}.
Environmental effects were modeled at two-levels: 1) through the Conductor-like Polarizable Continuum Model (C-PCM) \cite{barone1998quantum} by using a dielectric permittivity $\varepsilon=4$ to mimic electronic polarization within the protein~\cite{gilson1988calculation}; 2) using the full QM/MM model with explicit water solvent molecules and external classical charges and dispersion interactions (cf. Fig.~\ref{fig:michael}).

\begin{figure}[!ht]
    \centering
    \begin{minipage}{0.48\linewidth}
        \centering
        \includegraphics[width=\linewidth]{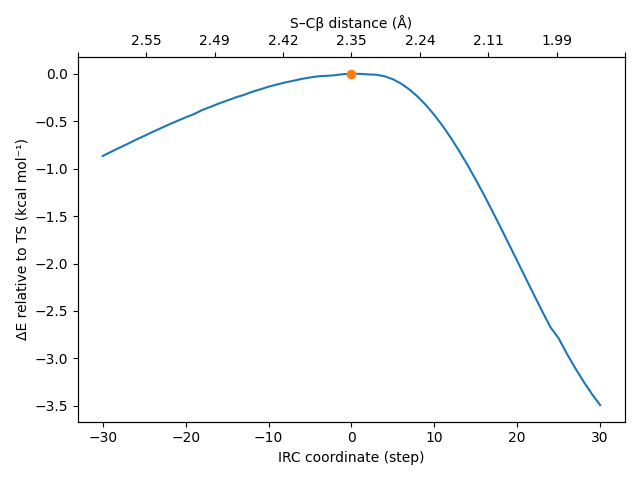}
    \end{minipage}
    \hfill
    \begin{minipage}{0.48\linewidth}
        \centering
        \includegraphics[width=0.7\linewidth]{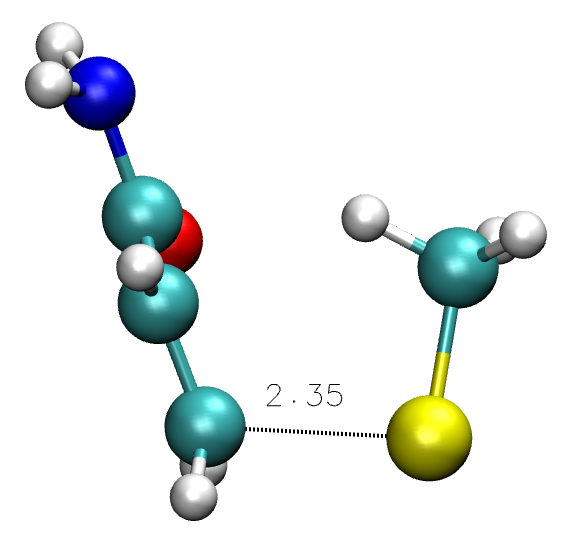}
    \end{minipage}
    \caption{Left: IRC energy profile for the model reaction as a function of the reaction coordinate (bottom axis) and S--C$_\beta$ distance (top axis). The transition state is marked at zero energy. Right: optimized transition state geometry.}
    \label{fig:irc_ts}
\end{figure}

In the C-PCM model, the reaction profile obtained from the IRC calculation is reported in Fig.~\ref{fig:irc_ts}. The energy is shown relative to the transition state, which is set to zero, and plotted as a function of both the IRC coordinate and the corresponding S--C$_\beta$ distance.

The IRC profile exhibits a single well-defined barrier, confirming that the identified transition state connects reactants and products along a continuous reaction pathway. The reaction coordinate can be directly mapped onto the S--C$_\beta$ distance, which decreases monotonically from the reactant region (ca.~2.5~\AA) to the product region (ca.~2.0~\AA), reflecting the progressive formation of the covalent bond.
At the transition state, the S--C$_\beta$ distance is 2.35~\AA, indicating a partially formed bond. This is consistent with a late transition state along the nucleophilic addition pathway. The IRC profile exhibits a single barrier with no intermediate minima, supporting a concerted reaction mechanism.

\begin{figure}[h]
    \centering

    \begin{subfigure}{0.3\textwidth}
        \centering
        \includegraphics[width=\linewidth]{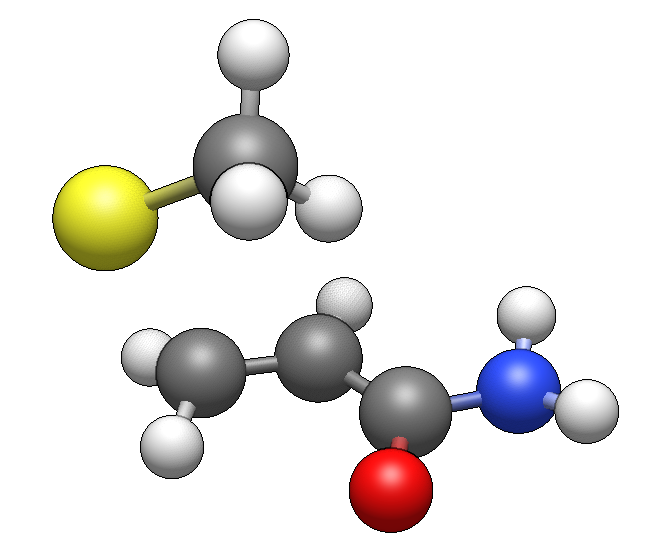}
        \caption{No explicit water}
    \end{subfigure}
    \begin{subfigure}{0.3\textwidth}
        \centering
        \includegraphics[width=\linewidth]{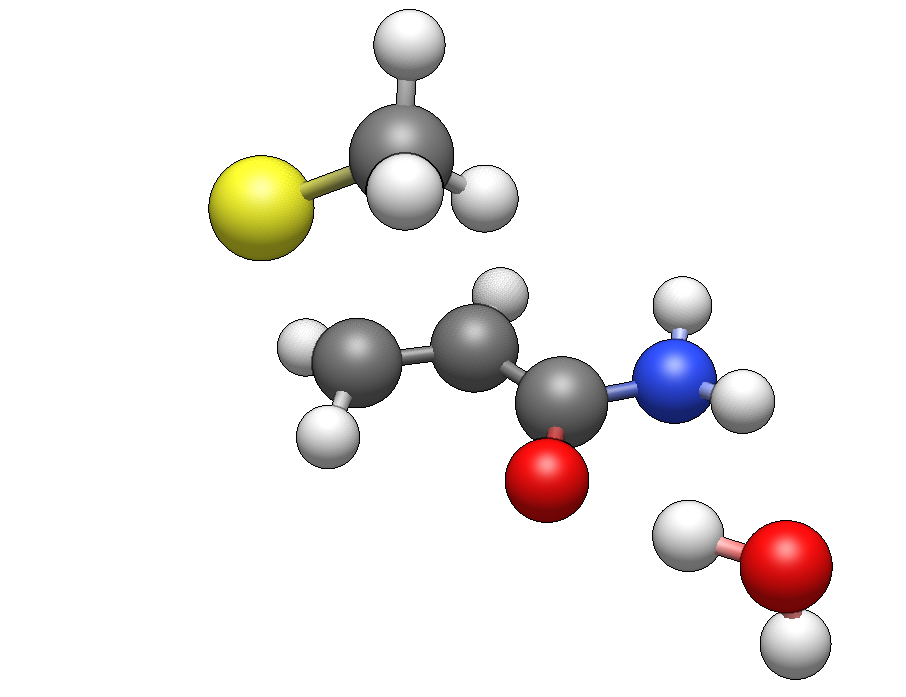}
        \caption{Explicit water-oxygen bond}
    \end{subfigure} 

    \vspace{0.2cm}

    \begin{subfigure}{0.3\textwidth}
        \centering
        \includegraphics[width=\linewidth]{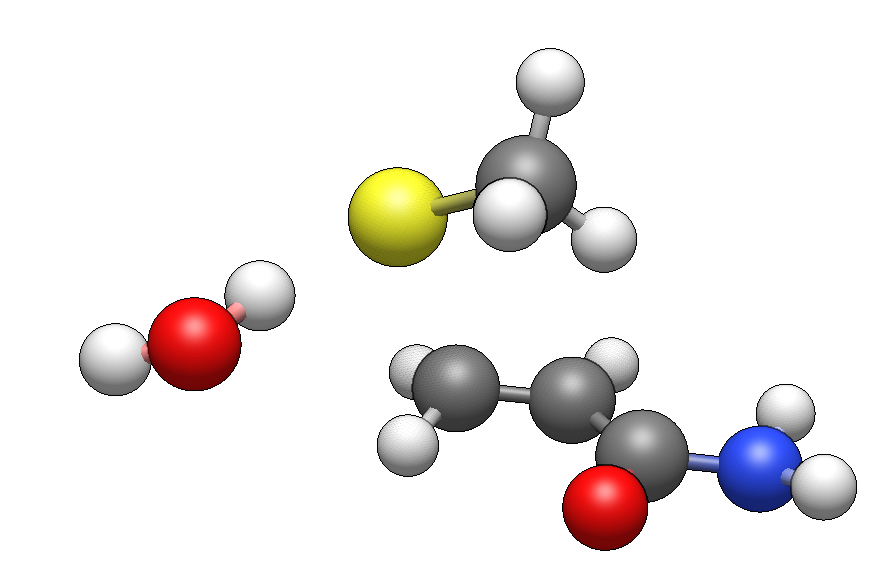}
        \caption{Explicit water-sulfur bond}
    \end{subfigure}
    \begin{subfigure}{0.3\textwidth}
        \centering
        \includegraphics[width=\linewidth]{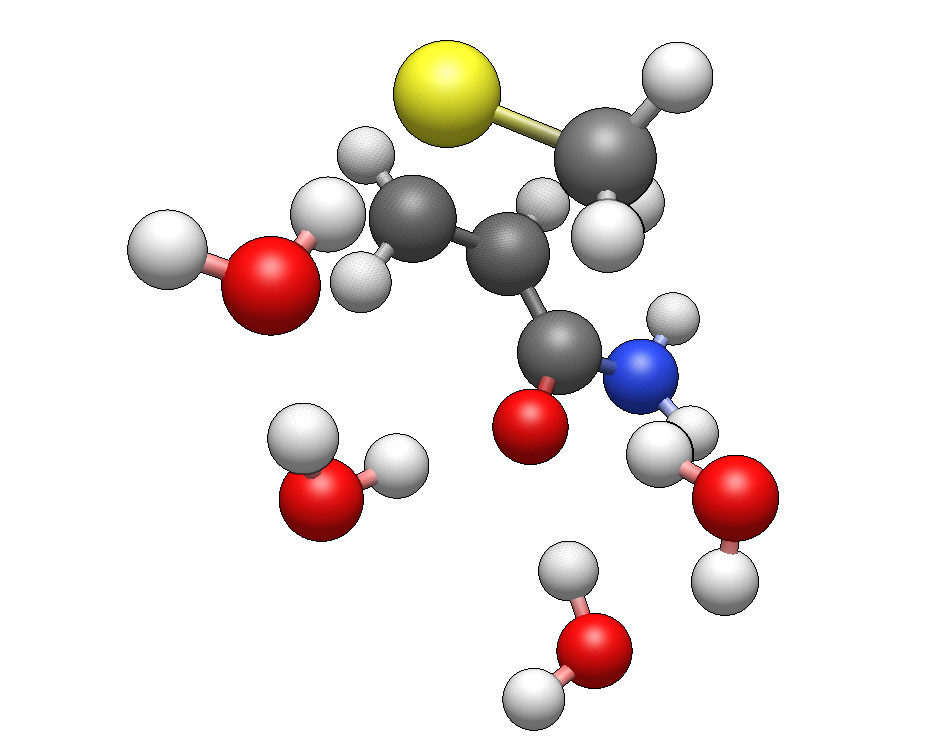}
        \caption{Explicit water cluster}
    \end{subfigure}

    \caption{Geometries of the tested systems for which the transition state search was calculated. The calculations were performed with DFT-$\omega$B97X-D3BJ using aug-cc-pVDZ basis with density fitting. No QM/MM interaction or implicit solvent models were used. Only for the system with an explicit water cluster (d) was the transition state search successful.}
    \label{fig:explicit_water_system_0}
\end{figure}

\begin{figure}[H]
    \centering
    \includegraphics[width=0.5\linewidth]{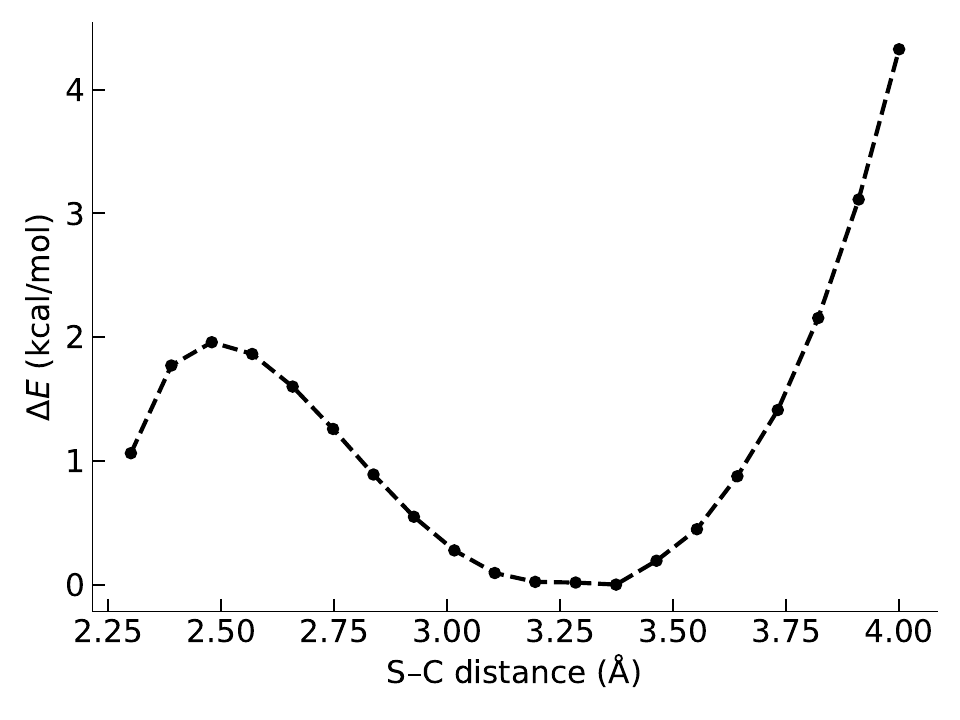}
    \caption{Energy profile obtained using our QM/MM model for acrylamide-cysteine complex with explicit water solvent. The energies were calculated with DFT-wB97X-D3BJ with aug-cc-pVDZ basis and aug-cc-pVDZ/C auxiliary basis used for resolution of identity approximation.}
    \label{fig:QM-MM-profile}
\end{figure}

Initial testing showed that it within the QM/MM model with Coulomb and dispersion interactions, is not sufficient to optimize the protein-ligand system alone in order to obtain accurate transition state energies. For this reason, we performed calculations on a reduced-size model composed of acrylamide and methanethiolate anion, with and without extending the quantum region by including explicit water molecules, as shown in Figure~\ref{fig:explicit_water_system_0}. For DFT-wB97X-D3BJ calculation with aug-cc-pVDZ basis, only the system with explicit water cluster yielded a transition state. As a result, by default, the quantum region was extended to include water molecules within 3~$\AA$ distance.  An example reaction profile for a representative MD snapshot for the acrylamide-cysteine complex is displayed in Fig.~\ref{fig:QM-MM-profile}.

\subsection{Reaction barrier within QM/QM/MM framework}
In order to obtain energy barrier with the QM/QM/MM framework, we first calculated the energy profile with QM/MM scheme using DFT, as shown in Fig.~\ref{fig:QM-MM-profile}. Transition state and pre-complex geometries correspond to the local maximum and minimum, respectively. For these two geometries QM/QM/MM calculation was performed, using the DMET method. Natural atomic orbitals (NAO) were used as a localized basis. Fragment orbitals where chosen based on the overlap with atomic orbital of the chemically important atoms (sulfur for cysteine, $\alpha$-carbon and $\beta$-carbon for acrylamide).  The number of orbitals in the bond formation region was set to match the number of orbitals in the reduced-size model composed of acrylamide and methanethiolate anion in cc-pVDZ basis. The results of the QM/QM/MM calculation for the same MD snapshot as Fig.~\ref{fig:QM-MM-profile}, as well as the corresponding reference calculation with Hartree-Fock and DFT methods, are shown in Table~\ref{tab:dmet_barrier_results}.
\begin{table}[]
    \centering
    \begin{tabular}{c|c|c|c}
         &$E^{TS}$ [a.u.]& $E^{pre-complex}$ [a.u.]&$\Delta E$ [kcal mol${^{-1}}$]\\
         \hline
         QM/MM - DFT& $-1276.36852$ & $-1276.37164$ & $1.96$\\
         \hline
         QM/MM - HF& $-1270.40294$ & $-1270.42095$ & $11.30$\\
         \hline
         QM/QM/MM - CCSD& $-1270.48056$ & $-1270.49026$ & $6.09$\\
    \end{tabular}
    \caption{Comparison between calculated energies within the QM/MM framework using Hartree-Fock and DFT for the entire quantum region and the QM/QM/MM results, where quantum region is treated with the DMET method. Single For the bond formation region, consisting of 146 active orbitals, CCSD solver was used, while the environment is treated at mean-field level for a single MD snapshot.}
    \label{tab:dmet_barrier_results}
\end{table}

\subsection{Results: transition state (ECC-DMET)}\label{sec:dmet-ts}
For finding single-point electronic energy at the transition state, we used our version of density matrix embedding theory. DMET requires choosing the fragment orbitals, as explained in sec.~\ref{sec:summary_contribution}.
These orbitals should be localized, and preferably, they should be carry information about electron correlation. Our ECC-DMET method (sec.~\ref{sec:ECC-DMET}) guides the selection of fragment orbitals using quantum-information metrics imposed on localized orbitals. As a consequence, fragment choice in DMET is less heuristic and achieves better accuracy using a lower number of orbitals.

We constructed the orbital optimization unitary defined in eq.~\eqref{eq:unitary} by accounting for the classification of localized Boys orbitals based on their support on specific atoms that participate in the chemical reaction. The admissible rotations have the block-diagonal form $\mathbf U = \bigoplus \mathbf U_i$,
where each $\mathbf U_i$ is defined as the set of orbitals associated with a given atomic equivalence class defined in app.~\ref{appendix:localization}. Our objective function is the Frobenius norm of the fragment-environment block of the 1-RDM. We minimize it to decouple the fragment from the environment, lowering entanglement between these subsystems ($\omega^{(1)}_3=1$ in eq.~\eqref{eq:leakage-functional}). To the best of our knowledge, no other dedicated method based on quantum-information properties has been proposed for selecting DMET fragment orbitals. For this reason, for the baseline reference in benchmarking our method, we chose a set of chemically-motivated orbitals, selected in a systematic way. Chemically-motivated orbitals include Boys orbitals having the largest overlap with the ``chemical orbitals'' obtained by performing the Pipek-Mezey localization on occupied and virtual orbitals separately.

\begin{figure}[H]
    \centering
    \includegraphics[width=0.9\linewidth]{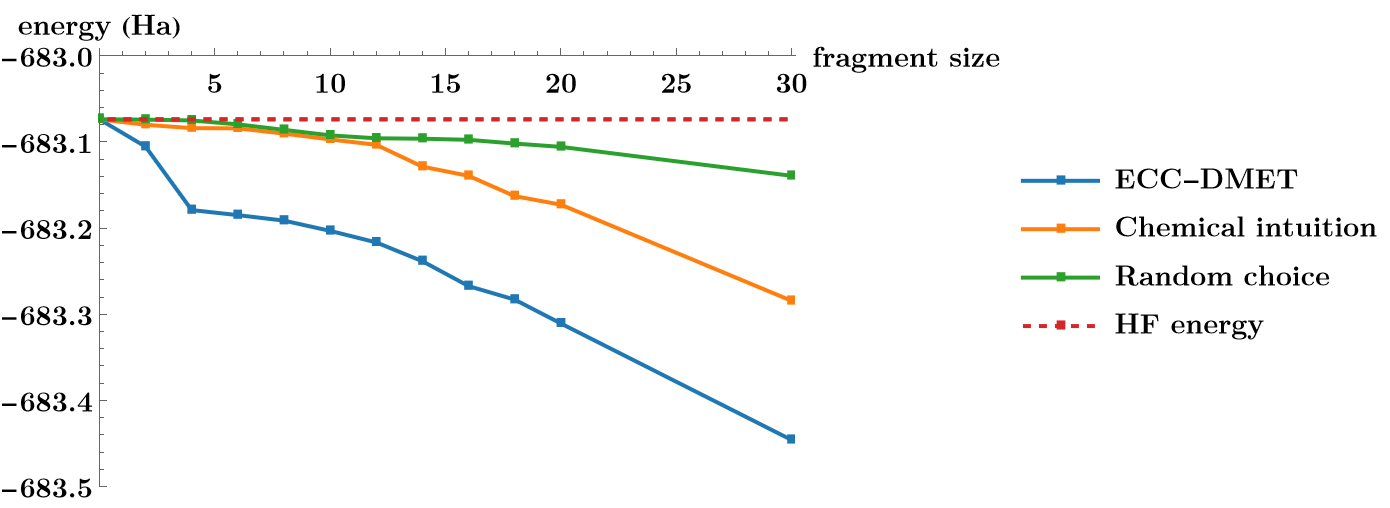}
    \caption{Electronic energies calculated at the TS geometry of acrylamide-methanethiol in aug-cc-pVDZ basis with QIO, chemically-motivated and random orbitals. The red dashed horizontal line represents HF energy.}
    \label{fig:MI_vs_random_selection}
\end{figure}

In Fig.~\ref{fig:MI_vs_random_selection}, we compare electronic energies at TS between the ECC-DMET selection, chemically-motivated selection and random selection of orbitals. 
Here, we used CCSD as the post-HF fragment solver with a non-correlated environment reference state, while the solvent was described using C-PCM ($\varepsilon = 4$).

\begin{figure}[H]
    \centering
    \includegraphics[width=0.7\linewidth]{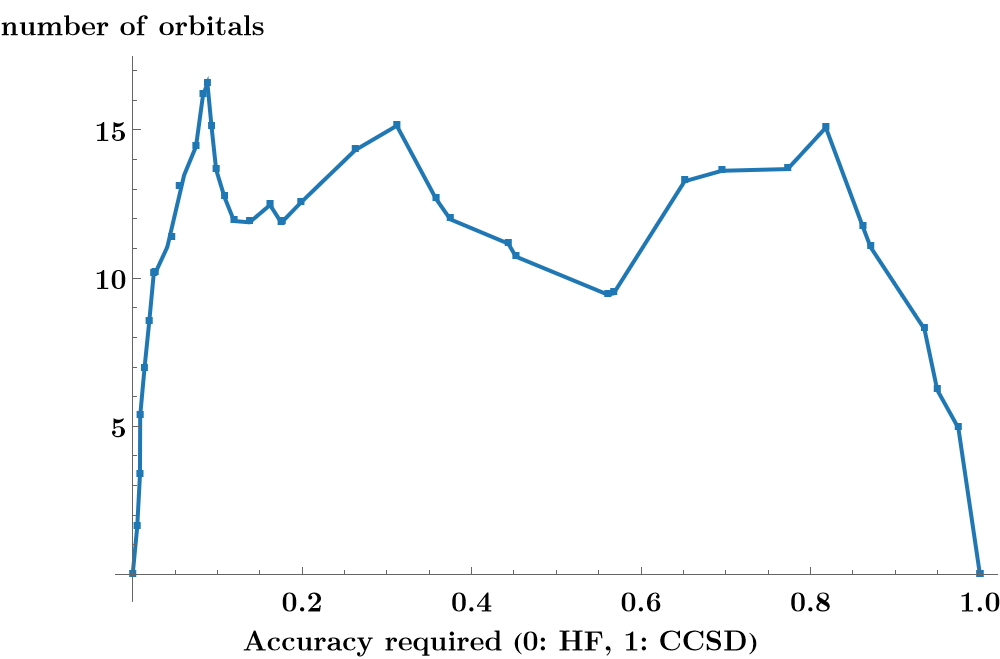}
    \caption{Reduction in the number of fragment orbitals as a function of energy accuracy for ECC-DMET vs. standard (chemically motivated) DMET. For matching a standard DMET energy at a given accuracy (horizontal axis), significantly fewer (10-15) QIO fragment orbitals are required.
    For accuracy 0 or 1, there is no improvement as then fragment + bath is trivial -- it is either empty or it spans all the orbitals. As a baseline, we consider the chemically-motivated orbitals that are involved in bond formation.
    The system we consider here is the same as that in Fig.~\ref{fig:MI_vs_random_selection}.}
    \label{fig:orbital_number_improvement}
\end{figure}

Our results show that in order to match the electronic energy calculated with 4 ECC-DMET orbitals, it is necessary to include more than 20 chemically-motivated orbitals. Thus our method gives 5-fold reduction of resources for this small-scale example. The improvement does not change significantly with the accuracy (or the fragment size), as shown in Fig.~\ref{fig:orbital_number_improvement}. We thus conclude that for a fixed fragment size, our method generally requires significantly fewer orbitals for a given accuracy. 
Even though the objective function we used for the zanubrutinib-BTK study is constructed from 1-RDMs alone, our results show that it can be successfully applied to refine the DMET-fragment selection in a way that improves the energy.

\paragraph{Insight into static correlation}
To quantify the extent of static correlation at the transition state, we carried out a CASSCF calculation using an AVAS-defined active space \cite{Sayfutyarova2017AVAS}. The AVAS selection was performed with a truncation threshold of 0.3, targeting S($3p$) on the methanethiolate and C/O($2p$) on the acrylamide acceptor, within an aug-cc-pVDZ basis and a C-PCM embedding to mimic the protein dielectric environment. All calculations were performed with \texttt{PySCF} \cite{pyscf1,pyscf2}.

AVAS returns an active space of (26e,17o), which we then used in CASSCF. The resulting natural orbital occupation numbers (NOONs) indicate a predominantly closed-shell electronic structure with moderate multiconfigurational character: the most occupied active orbitals remain very close to doubly occupied (8 orbitals with NOON $\ge 1.998$), while the least occupied ones are only weakly populated (NOON $\simeq 0.02$--$0.05$). Consistently, the Head--Gordon effective number of unpaired electrons \cite{HEADGORDON2003508} is $N_\mathrm{unpaired}=0.55$, pointing to non-negligible but not dominant open-shell character.
At the same time, the CI expansion is not dominated by a single determinant: the largest configuration weight is 0.62 (with the top-5 and top-10 configurations accounting for 0.80 and 0.85 of the wavefunction weight, respectively), supporting the presence of moderate static correlation within the chosen active space. 

Finally, comparison with correlated single-reference calculations highlights that dynamical correlation dominates the energetics: for the same basis set, CCSD/DF yields a substantially lower total energy ($E_\mathrm{CCSD/DF}=-684.2448$~Hartree), far exceeding the CASSCF energy ($E_{\mathrm{CASSCF}}=-683.1257$~Hartree). Therefore, while static correlation is detectable at the transition state, we expect it to provide a secondary contribution to the reaction energetics compared to dynamical correlation, and we treat the system as effectively single-reference for the purposes of benchmarking and energy comparisons.

\subsection{Results: reaction barrier calculations}
\label{sec:results}

We now move to the determination of the reaction barrier energy for our model. The reaction barrier $\Delta E$ is defined as the energy difference between the transition state geometry and the pre-complex geometry (see Eq. \ref{eq:energy_diff}), defined as the last step obtained from the IRC calculation described in Sec. \ref{sec:irc}, consistently with QM/MM calculations on MD snapshots. 

We first performed reference calculations using \texttt{PySCF} \cite{pyscf1,pyscf2} across different electronic structure methods, basis sets, and environments. In particular, we considered HF, $\omega$B97X-D3BJ, MP2, and CCSD levels of theory, with CCSD calculations carried out using the density fitting (DF) approximation to reduce computational cost \cite{dunlap2000robust}. 

A systematic exploration of basis set effects was performed using cc-pVDZ and cc-pVTZ basis sets, as well as their truncated diffuse (jun-cc-pVDZ, jun-cc-pVTZ)~\cite{papajak2011a} and augmented (aug-cc-pVDZ, aug-cc-pVTZ) variants. For CCSD-DF calculations, auxiliary basis sets were taken from the default \texttt{PySCF} implementation: even-tempered basis sets were used for the jun-cc-pVXZ family~\cite{pu2025enhancing}, while standard (aug-)cc-pVXZ-jkfit basis sets were employed in all other cases~\cite{weigend2002fully}.

\begin{figure}[!ht]
    \centering
    \includegraphics[width=\linewidth]{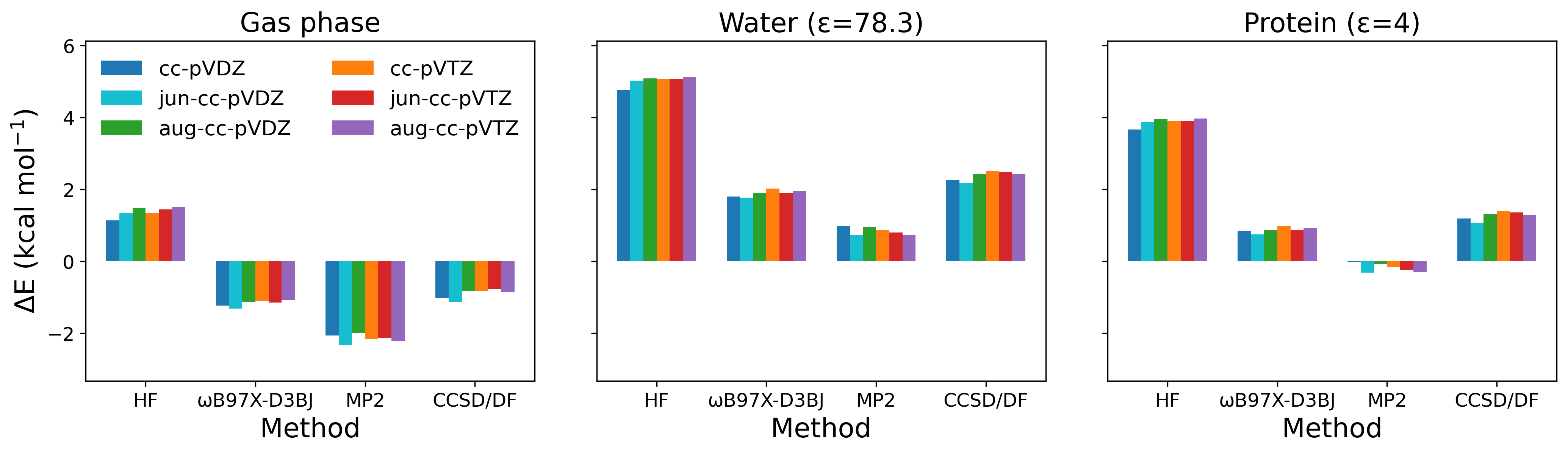}
    \caption{Reaction barrier $\Delta E$ computed with different electronic structure methods and basis sets in gas phase, aqueous solution ($\varepsilon=78.3$), and protein-like environment ($\varepsilon=4$). Colors denote different basis sets. 
    }
    \label{fig:benchmark}
\end{figure}

All calculations were performed in three environments: gas phase, aqueous solution ($\varepsilon=78.3$), and a protein-like medium ($\varepsilon=4$). The resulting barrier heights are summarized in Fig.~\ref{fig:benchmark}.

In the gas phase, the predicted barrier heights remain strongly dependent on the level of theory. HF is the only method that consistently predicts a positive barrier, with values ranging from 1.13 to 1.51 kcal/mol across the basis-set series. In contrast, all correlated methods predict negative barriers: $\omega$B97X-D3BJ gives values between $-1.32$ and $-1.08$ kcal/mol, MP2 between $-2.33$ and $-2.00$ kcal/mol, and CCSD/DF between $-1.13$ and $-0.78$ kcal/mol. This behavior reflects the lack of electron correlation in HF, which leads to an overestimation of the barrier height, whereas correlated methods stabilize the transition state relative to the precomplex.

The inclusion of an implicit environment has a pronounced effect on the energetics. Moving from the gas phase to aqueous solution leads to a substantial increase in the barrier height for all methods, and all computed barriers become positive. In water, HF predicts barriers between 4.76 and 5.13 kcal/mol, $\omega$B97X-D3BJ between 1.77 and 2.02 kcal/mol, MP2 between 0.73 and 0.98 kcal/mol, and CCSD/DF between 2.18 and 2.51 kcal/mol. This trend is consistent with a stronger stabilization of the charge-separated precomplex, in particular the thiolate-containing reactant complex, by the polar solvent. In the protein-like environment, the barriers are systematically reduced relative to water because of the weaker dielectric screening. 

A basis-set dependence is also observed, although its magnitude depends on the electronic-structure method. For HF, increasing the basis-set quality and adding diffuse functions generally lead to slightly larger barrier heights in all environments. For $\omega$B97X-D3BJ, the basis-set dependence is comparatively modest, with variations of only a few tenths of a kcal/mol across the full series. MP2 also shows relatively small but non-monotonic basis-set effects, particularly in the protein-like medium where the barrier remains close to zero. CCSD/DF exhibits a somewhat clearer basis-set trend, with triple-$\zeta$ basis sets generally giving slightly higher barriers than double-$\zeta$ ones, although the effect of diffuse augmentation remains moderate. Overall, the updated data indicate that the reaction barrier is governed primarily by the interplay between electronic-structure method and environment, while basis-set effects are secondary but still non-negligible.

Overall, the results highlight the strong interplay between electronic-structure method, basis set, and environment in determining the reaction barrier. In particular, at the $\omega$B97X-D3BJ/aug-cc-pVTZ level of theory, the updated calculations yield barrier heights of $\Delta E^\mathrm{DFT}_\mathrm{water} = 1.95$ kcal/mol and $\Delta E^\mathrm{DFT}_\mathrm{protein} = 0.93$ kcal/mol. These values should not be compared directly to the previously reported values for the same thio-Michael addition reactions ($\Delta E^\mathrm{DFT,lit}_\mathrm{water} = 9.27$ kcal/mol and $\Delta E^\mathrm{DFT,lit}_\mathrm{protein} = 4.84$ kcal/mol, Table 2 in Ref.~\cite{chaudhuri2024quantum}), because the two barrier definitions are based on different reference states. In the present work, the barrier is computed relative to the precomplex, whereas in Ref.~\cite{chaudhuri2024quantum} it is referenced to the isolated reactants.

In addition to the standard calculations discussed above, we also explored a DMET-based treatment following the scheme outlined in Sec.~\ref{sec:dmet-ts}. In the present case, however, the resulting reaction barriers did not prove sufficiently robust for quantitative discussion, due to an inconsistency in the choice of orbitals between the TS and the precomplex. As the orbital space is enlarged, this lack of consistency leads to an increasingly unbalanced description of the two endpoints, yielding barrier values that are both noisy and physically unreliable.  Addressing this issue through a more consistent orbital-selection protocol will be the focus of future investigations. These data are therefore reported only for completeness in Appendix~\ref{sec:dmet-barrier}.

\subsection{Results: fault-tolerant resource estimates for reaction-barrier simulations}
We next quantify the fault-tolerant overhead associated with the quantum-simulation stage of a reaction-barrier evaluation. For each embedded cluster Hamiltonian produced by the ECC-DMET/QM/MM workflow, we construct a qubitized quantum phase estimation (QPE) procedure from a double-factorized representation of the electronic Hamiltonian and estimate the resulting total T-gate count. We then compare two preprocessing choices: standard double factorization and the symmetry-optimized BEIT factorization introduced in Sec.~\ref{sec:FTQC}. Because both choices are benchmarked at the same target precision, this comparison isolates the impact of Hamiltonian preprocessing on the dominant fault-tolerant resource metric.

\begin{figure}[H]
    \centering
    \includegraphics[width=0.7\textwidth]{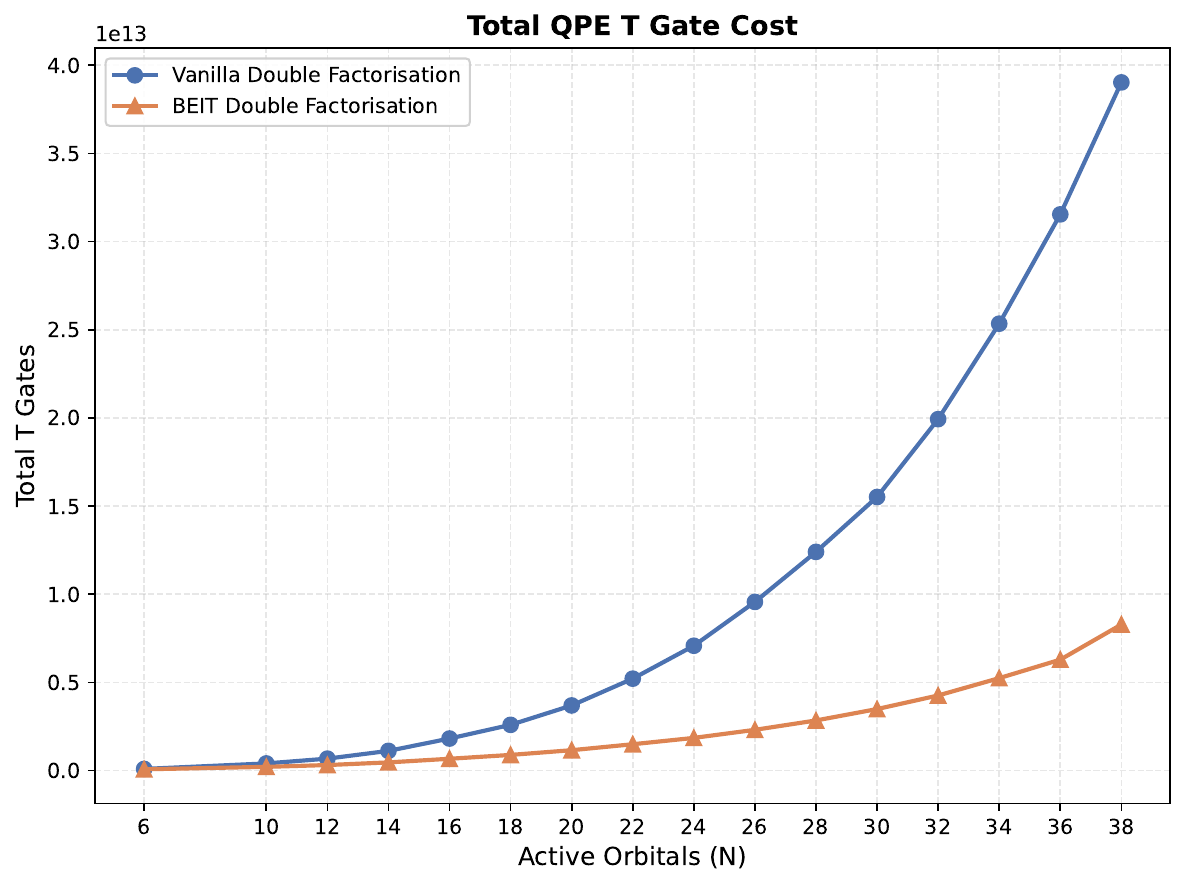}
    \caption{Estimated total T-gate cost for qubitized quantum phase estimation as a function of the number of active (fragment) orbitals in the embedded Hamiltonian. The curves compare standard double factorization with the symmetry-optimized BEIT factorization introduced in Sec.~\ref{sec:FTQC}. The optimized factorization lowers the block-encoding normalization and therefore reduces the number of walk-operator applications required to reach a fixed target precision. The benefit grows with active-space size and approaches a fivefold reduction for the largest instance shown.}
    \label{fig:ftqc_beit_df_t_gate_benchmark}
\end{figure}

The resulting T-gate estimates are shown in Fig.~\ref{fig:ftqc_beit_df_t_gate_benchmark} as a function of the number of active (fragment) orbitals. 
The optimized factorization yields a systematic reduction in T-count across the entire range studied. The improvement becomes increasingly pronounced as the active space grows. For the largest instance shown, the reduction is close to a factor of five.
This behavior follows directly from the structure of qubitized QPE. At fixed target energy precision and success probability, the number of gates scales linearly with the block-encoding normalization \(\lambda\). The symmetry-optimized factorization reduces this while representing the same electronic Hamiltonian in a way that is beneficial for the fault-tolerant implementations.

\section{Summary and Outlook}
In this work, we introduced a multiscale computational framework for the accurate modeling of reactive protein--ligand complexes, with particular emphasis on systems in which covalent bond formation and strong electronic correlation play a central role. The framework combines QM/QM/MM embedding, a correlated embedding treatment based on ECC-DMET, and quantum-information-guided orbital optimization (QIO) to construct compact embedded Hamiltonians that retain the essential correlation physics of the chemically active region while substantially reducing the dimensionality of the electronic problem. By construction, this embedded Hamiltonian serves as a common interface between classical quantum-chemistry solvers, GPU-accelerated circuit simulators, emerging quantum hardware, and fault-tolerant quantum-computing workflows.

This architecture addresses a central limitation of conventional computational drug-discovery pipelines: reaction barriers and bond-making/bond-breaking events are often treated using approximate scoring models or reduced empirical descriptions. In contrast, the present approach is rooted in first-principles electronic structure and is therefore better positioned to transfer across chemically diverse ligands, mechanisms, and protein environments. This feature is also important for the generation of high-fidelity datasets suitable for training next-generation machine-learning potentials and generative molecular-design models, where physical consistency and transferability are essential.
In addition, the choice of the orbitals for the active quantum site is determined based on the reference state alone, without the need for chemical insight into the reaction pathway. This enables automation for screening ligand databases. 
In other tools, the selection of the relevant chemical orbitals must be done manually on the basis of quantum chemical expertise, incurring additional costs.

The case study on covalent inhibition of Bruton's tyrosine kinase illustrates the practical value of the framework for modeling realistic biochemical reactions in complex environments. Our results show that QIO orbitals improve the compactness of the embedding and reduce the effective size of the correlated subsystem, thereby expanding the range of high-accuracy many-body solvers that can be brought to bear on the problem. We also find that an explicit quantum-chemical treatment of the local water network is essential for a reliable description of the transition state and the resulting barrier height.

Looking ahead, several directions will further expand the platform's capabilities. First, extensive benchmarking across diverse protein-ligand systems and reaction classes will be necessary to establish robust accuracy metrics for binding free energies and activation barriers. Second, further methodological developments in correlated embedding and orbital optimization may permit additional reductions in the size of the embedded cluster Hamiltonian, improving both classical and quantum computational efficiency. 
Third, treating one of the central problems of the energy barrier calculations, namely, the inconsistency of the active/fragment spaces across the reaction path.
Simultaneous rotation of the orbitals at all reaction path points will decrease the error associated with the estimation of the barrier energy.
Fourth, the integration of the framework with generative AI and machine-learning pipelines will enable the systematic generation of high-fidelity quantum-mechanical datasets for training next-generation molecular models. Finally, as fault-tolerant quantum computers mature, the presented architecture provides a natural interface for incorporating large-scale quantum electronic-structure calculations into practical drug-discovery workflows.

\section{Acknowledgments}
The authors gratefully acknowledge the support of the NVIDIA Inception Program for providing resources and technical assistance that contributed to this work. The authors thank the whole BEIT team for their support.

\printbibliography

\appendix

\section{Notation used}
\begin{table}[H]
    \centering
    \begin{tabular}{|c|c|}
    \hline
         symbol & meaning  \\ \hline
         $a_i$ & ann. operator for canonical molecular (CMO) orbitals  \\ \hline 
                 $\tilde{a}_i$ & ann. op. for W-rotated canonical molecular (CMO) orbitals \\ \hline 
         $d_i$ & ann. op. for localized molecular (LMO) orbitals \\ \hline  
         $b_i$ & annihilation operator for bath orbitals  \\ \hline   
         $\tilde{d}_i$ & ann. op. for quantum-information-optimized localized (QIO) orb.  \\ \hline   
         $\ket{f_i}$ & states in the Fock space of the fragment  \\ \hline  $\ket{e_i}$ & states in the Fock space of the environment  \\ \hline   
    \end{tabular}
    \caption{Notation for the annihilation operators and states in the second quantization, for the corresponding orbitals that we use across this publication.}
    \label{tab:notation}
\end{table}

\section{Entanglement-Consistent Correlated Density Matrix Embedding Theory}
\label{app:ECC-DMET-theory}
In this appendix section, we provide details of a new variant of DMET for electronic-structure calculations, which we call Entanglement-Consistent Correlated DMET (ECC-DMET). The central idea is to retain the DMET projector formalism while replacing purely mean-field bath selection with a correlated, orbital-entanglement-aware fragment+bath (cluster) space construction algorithm.

In summary, ECC-DMET combines three ingredients:
\begin{enumerate}
\item \textbf{Schmidt-consistent embedding}: preserve the exact fragment-environment Schmidt factorization property of DMET. 
\item \textbf{Correlated diagnostics}: use one- ($\gamma$) and 2-body RDMs ($\Gamma$), mutual information, and cumulant-derived metrics for cluster space construction and connect orbital entanglement with electron correlation and electronic energy.
\item \textbf{Controlled basis compression}: build a compact bath by SVD/rank truncation.
\end{enumerate}

The goal is to reduce basis size and improve robustness of calculations for strongly correlated systems while maintaining the favorable low-dimensional embedded electronic structure problem of standard DMET. 

\subsection{Preliminaries: notation and standard DMET}
We begin by setting up notation and with a pedagogical introduction to the standard version of DMET. For details of our new method, see sec.~\ref{app:ECC-DMET-theory-details} further on.

\subsubsection{Schmidt decomposition}
Any electronic state  $\lvert \Psi \rangle$ can be written as 
\begin{equation}
    \lvert \Psi \rangle
= \sum_{j=1}^{D_F} \sum_{k=1}^{D_E} \Psi_{jk}\, \lvert f_j \rangle \otimes \lvert e_k \rangle,
\end{equation}
where $\ket{f_j}$ and $\ket{e_k}$ are the basis states of the fragment and the environment, built from $D_F$ and $D_{E}$ orbitals, respectively. 
For the rest of this appendix, we will use $\lvert \Psi \rangle$ to denote the ground state of the system.
The coefficients matrix can be decomposed via singular value decomposition (SVD) 
\begin{equation}
\Psi_{jk} = \sum_{\ell=1}^{D_F} U_{j\ell}\,\lambda_\ell\,V_{k\ell},
\qquad
\Psi = U\,\Lambda\,V^\mathsf{T}.
\end{equation}
leading to the following form for the ground state:
\begin{equation}
\begin{aligned}
\lvert \Psi \rangle
&= \sum_{j=1}^{D_F}\sum_{k=1}^{D_E}
\left(\sum_{\ell=1}^{D_F} U_{j\ell}\lambda_\ell V_{k\ell}\right)
\lvert f_j\rangle\otimes\lvert e_k\rangle \\
&= \sum_{\ell=1}^{D_F} \lambda_\ell
\left(\sum_{j=1}^{D_F} U_{j\ell}\lvert f_j\rangle\right)
\otimes
\left(\sum_{k=1}^{D_E} V_{k\ell}\lvert e_k\rangle\right) \\
&= \sum_{\ell=1}^{D_F} \lambda_\ell \, \lvert \tilde f_\ell\rangle \otimes \lvert \tilde e_\ell\rangle.
\end{aligned}  
\end{equation}
The last expression is the Schmidt decomposition of the state. 
Instead of using a matrix of coefficients $\Psi_{jk}$, we describe the state by a single vector $\lambda_\ell$ in a special basis of the fragment and environment space. 
The rotated environment basis is defined by
\[
\lvert \tilde e_\ell\rangle = \sum_{k=1}^{D_E} V_{k\ell}\lvert e_k\rangle,
\]
so the set $\{\lvert \tilde e_\ell\rangle\}_{\ell=1}^{D_F}$ spans a $D_F$-dimensional subspace of the $D_E$-dimensional environment space.

\subsubsection{DMET theorem}
The DMET theorem states that if $\ket{\Psi}$ is an eigenstate of the Hamiltonian
\begin{equation}
    \hat H\lvert\Psi\rangle = E\lvert\Psi\rangle,
\end{equation}
then it is also an eigenstate of the embedded Hamiltonian 
\begin{equation}
\hat H_{\mathrm{emb}}\lvert\Psi\rangle = E\lvert\Psi\rangle,
\end{equation}
where $\hat H_{\mathrm{emb}} = \hat P\,\hat H\,\hat P$ is defined through a projector onto $2D_F$ orbitals
\begin{equation}
\hat P
= \sum_{j=1}^{D_F} \lvert f_j\rangle\langle f_j\rvert
\otimes
\sum_{\ell=1}^{D_F} \lvert \tilde e_\ell\rangle\langle \tilde e_\ell\rvert.
\end{equation}

\subsubsection{DMET with a mean-field reference state}
\paragraph{Preliminaries and notation.}
We use the notation $\ket{\Psi_0}$ for reference (mean-field) state with $\eta$  electrons in  $N$  orbitals.
In second-quantization, the reference state can be expressed by spin-orbital creation operators acting on the vacuum state:
\begin{equation}
    \lvert \Psi_0\rangle = \prod_{j\sigma \in I_{\mathrm{occ}}} a_{j\sigma}^{\dagger}\,\lvert 0\rangle,
\end{equation}
where
\begin{equation}
    a_{j\sigma}^{\dagger}\lvert 0\rangle = \lvert \phi_{j\sigma}\rangle,
\qquad
j=1,\ldots,N,
\qquad
\sigma\in\{0,1\},
\end{equation}
and $\lvert \phi_{j\sigma}\rangle$ are Hartree-Fock \textit{canonical molecular orbitals} (CMO).

In coordinate representation, the reference state is given by the Slater determinant formed from CMOs:
\begin{equation}
\langle r\mid\Psi_0\rangle
= \det\!\begin{pmatrix}
\phi_1(r_1) & \phi_1(r_2) & \cdots & \phi_1(r_\eta) \\
\vdots & \vdots & \ddots & \vdots \\
\phi_N(r_1) & \phi_N(r_2) & \cdots & \phi_N(r_\eta)
\end{pmatrix}.
\end{equation}
\paragraph{LCAO-MO expansion.}
The CMOs can be expanded on an atomic basis:
\begin{equation}
    \lvert \phi_{j\sigma}\rangle = \sum_{u=1}^{N} C_{uj}\,\lvert \chi_u\rangle,
\end{equation}
where $C\in\mathbb{R}^{N\times N}$ is the linear combination of atomic orbitals (LCAO-MO) coefficient matrix.
Here, we assume \textit{restricted}-formalism, where the spatial orbital is shared for both electron spin states $\sigma=\alpha,\beta$.
The set of atomic basis orbitals is defined as
\begin{equation}
    B_A = \{\lvert \chi_u\rangle\mid u=1,2,\ldots,N\},
\end{equation}
and
\begin{equation}
    B_{\mathrm{MO}} = \{\lvert \phi_{j\sigma}\rangle\mid j=1,2,\ldots,N;\ \sigma=\alpha,\beta\},
\end{equation}
is the set of molecular spin-orbitals. For Fock representation we define
\begin{equation}
  I_{\mathrm{occ}} = (n_{1\alpha},n_{1\beta},n_{2\alpha},n_{2\beta},\ldots,n_{N\alpha},n_{N\beta}),
\qquad
n_{u\sigma}\in\{0,1\},  
\end{equation}
the spin-orbital occupancy vector, with the property that $\lvert I_{\mathrm{occ}}\rvert = \eta$.
Orbitals can have occupations
\begin{equation}
    \begin{aligned}
     \text{empty: }(0,0)\rightarrow \ket{0}&, \qquad\,\,\,\,\,\,\,\,\,\,\,\,\,
\alpha\text{-occupied: }(1,0) \rightarrow \ket{\alpha},  \\
\beta\text{-occupied: }(0,1)\rightarrow \ket{\beta}&,  \qquad
\text{doubly occupied: }(1,1)\rightarrow \ket{2},   
    \end{aligned}
\end{equation}
such that there are $N_{occ} = \frac{\eta}{2}$ occupied orbitals in total for closed-shell systems we consider. Generalizations to open-shell systems are straightforward.

\subsubsection{Localized orbitals and fragment partitions}
Embedding methods typically utilize a form of spatially localized basis. We thus define localized molecular orbitals (LMO) as linear combinations of CMOs written as follows:
\begin{equation}
    \lvert \phi^{(L)}_j\rangle = \sum_{k=1}^{N} \mathbf{U}^{(L)}_{kj}\,\lvert \phi_k\rangle,
\end{equation}
where $\mathbf{U}^{(L)}\in\mathbb{R}^{N\times N}$ is a localization unitary matrix, and the associated localized orbitals set is given as
\begin{equation}
    B_{LMO}=\{\lvert\phi^{(L)}_j\rangle\mid j=1,2,\ldots,N\}.
\end{equation}
We note that CMOs are orthonormal, while AOs are not, making the overlap matrix $S^{AO}$ non-diagonal
\[
S^{\mathrm{AO}}_{\mu\nu}=\langle\chi_\mu\mid\chi_\nu\rangle \neq \delta_{\mu \nu},
\]
while the other overlap matrix $S^{MO}$ is
\[
S^{\mathrm{MO}}_{jk}=\langle\phi_j\mid\phi_k\rangle
=\sum_{\mu,\nu} C^*_{j\mu}\,C_{k\nu}\,S_{\mu\nu}
\;\Rightarrow\;
S^{\mathrm{MO}}=C\,S^{AO}\,C^{\dagger}=\mathbb{I}.
\]
For numerical convenience, orthogonal LMOs are recommended.

\subsubsection{System partition}
DMET requires system partitioning. For each possible partition $P_i$ of system $\mathcal{S}$, we define fragments $\mathcal F^{(P_i)}_{k_i}$ as subsets associated with that partition.  
The full set of possible partitions is denoted by
\begin{equation}
    \mathcal{P}=\{P_i\}_{i=1}^{N_P}.
\end{equation}
Each partition $P_i$ is composed of $K_i$ fragment-environment pairs $(\mathcal F^{(i)}_l,\mathcal{E}^{(i)}_l)$ covering system $\mathcal{S}$.
For a selected partition, we denote $\bigl(\mathcal F_l^{(P_i)},\mathcal{E}_l^{(P_i)}\bigr)\equiv(\mathcal F,\mathcal{E})$ and define the fragment-orbital set $
B_F=\{\lvert\phi^{(F)}_j\rangle\mid j\in F\}$ with size
$\lvert B_F\rvert=D_F,$
where each orbital $\lvert\phi^{(F)}\rangle=U^{(F)}\lvert\phi\rangle.$

Typical choices for $\mathbf{U}^{(F)}$ are:
\[
\begin{aligned}
&\text{(a) }\mathbf{U}^{(L)} &&\text{when localized orbitals are used,}\\
&\text{(b) }\mathbf{U}^{(F)}\mathbf{U}^{(L)} &&\text{when they are further transformed,}\\
&\text{(c) }\mathbf{U}^{(F)} &&\text{when localization and other unitary rotations are applied simultaneously.}
\end{aligned}
\]
The environment orbital set is defined as
\begin{equation}
    B_{\mathcal{E}}=\{\lvert\phi^{(E)}_j\rangle\mid j\notin \mathcal F\}.
\end{equation}
The central object utilized in our method is the orbital-localization matrix, which can be partitioned as follows
\begin{equation}
 U=
\begin{pmatrix}
U^{FO} & U^{FV}\\
U^{EO} & U^{EV}
\end{pmatrix},   
\end{equation}
with row blocks of sizes $D_F$ and $D_{E}$ respectively, and column blocks associated respectively with occupied ($N_{\mathrm{occ}}$) and virtual orbitals' ($N_{\mathrm{virt}}$) sectors. The total orbital dimension satisfies $D_F+D_{E} = N_{\mathrm{occ}} + N_{\mathrm{virt}} =N$.
For constructing bath orbitals for DMET and ECC-DMET, we define an auxiliary set of orbitals that are occupied and have non-zero overlap
with the selected fragment $\mathcal{F}$.
For this purpose, we first apply an SVD to the fragment-occupied block:
\begin{equation}
    \mathbf{U}^{FO}=\mathbf Q\,\mathbf \Sigma\,\mathbf V^{\dagger}.
    \label{eq:SVD-FO}
\end{equation}
where $\mathbf{Q}\in\mathbb{R}^{D_F\times D_F}$ , $\mathbf{\Sigma}\in\mathbb{R}^{D_F\times D_F}$ and $\mathbf{V^{\dagger}}\in\mathbb{R}^{D_F\times N_{occ}}$.

From the right singular vectors we form the \textit{occupied-orbital fragmentation rotation matrix} defined as follows
\begin{equation}
  \mathbf W=[\mathbf V\;\mathbf V_\perp],  
\end{equation}
with $\mathbf W\in\mathbb{R}^{N_{\mathrm{occ}}\times N_{\mathrm{occ}}}$ and
\begin{equation}
    \mathbf V_\perp^{\dagger}\mathbf V=\mathbf V\mathbf V_\perp^{\dagger}=\mathbf 0,
\end{equation}
so that columns in $\mathbf V_\perp$ are orthonormal and orthogonal to those in $\mathbf V$.

The corresponding transformed occupied-CMO creation operators are
\[
\tilde a_p^{\dagger}=\sum_{m=1}^{N_{\mathrm{occ}}}\mathbf W_{mp}\,a_m^{\dagger},
\qquad
a_m^{\dagger}\lvert0\rangle=\lvert\phi_m\rangle,
\qquad
p=1,\ldots,N_{\mathrm{occ}}.
\]
This transformation divides occupied CMOs into two sets:
\[
B_{OF}=\{\tilde a_p^{\dagger}\ket{0}\mid p=1,\ldots,D_F\},
\]
occupied CMOs with support on fragment $\mathcal F$, which we shall call $W$-orbitals,
and
\[
B_{OE}=\{\tilde a_p^{\dagger}\ket{0}\mid p=D_F+1,\ldots,N_{\mathrm{occ}}\},
\]
occupied CMOs with no overlap on fragment $\mathcal F$.
The occupied orbitals are related to fragment orbitals $d_k^{\dagger}$ through the following equation:
\begin{equation}
    a_m^{\dagger}=\sum_{k=1}^{N}U_{km}\,d_k^{\dagger}.
\end{equation}
Then
\begin{equation}
    \begin{aligned}
\tilde a_p^{\dagger}
&=\sum_{m=1}^{N_{\mathrm{occ}}}W_{mp}\sum_{k=1}^{N}U_{km}\,d_k^{\dagger} \\
&=\sum_{k=1}^{N}\left(\sum_{m=1}^{N_{\mathrm{occ}}}U_{km}W_{mp}\right)d_k^{\dagger}
=\sum_{k=1}^{N}G_{kp}\,d_k^{\dagger}.
\end{aligned}
\end{equation}

Hence
\begin{equation}
 \mathbf G=\mathbf U \mathbf W,
\qquad
\tilde a^{\dagger}=d^{\dagger} \mathbf G.   
\end{equation}

In the $W$-orbital basis the ground-state Slater determinant factorizes as follows:
\begin{equation}
  \lvert\Psi_0\rangle
=\prod_{p=1}^{D_F}\tilde a_p^{\dagger}
\prod_{p'=D_F+1}^{N_{\mathrm{occ}}}\tilde a_{p'}^{\dagger}\lvert0\rangle.  
\end{equation}
Only $D_F$ rotated occupied orbitals have non-zero overlap with the fragment; therefore, occupied orbitals with zero fragment overlap are not used to build bath orbitals.
In summary, the three relevant transformations are
\[
a^{\dagger}=d^{\dagger}U,
\qquad
\tilde a^{\dagger}=a^{\dagger}W,
\qquad
\tilde a^{\dagger}=d^{\dagger}G,
\]
all of which are $U(N)$-type rotations, and with an accompanying operator map
\[
a_m^{\dagger}\xrightarrow{\;\mathbf W'\;}\tilde a_p^{\dagger},
\qquad
d_k^{\dagger} \xrightarrow{\;\mathbf U\;}\ a_m^{\dagger},
\qquad
d_k^{\dagger}\xrightarrow{\;\mathbf G\;}\tilde a_p^{\dagger}.
\]

\subsubsection{Bath orbitals}
Bath orbitals in standard DMET can be chosen in at least three equivalent ways: through diagonalization of the environment block of 1-RDM, through SVD of the fragment-environment block of 1-RDM, or by eigendecomposition of overlap matrix for localized orbitals. 

\paragraph{Bath orbitals from $\mathbf{G}^{EF}$.}
Bath orbitals can be constructed from the $\mathbf{G}^{EF}$ block of $\mathbf{G}$:
\begin{equation}
 \mathbf G=
\begin{pmatrix}
\mathbf G^{FF} & \mathbf 0 & \mathbf G^{FV}\\
\mathbf G^{EF} & \mathbf G^{EF'} & \mathbf G^{EV}
\end{pmatrix},
\qquad
\mathbf G^{EF'}=\mathbf U^{EO}V_{\perp}, \;\; \\
\mathbf G^{EF}=\mathbf U^{EO}\mathbf W\big|_{D_F},
\end{equation}
where $\mathbf G^{EF}$ is a $D_E\times D_F$ matrix. Accordingly, bath orbitals are defined through projection of $W$-orbitals onto the environment:
\begin{equation}
b_p^{\dagger}=\hat P_{E}\,\tilde a_p^{\dagger},
\,\,\, \text{with} \,\,\,\,
\hat P_{E}=\sum_{j=D_F+1}^{D}\lvert\phi_j^{(L)}\rangle\langle\phi_j^{(L)}\rvert,
\end{equation}
and
\begin{equation}
    b_p^{\dagger}=\alpha\sum_{q=1}^{D_E}G^{EF}_{qp}a_q^{\dagger},
\,\,\, \text{where} \,\,\,\,
p=1,2,\ldots,D_F.
\end{equation}

From normalization we get
\[
\lVert G_p^{FF}\rVert=\sigma_p
\quad\Rightarrow\quad
\lVert G_p^{EF}\rVert=\sqrt{1-\sigma_p^2},
\]
so
$
\alpha=\frac{1}{\sqrt{1-\sigma_p^2}}.
$
Hence,
\[
b_p^{\dagger}
=\frac{1}{\sqrt{1-\sigma_p^2}}
\sum_{q=1}^{D_E}(\mathbf  U^{EO}\mathbf W)_{qp}\,a_q^{\dagger},
\,\,\, \text{where} \,\,\,\,
p=1,\ldots,D_F.
\]
In the above equation, $\sigma_p$ are elements of the diagonal matrix $\mathbf{\Sigma}$ defined in eq.~\eqref{eq:SVD-FO}.

\paragraph{Bath orbitals from occupied-overlap matrix.}
An alternative way of getting bath orbitals is through the diagonalization of the overlap matrix between the $W$-orbitals and localized-orbitals spaces:
\begin{equation}
    \mathbf S^{WL}=\langle\tilde\phi\mid\phi^{(L)}\rangle.
\end{equation}
For occupied index $p$ and localized-fragment index $k$ we can write
\begin{equation}    
\begin{aligned}
S^{WL}_{pk}
&=\langle\tilde\phi_p\mid\phi_k^{(L)}\rangle
=\sum_{q=1}^{N}G_{qp}\,\langle0|\, d_q\,d_k^{\dagger}\,|0\rangle \\
&=\sum_{q=1}^{N}G_{qp}\,\delta_{qk}
=G_{kp}
=\begin{cases}
\alpha\,\sigma_{kp}\,V_{pk}, & p\le D_F,\\
0, & p>D_F,
\end{cases}
\end{aligned}
\end{equation}
so only $D_F$ rotated occupied orbitals have non-zero overlap with fragment orbitals.
Also,
\[
S^{OO}_{mn}=\langle\tilde{\phi}_m\mid\tilde{\phi}_n\rangle,
\qquad
m,n\in I_{\mathrm{occ}}.
\]

Using the fragment block,
\[
\begin{aligned}
S^{OO}_{mn}
&=\sum_{\ell=1}^{D_F}\langle\phi_m^{(L)}\mid U^{FO\dagger}_{m\ell}U^{FO}_{\ell n}\mid\phi_n^{(L)}\rangle \\
&=\sum_{\ell=1}^{D_F}U^{FO\dagger}_{m\ell}U^{FO}_{\ell n}
=\sum_{\ell=1}^{D_F}(Q\Sigma V^{\dagger})^{\dagger}_{m\ell}(Q\Sigma V^{\dagger})_{\ell n} \\
&=\sum_{\ell=1}^{D_F}V_{m\ell}\,\Sigma_{\ell\ell}^{2}\,V^{\dagger}_{\ell n}
=(V\Sigma^2V^{\dagger})_{mn}.
\end{aligned}
\]

So
\[
\mathbf S^{OO}=\mathbf V\mathbf \Sigma^2\mathbf V^{\dagger}=(\mathbf U^{FO})^{\dagger}\mathbf U^{FO},
\]
which quantifies how much each occupied orbital overlaps the fragment-orbital subspace.

The eigendecomposition of $\mathbf S^{OO}$ yields $\mathbf V$ and $\mathbf \Sigma$, from which bath orbitals are constructed:
\begin{equation}
  b_p^{\dagger}
=\frac{1}{\sqrt{1-\sigma_p^2}}
\sum_{q=1}^{D_E}\sum_{m=1}^{N_{\mathrm{occ}}}U^{EO}_{qm}V_{mp}\,d_q^{\dagger},
\,\,\, \text{for} \,\,\,\,
p=1,2,\ldots,D_F.  
\end{equation}

Then the DMET state is
\begin{equation}
    \lvert\Psi\rangle=\sum_{k=1}^{D_F}\sigma_k\,\lvert\phi_k^{(L)}\rangle\otimes\lvert b_k\rangle,
\end{equation}
with $b_k^{\dagger}\lvert0\rangle=\lvert b_k\rangle$.

\paragraph{Bath orbitals from environment block of the 1-RDM.}
Finally, the most popular way of obtaining the DMET bath is through diagonalization of the environment block of 1-RDM.
Using the block form for the 1-RDM
\begin{equation}
    \mathbf \gamma=
\begin{pmatrix}
\mathbf \gamma^{FF} & \mathbf \gamma^{FE}\\
\mathbf \gamma^{EF} & \mathbf \gamma^{EE}
\end{pmatrix},
\end{equation}
the environment block can be expressed by the block of the unitary transformation that corresponds to initial occupied orbitals
\[
\mathbf \gamma^{EE}=\mathbf U^{EO}(\mathbf U^{EO})^{\dagger}.
\]
With the following identities:
\[
\mathbf S^{OO}=(\mathbf U^{FO})^{\dagger}\mathbf U^{FO},
\qquad
\mathbf \gamma^{FF}=\mathbf U^{FO}(\mathbf U^{FO})^{\dagger},
\]
and
\[
(\mathbf U^{EO})^{\dagger}(\mathbf U^{EO})+(\mathbf U^{FO})^{\dagger}\mathbf U^{FO}=\mathbb{I}.
\]
it is straightforward to observe that this method of finding bath orbitals is equivalent to the other two ways.

\paragraph{Core and virtual environment orbitals}
Core and virtual environment orbitals are given by
\[
\varepsilon_p^{\dagger}=\sum_{q=1}^{D-2D_F}U^{EV}_{qp}\,d_q^{\dagger}.
\]
In compact notation,
\begin{equation}
    \varepsilon^{\dagger}
=d^{\dagger}\mathbf U^{EV},
\end{equation}
which spans a $(N-2D_F)$-orbital space.

\paragraph{Why are there $D_F$ bath orbitals in standard DMET?}
Following MacDonald's theorem~\cite{MacDonald_1933}, if the one-particle density matrix is an idempotent projector, then when restricted to a $D_F$-dimensional fragment subspace,
at most $D_F$ eigenvalues can lie in $(0,1)$.
These partially occupied modes are precisely the bath orbitals, and at most $D_F$ bath orbitals are needed.
The interpretation within the standard version of DMET is that we only need to provide $D_F$ orbitals and eigenvalues (information) to reproduce the full state of the system, with mean field reference.

\subsection{ECC-DMET}
\label{app:ECC-DMET-theory-details}
While in standard DMET the reference state is typically a mean-field wavefunction, fragment orbitals are chosen manually, and the bath is constructed by diagonalizing the environment block of the one-body reduced density matrix (1-RDM), our proposal differs in several important aspects. We aim to construct an approximate electronic Hamiltonian of the system, for which the electronic Schr\"odinger equation is solved while incorporating the effects of the solvent and protein environment:
\begin{equation}
\hat H\lvert\Psi\rangle = E\lvert\Psi\rangle .
\end{equation}
Within ECC-DMET, the Schr\"odinger equation is solved for an embedded electronic Hamiltonian:
\begin{equation}
\hat H_{\mathrm{emb}}\lvert\Psi\rangle = E\lvert\Psi\rangle,
\qquad
\hat H_{\mathrm{emb}} = \hat P\hat H\hat P ,
\label{eq:projection}
\end{equation}
where $
\hat P
= \sum_{j=1}^{D_F} \lvert f_j\rangle\langle f_j\rvert
\otimes
\sum_{\ell=1}^{D_B} \lvert \tilde e_\ell\rangle\langle \tilde e_\ell\rvert
$ is a projector onto a $(D_F + D_B)$ orbitals of the fragment ${|f_j\rangle}$ and the bath ${|\tilde e_\ell\rangle}$. 
Formally, DMET partitions the $N$-dimensional Hilbert space of the quantum system into two parts: a fragment and its environment. The fragment corresponds to a small subspace of dimension $D_F \ll N$, while the remaining degrees of freedom constitute the environment with dimension $D_E = N - D_F$. The exact ground state of the full Hamiltonian $\hat H$ can be written as $|\Psi\rangle = \sum_{l=1}^{D_F}\sum_{k=1}^{D_E}\Psi_{lk}|f_l\rangle \otimes |e_k\rangle.$ Applying the Schmidt decomposition to the coefficient matrix $\Psi_{lk}$ yields $|\Psi\rangle = \sum_{l=1}^{D_F}\lambda_l |f_l\rangle \otimes |b_l\rangle,$
where $|f_l\rangle$ are fragment states and $|b_l\rangle$ are bath states that encode the entanglement between the fragment and the environment. Although the environment Hilbert space may be extremely large, the Schmidt decomposition demonstrates that only $D_F$ bath states are required to represent the entanglement with the fragment exactly. Consequently, the electronic structure of the entire system can, in principle, be reproduced exactly by solving a much smaller embedded problem consisting only of the fragment and its associated bath. Therefore, if the exact reference state were known, the Hamiltonian projected onto the exact fragment and bath subspaces, as defined above, would have the ground state energy as the full electronic Hamiltonian.

\paragraph{Correlated Reference State.}
In practice, the exact ground state required for the Schmidt decomposition is not available. Therefore, practical DMET implementations approximate the bath construction using a mean-field reference wavefunction, typically obtained from a Hartree–Fock calculation. For a closed-shell system with $\eta$ electrons in $N$ spatial orbitals, the mean-field reference can be expressed using spin-orbital creation operators acting on the vacuum state:

\begin{equation}
\lvert \Psi_0\rangle = \prod_{j\sigma \in I_{\mathrm{occ}}} a_{j\sigma}^{\dagger},\lvert 0\rangle,
\end{equation}
where $a_{j\sigma}^{\dagger}\lvert 0\rangle = \lvert \phi_{j\sigma}\rangle$ with $j=1,\ldots,N$ and $\sigma\in{0,1}$ define the Hartree–Fock CMOs $\lvert \phi_{j\sigma}\rangle$. The mean-field ground state can thus be written in the occupation-number representation as $\lvert 2_1 2_2 2_3 \dots 2_{\eta/2} 0_{\eta/2+1} \dots 0_{N}\rangle$.
In place of the standard mean-field reference we adopt a correlated reference state for DMET (correlated-DMET, or C-DMET), written in the second-quantized form as
\begin{equation}
\lvert\Psi_0\rangle
=\sum_{n_1,\ldots,n_N}\psi_{n_1\cdots n_N}\lvert n_1\cdots n_N\rangle,
\end{equation}
which in our case is obtained from DMRG, CCSD, or MP2 calculations. For the DMRG calculations, which can be highly accurate, an appropriate method for orbital ordering at the DMRG sites is required. We discuss the details of this choice in Appendix section~\ref{app:DMRG}.
For a correlated reference state $\lvert\Psi_0\rangle$, the most straightforward route to bath orbital construction is a Schmidt decomposition onto fragment orbitals:
\begin{equation}
    \lvert\Psi_0\rangle
=\sum_j \lambda_j\,\lvert \phi_j^{(L)}\rangle\otimes\lvert b_j\rangle,
\label{eq:corr-ref}
\end{equation}
possibly obtained via diagonalization of the corresponding $\gamma^{EE}$ block of the 1-RDM. In such a case, there are no longer only $D_F$ bath orbitals. The multireference character of the wavefunction causes the 1-RDM to be no longer idempotent, its spectrum is more diffuse than for the mean-field case, breaking the assumptions of MacDonald's theorem. For this reason, in the case of correlated reference, a new method for choosing bath in the correlated case is needed. We propose below a procedure for selecting fragment orbitals $F_i$ and the associated bath for a given system partitioning $P_i$.

\subsection{Orbital localization for QIO orbitals}
\label{appendix:localization}
\begin{figure}[H]
    \centering
    \includegraphics[width=0.95\linewidth]{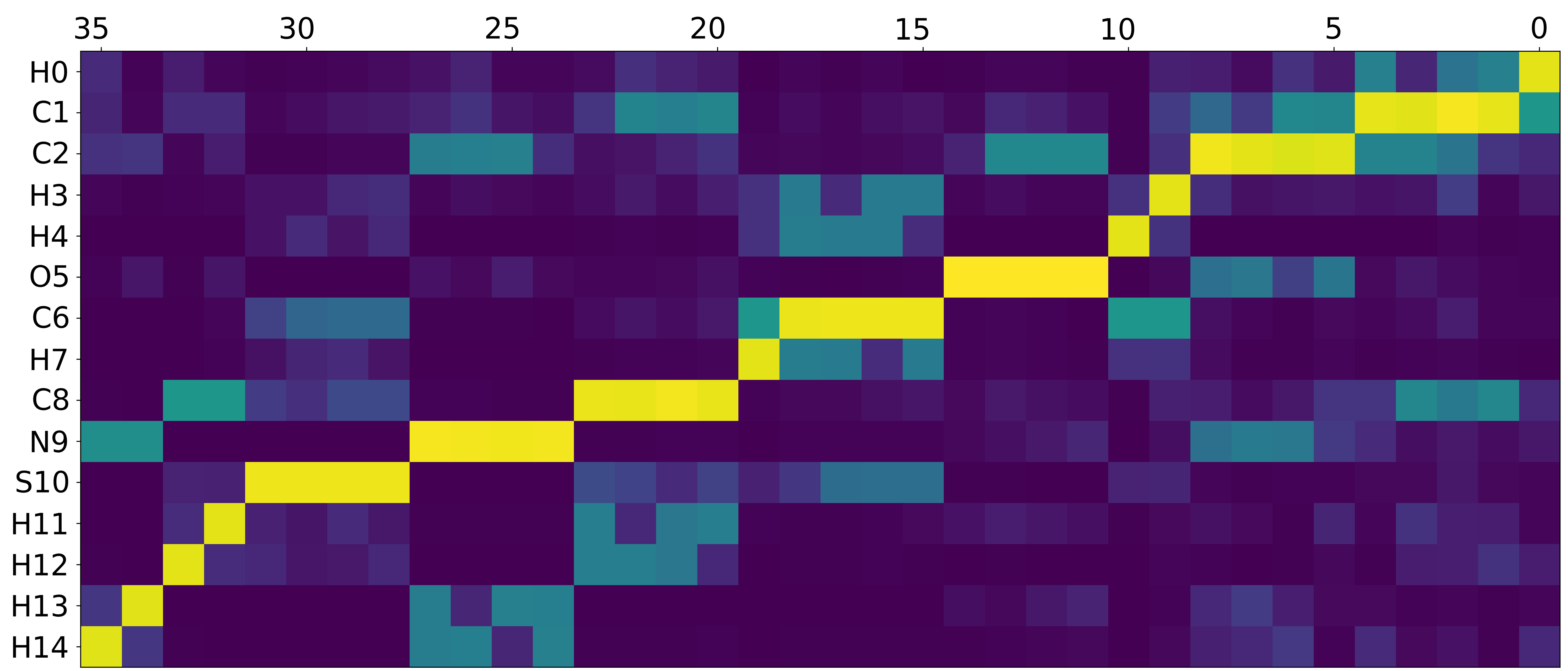}
    \caption{The support of given localized orbitals (columns) with atoms, more concretely, with their atomic orbitals in STO-3G basis (rows), excluding the core orbitals.
    The ordering of the atoms is arbitrary, and the geometry of the system is the transition state.
    We have used the smallest basis to visualize the general idea; for the results, we were using aug-cc-pVDZ basis.
    }
    \label{fig:atom_affinity}
\end{figure}

\begin{figure}[H]
    \centering
    \includegraphics[width=0.5\linewidth]{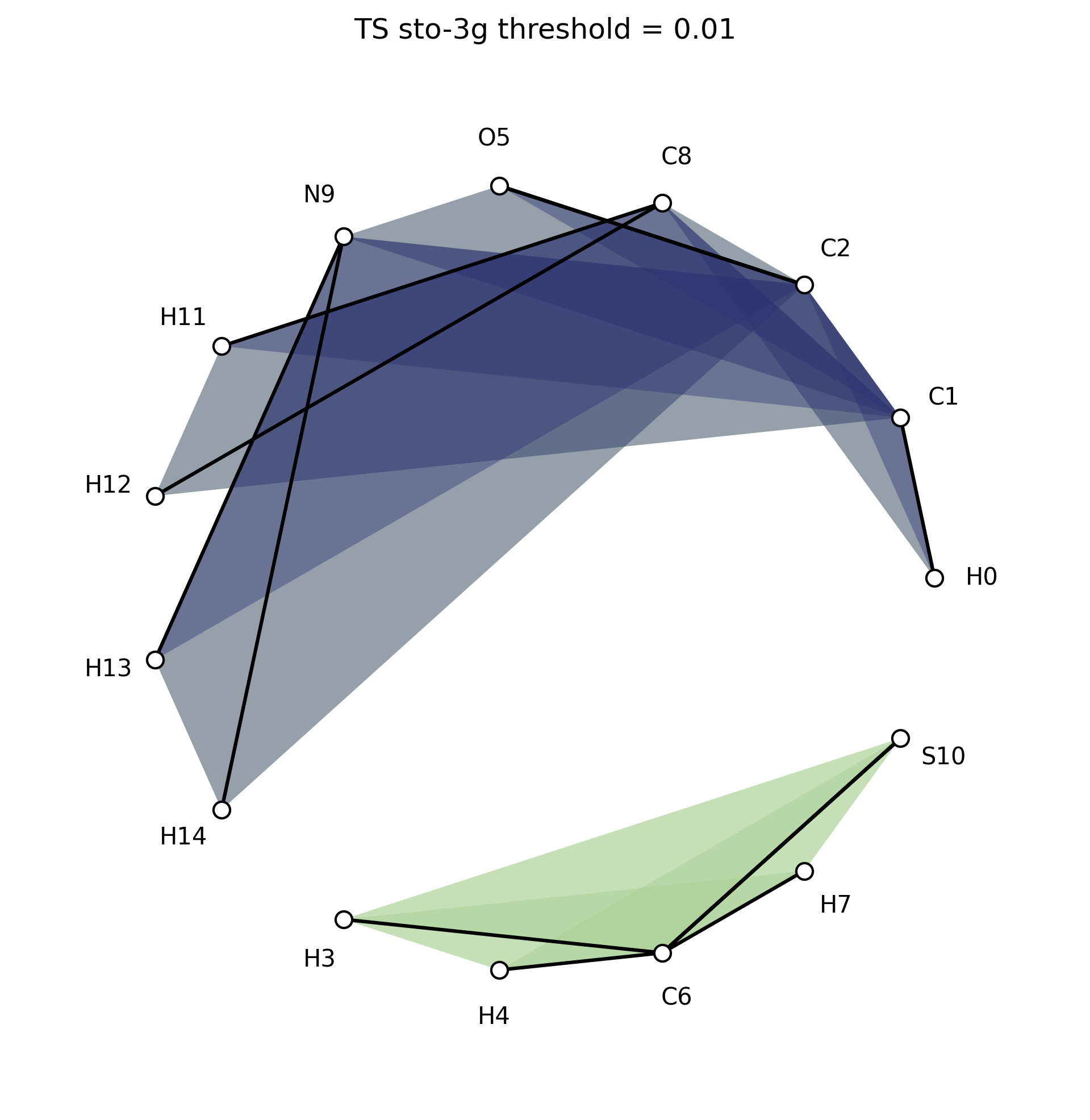}
    \caption{A hypergraph visualizing the connections between the atoms -- for each nontrivial orbital, we draw a hyperedge with the vertices (atom) that are above the threshold. 
    In other words, this corresponds to selecting the columns from Fig.~\ref{fig:atom_affinity} that have more than 1 element above the threshold.
    A hyperedge with two vertices is depicted as a black line.
    The system is the same as in previous figure.
    }
    \label{fig:atom_affinity_hypergraph}
\end{figure}

For the case study of zanubrutinib+BTK, we focused on the acrylamide-cysteine fragment of the whole system. For this system, we have divided the set of these localized orbitals into disjoint equivalence classes, where two orbitals $\ket{\phi^{(L)}_i} \sim \ket{\phi^{(L)}_j}$ iff two atoms that maximally support orbital $i$ and $j$ are the same. 
By support on a given atom, we mean that the absolute value squared
\begin{equation}
    \sum_{k \in \text{AO}}|C_{ki}|^2
\end{equation}
of the weights orbital decomposition in terms of the atomic orbital (AO) basis $\ket{\phi^{(L)}_i} = \sum_{k}C_{ki} \ket{\chi_k}$.
For the exemplary visualization of the atom support, see Figs.~\ref{fig:atom_affinity} and~\ref{fig:atom_affinity_hypergraph}.

Subsequently, we explored the unitaries of the block-diagonal form $U=\bigoplus_i U_i$, where each $U_i$ acts within a single equivalence class.
As a result, unitary $U$ will not change the localization property of the orbitals.

\paragraph{Single-orbital entropy}
For choosing fragment orbitals at the single-orbital level, it is convenient to use single-orbital entropy~\cite{Boguslawski_2013, Boguslawski_2014}. The single-orbital reduced density matrix $\rho^{(1)}_i$ is a $4\times 4$ diagonal matrix with the following elements:
\begin{equation}
p_3=k_i,
\qquad
p_1 = p_2 =\langle n_i\rangle-2k_i,
\qquad
p_0=1-\langle n_i\rangle+k_i,
\label{eq:1-orbital_rdm}
\end{equation}
where $n_{i\alpha} = a^{\dagger}_{i\alpha}a_{i\alpha}$, $k_i=\langle n_{i\alpha} n_{i\beta}\rangle$, and $n_i = n_{i\alpha} + n_{i\beta}$. 
The 1-orbital entropy can be then calculated as
\[
S_i=-p_0\log p_0-p_2\log p_2-2\Bigl(\frac{p_1}{2}\Bigr)\log\Bigl(\frac{p_1}{2}\Bigr).
\]

At the single-orbital level, fragment orbitals can be chosen by a combination of spatial localization in the active center supported by 1-orbital entropy, with large entropy indicating strong correlation, supporting including the orbitals in the active embedded space.

\paragraph{Two-particle entropy and mutual information}
The two-orbital entropy can be expressed as
\begin{equation}
   S_{ij}=-\mathrm{Tr}\bigl(\rho^{(2)}_{ij}\log \rho^{(2)}_{ij}\bigr),
\end{equation}
with reduced two-orbital density operator
\[
\rho^{(2)}_{ij}=\mathrm{Tr}_{\ell\ne(i,j)}\hat\rho,
\qquad
\hat\rho=\lvert\Psi_0\rangle\langle\Psi_0\rvert.
\]

The mutual information, measuring the correlations between a pair of orbitals $i$ and $j$, is written as
\begin{equation}
    I_{ij}=S_i+S_j-S_{ij}.
\end{equation}

The exact $16 \times 16 $ matrix can be written as~\cite{Boguslawski_2014}
\begin{equation}
\rho_{ij}^{(2)} =
\scalebox{0.7}{$\displaystyle
\begin{pmatrix}
\rho_{1,1} & 0 & 0 & 0 & 0 & 0 & 0 & 0 & 0 & 0 & 0 & 0 & 0 & 0 & 0 & 0 \\
0 & \rho_{2,2} & \rho_{2,3} & 0 & 0 & 0 & 0 & 0 & 0 & 0 & 0 & 0 & 0 & 0 & 0 & 0 \\
0 & \rho_{3,2} & \rho_{3,3} & 0 & 0 & 0 & 0 & 0 & 0 & 0 & 0 & 0 & 0 & 0 & 0 & 0 \\
0 & 0 & 0 & \rho_{4,4} & \rho_{4,5} & 0 & 0 & 0 & 0 & 0 & 0 & 0 & 0 & 0 & 0 & 0 \\
0 & 0 & 0 & \rho_{5,4} & \rho_{5,5} & 0 & 0 & 0 & 0 & 0 & 0 & 0 & 0 & 0 & 0 & 0 \\
0 & 0 & 0 & 0 & 0 & \rho_{6,6} & 0 & 0 & 0 & 0 & 0 & 0 & 0 & 0 & 0 & 0 \\
0 & 0 & 0 & 0 & 0 & 0 & \rho_{7,7} & 0 & 0 & 0 & 0 & 0 & 0 & 0 & 0 & 0 \\
0 & 0 & 0 & 0 & 0 & 0 & 0 & \rho_{8,8} & \rho_{8,9} & \rho_{8,10} & \rho_{8,11} & 0 & 0 & 0 & 0 & 0 \\
0 & 0 & 0 & 0 & 0 & 0 & 0 & \rho_{9,8} & \rho_{9,9} & \rho_{9,10} & \rho_{9,11} & 0 & 0 & 0 & 0 & 0 \\
0 & 0 & 0 & 0 & 0 & 0 & 0 & \rho_{10,8} & \rho_{10,9} & \rho_{10,10} & \rho_{10,11} & 0 & 0 & 0 & 0 & 0 \\
0 & 0 & 0 & 0 & 0 & 0 & 0 & \rho_{11,8} & \rho_{11,9} & \rho_{11,10} & \rho_{11,11} & 0 & 0 & 0 & 0 & 0 \\
0 & 0 & 0 & 0 & 0 & 0 & 0 & 0 & 0 & 0 & 0 & \rho_{12,12} & \rho_{12,13} & 0 & 0 & 0 \\
0 & 0 & 0 & 0 & 0 & 0 & 0 & 0 & 0 & 0 & 0 & \rho_{13,12} & \rho_{13,13} & 0 & 0 & 0 \\
0 & 0 & 0 & 0 & 0 & 0 & 0 & 0 & 0 & 0 & 0 & 0 & 0 & \rho_{14,14} & \rho_{14,15} & 0 \\
0 & 0 & 0 & 0 & 0 & 0 & 0 & 0 & 0 & 0 & 0 & 0 & 0 & \rho_{15,14} & \rho_{15,15} & 0 \\
0 & 0 & 0 & 0 & 0 & 0 & 0 & 0 & 0 & 0 & 0 & 0 & 0 & 0 & 0 & \rho_{16,16}
\end{pmatrix}$}.
\label{eq:2-orbital_rdm}
\end{equation}

The nonzero matrix elements of $\rho_{kl} \coloneqq \big[\rho_{ij}^{(2)}\big]_{kl}$, i.e., $k^\text{th}$ row and $l^\text{th}$ column, are
\begin{equation}
\begin{aligned}
&\rho_{1,1}
=
1 - 2\gamma_{ii} - 2\gamma_{jj}
+ \gamma_{ii}^{2} + \gamma_{jj}^{2} - 2\Gamma_{ijij}
+ 2\gamma_{ii}\gamma_{jj}
+ 2\gamma_{ii}\Gamma_{ijij} + 2\gamma_{jj}\Gamma_{ijij}
+ \Gamma_{ijij}^{2}, \\[0.3em]
&\rho_{2,2}
=
\gamma_{jj} + \Gamma_{ijij} - \gamma_{ii}\gamma_{jj} - \gamma_{jj}^{2}
- \gamma_{ii}\Gamma_{ijij} - 2\gamma_{jj}\Gamma_{ijij}
- \Gamma_{ijij}^{2}, \\[0.3em]
&\rho_{3,3}
=
\gamma_{ii} - \gamma_{ii}^{2} - \gamma_{ii}\gamma_{jj}
+ \Gamma_{ijij}
- 2\gamma_{ii}\Gamma_{ijij} - \gamma_{jj}\Gamma_{ijij}
- \Gamma_{ijij}^{2}, \\[0.3em]
&\rho_{4,4}
=
\gamma_{jj} - \gamma_{ii}\gamma_{jj} + \Gamma_{ijij} - \gamma_{jj}^{2}
- \gamma_{ii}\Gamma_{ijij} - 2\gamma_{jj}\Gamma_{ijij}
- \Gamma_{ijij}^{2}, \\[0.3em]
&\rho_{5,5}
=
\gamma_{ii} - \gamma_{ii}^{2} - \gamma_{ii}\gamma_{jj}
+ \Gamma_{ijij}
- 2\gamma_{ii}\Gamma_{ijij} - \gamma_{jj}\Gamma_{ijij}
- \Gamma_{ijij}^{2}, \\[0.3em]
&\rho_{6,6} = \rho_{7,7}
=
-\Gamma_{ijij} + \gamma_{ii}\Gamma_{ijij} + \gamma_{jj}\Gamma_{ijij}
+ \Gamma_{ijij}^{2}, \\[0.3em]
&\rho_{8,8}
=
\gamma_{jj}^{2} + 2\gamma_{jj}\Gamma_{ijij} + \Gamma_{ijij}^{2}, \\[0.3em]
&\rho_{9,9} = \rho_{10,10}
=
\gamma_{ii}\gamma_{jj} + \gamma_{ii}\Gamma_{ijij} + \gamma_{jj}\Gamma_{ijij}
+ \Gamma_{ijij}^{2}, \\[0.3em]
&\rho_{11,11}
=
\gamma_{ii}^{2} + 2\gamma_{ii}\Gamma_{ijij} + \Gamma_{ijij}^{2}, \\[0.3em]
&\rho_{12,12} = \rho_{14,14}
=
-\gamma_{jj}\Gamma_{ijij} - \Gamma_{ijij}^{2}, \\[0.3em]
&\rho_{13,13} = \rho_{15,15}
=
-\gamma_{ii}\Gamma_{ijij} - \Gamma_{ijij}^{2}, \\[0.3em]
&\rho_{16,16} = \Gamma_{ijij}^{2}, \\[0.7em]
&\rho_{2,3} = \rho_{3,2}
=
\gamma_{ij} - \gamma_{ij}\gamma_{ii} - \gamma_{ij}\gamma_{jj}
- \gamma_{ij}\Gamma_{ijij}, \\[0.3em]
&\rho_{4,5} = \rho_{5,4}
=
\gamma_{ij} - \gamma_{ij}\gamma_{ii} - \gamma_{ij}\gamma_{jj}
- \gamma_{ij}\Gamma_{ijij}, \\[0.3em]
&\rho_{8,9} = \rho_{9,8}
=
\gamma_{ij}\gamma_{jj} + \gamma_{ij}\Gamma_{ijij}, \\[0.3em]
&\rho_{8,10} = \rho_{10,8}
=
-\gamma_{ij}\gamma_{jj} - \gamma_{ij}\Gamma_{ijij}, \\[0.3em]
&\rho_{8,11} = \rho_{11,8}
=
\gamma_{ij}^{2}, \\[0.3em]
&\rho_{9,10} = \rho_{10,9}
=
-\gamma_{ij}^{2}, \\[0.3em]
&\rho_{9,11} = \rho_{11,9}
=
\gamma_{ij}\gamma_{ii} + \gamma_{ij}\Gamma_{ijij}, \\[0.3em]
&\rho_{10,11} = \rho_{11,10}
=
-\gamma_{ij}\gamma_{ii} - \gamma_{ij}\Gamma_{ijij}, \\[0.3em]
&\rho_{12,13} = \rho_{13,12}
=
\gamma_{ij}\Gamma_{ijij}, \\[0.3em]
&\rho_{14,15} = \rho_{15,14}
=
\gamma_{ij}\Gamma_{ijij}, \\[0.7em]
\end{aligned}
\end{equation}
and all remaining elements of $\rho_{ij}^{(2)}$ vanish.

\paragraph{Approximation to mutual information.}
\label{sec:MI_approximation}
Above, we have assumed that the mutual information of the reference post-HF state is accessible and can be used as a guide for the choice of the fragment orbitals. 
This can be very costly in practice -- to obtain the orbital entropies, we require up to 4-body RDMs~\cite{Boguslawski_2014}. For practical, virtual screening purposes, we instead relax this requirement by utilizing substantially less information about the post-HF state, while maintaining qualitative information about electron correlation. 
Various levels of approximations are accessible here; the most natural one is to use Wick's theorem to the post-HF state, i.e., to derive all the higher-body reduced density matrices from 1-body RDM.

To use the many-body physics language, we treat the post-HF state as a single-determinant state, describing it through the lowest moments possible. 
It is known~\cite{Marian_2013,Coffman_2025} that the state obtained with this approximation yields the closest single-determinant state to the original post-HF state. 
Nonetheless, in general, the closest state will be a mixture of single-determinant states. 

To determine the most correlated orbitals to include in the fragment, we have used the Wick's formula for the post-HF state to obtain the approximate 1- and 2-body orbital density matrices, $\rho^{(1)}_i$ and $\rho^{(2)}_{ij}$, respectively. 
Orbital density matrices are obtained by tracing out the rest of the orbitals.
These can be used to derive the mutual information between the $i^{\text{th}}$ and $j^{\text{th}}$ orbital using the following formula
\begin{equation}\label{eq:mutual_information}
    I_{ij} = S(\rho^{(1)}_i) + S(\rho^{(1)}_j) - S(\rho^{(2)}_{ij}),
\end{equation}
with $S$ denoting the von Neumann entropy, $S(\rho) = -\mathrm{Tr}\left(\rho\ln\rho\right)$.

In the course of our analysis, we have found that such an approximation is sufficient for the determination of the most entangled orbitals, as discussed in app.~\ref{app:approximation}.

\subsection{Wick's approximation.}\label{app:approximation}
As the total number of elements of the $k$-body RDM scales as $\mathcal{O}(N^{2k})$ with the number of orbitals $N$, we decided to use an approximation for the orbital reduced density matrices. 
In particular, we treat the state at our disposal as if it was single-determinant.
Then, by using the single orbital basis of $\mathcal{B}_i=\Bigl\{
|0\rangle,\ |\alpha\rangle,\ |\beta\rangle,\ |2\rangle
\Bigr\}$, we can express the single orbital density matrices from eq.~\eqref{eq:1-orbital_rdm} as
    \begin{equation}
        \rho_i^{(1)}=
\begin{pmatrix}
1-2\gamma_{ii}+\gamma_{ii}^2 & 0 & 0 & 0\\
0 & \gamma_{ii}-\gamma_{ii}^2 & 0 & 0\\
0 & 0 & \gamma_{ii}-\gamma_{ii}^2 & 0\\
0 & 0 & 0 & \gamma_{ii}^2
\end{pmatrix}.
    \end{equation}

In the same vein, the 2-orbital reduced density matrices given by  eq.~\eqref{eq:2-orbital_rdm} can be approximated via Wick's theorem: 2-RDM element are expressed as
\begin{equation}
    \Gamma_{ijij} \coloneqq  \langle a_{i}^\dagger a_{j}^\dagger a_{i} a_{j}\rangle
\approx \gamma_{i j}^{2} - \gamma_{i i }\gamma_{jj} ,
\end{equation}
with $i$ and $j$ denoting spin-orbitals.
Our approximation is valid with spin-0 wavefunction $\gamma_{ii} = \gamma_{jj}$ for $i$ and $j$ being the same spatial orbital with different spin.
This assumption is valid for the molecule at our disposal.
For spin-orbitals of different spin for the same spatial one, the above equation simplifies even more, $\Gamma_{ijij} \approx \gamma^2_{ii}$, under the assumption that the state is of single determinant, as then $\gamma_{ij}=0$.
Using this approach, we can approximately reconstruct the mutual information matrix $I_{ij}$ with much less resource than the exact matrix.

\paragraph{Example: selecting fragment orbitals from DMRG reference state.}
\label{app:DMRG}
An example procedure for selecting fragment orbitals with a correlated DMRG reference is discussed below. For DMRG-like references, the state is expressed in Matrix-Product State form, written as
\[
\lvert\Psi_0\rangle
=\sum_{n_1,\ldots,n_N}\psi_{n_1\cdots n_N}\,\lvert n_1\cdots n_N\rangle,
\]
with tensor-factorized coefficients
\[
\psi_{n_1\cdots n_N}
=\sum_{k_1,\ldots,k_{N-1}}
A^{(1)}_{k_1}(n_1)
A^{(2)}_{k_1k_2}(n_2)\cdots
A^{(N)}_{k_{N-1}}(n_N).
\]

Local occupation states are decomposed as
\[
\lvert n_i\rangle\in\{\lvert0\rangle,\lvert\alpha\rangle,\lvert\beta\rangle,\lvert2\rangle\}.
\]
We search for an orbital-permutation map:
\[
\pi(1,2,\ldots,N)=(\pi(1),\pi(2),\ldots,\pi(N)).
\]
that captures strong local correlation. The generic metric that such map may minimize is maximization of local overlap between orbitals, that is
\begin{equation}
     S_{ij} =\int dr |\phi_i(r)\phi_j(r)|-1,
\end{equation}
i.e., we are searching for the optimal permutation $\pi$
\[
\max_{\pi}\sum_i S_{i,i+1},
\]
with weakly connected pairs giving $S_{ij}\approx0$.
Our goal is to find such permutation that will keep the orbitals with large mutual information close.
We use graph-theoretic approach, which is a known technique for optimizing the DMRG orbital ordering~\cite{Rissler_2006,Barcza_2011}.
For finding the appropriate map, we construct a weighted graph $G_z$ with edge weights given by mutual information, $\omega_{ij}=I_{ij}$. 
We then choose the ordering $\pi$ to minimize
\begin{equation}
   C(\pi)=\sum_{i<j}\omega_{ij}\bigl(\mathrm{id}_{\pi}(i)-\mathrm{id}_{\pi}(j)\bigr)^2, 
\end{equation}
where $\mathrm{id}_{\pi}(u)$ is the position of orbital $u$ in ordering $\pi$.
After the start graph $G_{z_1}$ is chosen, we follow the following steps:
\begin{itemize}
    \item First, we prune the graph edges by the a filter function keeping only those edges that have weight above a given threshold value $\tau$:
\[
f_{\tau}(G_z)=G_{z,\tau},
\]
with only edges satisfying a threshold condition $\omega_{ij}<\tau$.
\item 
Next we build graph Laplacian
\begin{equation}
L=D-\Omega,
\qquad
D_{ii}=\sum_j\omega_{ij},
\qquad
\Omega_{ij}=\omega_{ij}.
\end{equation}
\item Compute Fiedler vector $\vec f$.
\item  Order orbitals by sorting components of $\vec f$; this defines permutation $\pi$.
\item Using $\pi$, divide the system into initial candidate sets for fragment $F$ and environment $\mathcal E$, using fragment-selection method described in sec.~\ref{sec:ECC-DMET}. Fragment orbitals are indexed by $i=1,\ldots,D_F$, with cut at $(D_F,D_F+1)$.
\item For the DMRG-reference one can perform Schmidt decomposition of $\lvert\Psi_0\rangle$ for the partition given by $\pi$ partition,
\[
\lvert\Psi_0\rangle=\sum_{\ell=1}^{M_A}\lambda_{\ell}\,\lvert f_{\ell}\rangle\otimes\lvert e_{\ell}\rangle.
\]
\end{itemize}

\section{Molecular dynamics parameters}
\label{app:md_details}
During the study of zanubrutinib, we used AmberTools with the following parameters:
\begin{itemize}
    \item The protein structure was processed using \texttt{MDAnalysis}~\cite{MichaudAgrawal2011,Gowers2016} to remove crystal water molecules, ions, and cofactors, and the \texttt{ff19SB} force field~\cite{Tian2020} was applied for protein parameterization. 
    
    \item The topology of the covalently bound zanubrutinib-CYS481 complex and the corresponding partial AM1-BCC atomic charges~\cite{jakalian2000fast,jakalian2002fast} were generated using the \texttt{antechamber} utility from the AmberTools suite~\cite{Case2023}. The modified residue was further processed with \texttt{prepgen} to ensure proper connectivity with neighboring protein residues. The cysteine residue (CYS481) was parameterized using the \texttt{ff14SB} protein force field~\cite{Maier2015}, while the zanubrutinib ligand was described using the \texttt{GAFF2} general force field~\cite{Wang2004}. Missing bonded parameters at the covalent interface between zanubrutinib and CYS481 were generated using \texttt{parmchk2}.
    
    \item The system was solvated using the OPC water model~\cite{Izadi2014} in a simulation box extending 10~\AA\ from the protein in all directions.
    
    \item \ce{Na+} and \ce{Cl-} counterions were added to achieve a salt concentration of approximately 0.15~M.
    
    \item The final solvated and ionized system was assembled and protonated at physiological pH using \texttt{tLEaP}~\cite{Case2023}. 
\end{itemize}

We then carried out a sequence of simulations:
\begin{itemize}
    \item Energy minimization for 50\,000 steps to remove steric clashes and relax the system.
    
    \item NVT equilibration for 250~ps, during which the system temperature was gradually increased to 300~K.
    
    \item NPT equilibration for 250~ps, allowing the system pressure to equilibrate at 1~bar.
    
    \item A production MD run of 500~ps.
\end{itemize}

\section{DMET barrier energies for acrylamide-methanethiol complex within PCM solvation model}\label{sec:dmet-barrier}

\begin{figure}[!ht]
    \centering
    \includegraphics[width=0.6\linewidth]{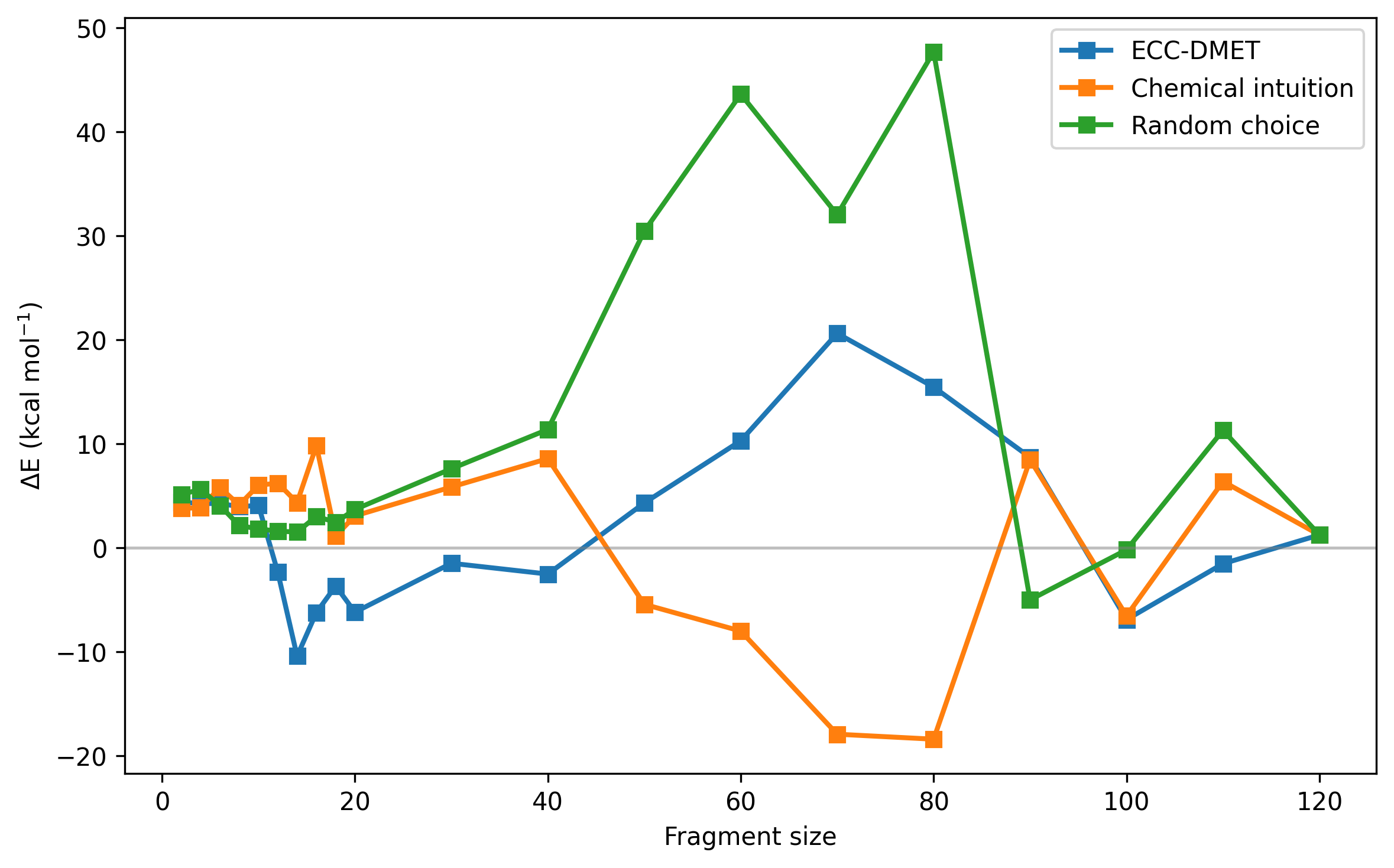}
    \caption{Barrier energy $\Delta E$ (kcal mol$^{-1}$) as a function of fragment size for three orbital-selection strategies: ECC-DMET (blue), chemical intuition (orange), and random choice (green). The horizontal gray line marks $\Delta E = 0$.}
    \label{fig:barrier_fragment_size}
\end{figure}

We reported the barrier energy as a function of fragment size for the three orbital-selection strategies considered in Fig.~\ref{fig:barrier_fragment_size}. For small orbital subspaces, all methods essentially recover the HF limit, and only limited differences are observed among them. In particular, the chemically selected orbitals closely follow the MI-based selection, whereas the random choice shows slightly larger deviations.

As the fragment size increases, the three curves start to diverge significantly. This behavior can be attributed to the loss of error cancellation between TS and pre-complex, due to the different effective electronic treatment of the two endpoints. As a consequence, the computed barrier becomes negative in several regions of the orbital space for all three strategies. A possible way to mitigate this issue would be to define the DMET fragment in a consistent manner for both endpoints, although this was not attempted in the present work.
Finally, in the large-fragment limit, all three approaches recover the same CCSD result, as expected when the full orbital space is included.

\end{document}